%% file: main.tex
\documentclass[11pt, preprintnumbers, superscriptaddress, nofootinbib]{revtex4}
\pdfoutput=1

\usepackage{ifthen}
\usepackage{amsmath,amssymb}
\usepackage{mathtools}
\numberwithin{equation}{section}
\usepackage{graphicx}
\usepackage{physics}
\usepackage{slashed}
\usepackage{color}

\makeatletter
\let\MYcaption\@makecaption
\makeatother

\usepackage{subcaption}

\makeatletter
\let\@makecaption\MYcaption
\makeatother

\begin{document}

\title{Next-to-leading-order corrections to the Higgs strahlung process\\ from electron--positron collisions in extended Higgs models}

\preprint{OU-HET-1099}

\author{Masashi Aiko}
\email{m-aikou@het.phys.sci.osaka-u.ac.jp}
\affiliation{Department of Physics, Osaka University, Toyonaka, Osaka 560-0043, Japan}

\author{Shinya Kanemura}
\email{kanemu@het.phys.sci.osaka-u.ac.jp}
\affiliation{Department of Physics, Osaka University, Toyonaka, Osaka 560-0043, Japan}

\author{Kentarou Mawatari}
\email{mawatari@iwate-u.ac.jp}
\affiliation{Faculty of Education, Iwate University, Morioka, Iwate 020-8550, Japan}

\begin{abstract}
We present the cross section for $e^{+}e^{-}\to hZ$ with arbitrary sets of electron and $Z$ boson polarizations at the full next-to-leading order in various extended Higgs models, such as the Higgs singlet model (HSM), the inert doublet model (IDM) and the two Higgs doublet model (2HDM).
We systematically perform complete one-loop calculations to the helicity amplitudes in the on-shell renormalization scheme, and present the full analytic results as well as numerical evaluations.
The deviation $\Delta R^{hZ}$ in the total cross section from its standard model (SM) prediction is comprehensively analyzed, and the differences among these models are discussed in details.
We find that new physics effects appearing in the renormalized $hZZ$ vertex almost govern the behavior of $\Delta R^{hZ}$, and it takes a negative value in most cases.
The possible size of $\Delta R^{hZ}$ reaches several percent under the theoretical and experimental bounds.
We also analyze the deviation $\Delta R^{hZ}_{XY}$ in the total cross section times decay branching ratios of the discovered Higgs boson by utilizing the \texttt{H-COUP} program.
It is found that the four types of 2HDMs can be discriminated by analyzing the correlation between $\Delta R^{hZ}_{\tau\tau}$ and $\Delta R^{hZ}_{bb}$ and those between $\Delta R^{hZ}_{\tau\tau}$ and $\Delta R^{hZ}_{cc}$.
Furthermore, the HSM and the IDM can be discriminated from the 2HDMs by measuring $\Delta R^{hZ}_{WW}$.
These signatures can be tested by precision measurements at future Higgs factories such as the International Linear Collider.
\end{abstract}

\maketitle
\newpage

\tableofcontents
\newpage

\input{01Introduction}
%
\input{02Models}
%
\input{03Process}
%
\input{04Numerical_results}
%
\input{05Conclusion}
%
\input{06Acknowledgements}
%
\appendix
\input{A01inputs}
\input{A02Loop_functions}

\bibliographystyle{junsrt}
\bibliography{refs}

\end{document}

%% file: 01Introduction.tex


\section{Introduction} \label{sec: introduction}
Since the discovery of the new particle with the mass of 125 GeV at the LHC in 2012 \cite{Aad:2012tfa, Chatrchyan:2012ufa}, it has turned out that its properties are in agreement with those of the Higgs boson in the standard model (SM) within theoretical and experimental uncertainties \cite{Aad:2019mbh, CMS:2020gsy}.
While no signal for new physics (NP) beyond the SM has been observed at the LHC up to now, there are phenomena that cannot be explained within the SM such as dark matter, baryon asymmetry of the universe and tiny neutrino masses. 
In addition to these phenomenological problems, there are conceptual problems in the SM such as the hierarchy problem, no unified description for the gauge group and the flavor structure and so on.
Therefore, the SM must be replaced by a more fundamental theory.

While the Higgs boson was found, the structure of the Higgs sector remains unknown.
There is no theoretical principle to insist on the minimal structure of the Higgs sector as introduced in the SM, and the possibility that the Higgs sector takes a non-minimal form is not excluded experimentally.
Furthermore, such non-minimal Higgs sectors are often introduced in various new physics models, where the above-mentioned problems are tried to be solved.
Therefore, unraveling the structure of the Higgs sector is one of the central interests of current and future high-energy physics, and the direction of new physics can be determined by reconstructing the Higgs sector experimentally.

The discovery of additional scalar bosons is a clear evidence of extended Higgs sectors, and enormous efforts have been devoted to discover such new particles in a wide variety of the search channels \cite{Aad:2020zxo, Aad:2019zwb, Aaboud:2018mjh, Aaboud:2018knk, Aaboud:2017gsl, Aad:2020fpj, Aad:2020tps, Aad:2020ncx, Aad:2021xzu, Sirunyan:2018taj, Sirunyan:2019wph, Sirunyan:2017uhk, Sirunyan:2017isc, Sirunyan:2019pqw, Sirunyan:2019xjg, Sirunyan:2020hwv, Sirunyan:2019hkq}.
However, no observation of such new particles has been reported, leading to constraints on parameters of extended Higgs models such as masses and coupling constants.
Direct searches of new particles are one of the key programs at the LHC as well as at the high-luminosity LHC (HL-LHC) \cite{ApollinariG.:2017ojx}.

In addition to the direct searches, extended Higgs sectors can be indirectly explored by measuring various properties of the discovered Higgs boson such as cross sections, the width and decay branching ratios because mixings of Higgs bosons and/or radiative corrections of additional Higgs bosons would modify them from their SM values.
If deviations from the SM are detected, the magnitude of deviations tells us upper limits on the mass scale of the second Higgs boson by taking into account theoretical consistencies \cite{Kanemura:2014bqa, Blasi:2017zel, Aiko:2020ksl}.
In addition, the pattern of deviations gives us information on the structure of the Higgs sector such as the representation of the weak isospin, the number of Higgs fields and the structure of Yukawa interactions \cite{Kanemura:2014bqa}.
Therefore, the discovered Higgs boson is a probe of new physics.

Precision measurements of the properties of the discovered Higgs boson are also one of the main targets at current and future collider experiments.
At the LHC, the Higgs boson couplings have been measured with typically order ten percent accuracy.
Most extended Higgs models can accommodate this SM-like situation at the lowest order of the perturbation.
Therefore, measurements with a few percent accuracies become more important where quantum corrections play an essential role, and it enables us to extract the signature of new physics.
Accuracy of the measurement of Higgs boson couplings is expected to be improved at the HL-LHC and further significantly at future lepton colliders; e.g., the International Linear Collider (ILC) \cite{Baer:2013cma, Fujii:2017vwa, Asai:2017pwp, Fujii:2019zll}, the Future Circular Collider (FCC-ee) \cite{Gomez-Ceballos:2013zzn} and the Circular Electron Positron Collider (CEPC) \cite{CEPC-SPPCStudyGroup:2015csa}.

In order to compare theoretical predictions with future precision measurements, theoretical calculations compatible with expected experimental accuracy are inevitable.
Radiative corrections to the SM-like Higgs boson vertices have been studied in various Higgs sectors such as the model with a real isospin singlet Higgs field (HSM) \cite{Bojarski:2015kra, Kanemura:2015fra, Kanemura:2016lkz, He:2016sqr}, two Higgs doublet models (2HDMs) \cite{Arhrib:2003ph, Kanemura:2004mg, Kanemura:2014dja, Kanemura:2015mxa, Castilla-Valdez:2015sng, Krause:2016oke, Arhrib:2016snv, Kanemura:2017wtm, Altenkamp:2017ldc, Altenkamp:2017kxk, Altenkamp:2018bcs, Gu:2017ckc, Chen:2018shg, Han:2020lta}, the inert doublet model (IDM) \cite{Arhrib:2015hoa, Kanemura:2016sos} and so on.
In order to see differences in the prediction among these models, it is quite important to calculate the renormalized SM-like Higgs boson vertices in a consistent and systematic way. 
The \texttt{H-COUP} program \cite{Kanemura:2017gbi, Kanemura:2019slf} enables us to evaluate the decay rates including higher-order corrections for the SM-like Higgs boson in the HSM, the IDM and the 2HDM with four types of Yukawa interactions classified under the softly-broken $Z_{2}$ symmetry \cite{Kanemura:2018yai, Kanemura:2019kjg}.
Also, other numerical tools to evaluate the decay of the Higgs boson with radiative corrections are available; e.g., \texttt{2HDECAY} \cite{Krause:2018wmo} and \texttt{Prophecy4f} \cite{Denner:2019fcr}.

In this paper, we present the cross section for $e^{+}e^{-}\to hZ$ with arbitrary sets of electron and $Z$ boson polarizations at the full next-to-leading order (NLO) in the HSM, the IDM and the 2HDMs.
At the future lepton colliders, not only the decay properties of the discovered Higgs boson but also its production cross section can be precisely measured.
Especially, $e^{+}e^{-}\to hZ$ is the dominant production process at the center-of-mass (CM) energy of 240--250 GeV,\footnote{For higher energies, the Higgs boson production via $W$ boson fusion is getting more important \cite{Denner:2003yg, Denner:2003iy, Belanger:2002ik}.} and precise calculations of its cross section in various extended Higgs models are quite important.
The NLO calculation has been performed in the SM \cite{Fleischer:1982af, Kniehl:1991hk, Denner:1992bc}, the IDM \cite{Abouabid:2020eik} and the 2HDM \cite{LopezVal:2010vk, Xie:2018yiv}.\footnote{In Refs.~\cite{Hempfling:1993ru, Driesen:1995ew, Heinemeyer:2001iy, Heinemeyer:2015qbu}, the NLO calculations for $e^{+}e^{-}\to hZ$ have been performed in the minimal supersymmetric standard model.}
We systematically perform complete NLO calculations to the helicity amplitudes in each extended Higgs model based on the on-shell renormalization scheme \cite{Bohm:1986rj, Hollik:1988ii, Kanemura:2004mg, Kanemura:2017wtm}, and the full analytic results, as well as numerical evaluations, are presented.
We comprehensively analyze the deviations in the cross section from the SM prediction in each model under the constraints of perturbative unitarity \cite{Cynolter:2004cq, Kanemura:2016lkz, Kanemura:1993hm, Akeroyd:2000wc, Ginzburg:2005dt, Kanemura:2015ska} and vacuum stability \cite{Pruna:2013bma, Deshpande:1977rw, Klimenko:1984qx, Sher:1988mj, Nie:1998yn, Kanemura:1999xf}, conditions to avoid wrong vacua \cite{Espinosa:2011ax, Chen:2014ask, Lewis:2017dme, Barroso:2013awa, Ginzburg:2010wa} and experimental constraints.
We discuss the differences in the predictions of the cross section among these models in detail.
We also show correlations of the deviation in the cross section times the decay branching ratios of the SM-like Higgs boson from the SM predictions and discuss the discrimination of the extended Higgs models.

This paper is organized as follows.
In Sec.~\ref{sec: models} we briefly introduce the HSM, the THDMs and the IDM.
In Sec.~\ref{sec: process} we present the helicity amplitude of $e^{+}e^{-}\to hZ$ including the electroweak (EW) radiative corrections.
In Sec.~\ref{sec: Numerical_results} we show numerical results of deviations in the cross section from the SM prediction in each model.
In addition, we show the correlations of the deviation in the cross section times the decay branching ratios of the SM-like Higgs boson from the SM predictions.
Conclusions are given in Sec.~\ref{sec: conclusion}.
In Appendix, the input parameters and explicit formulae for the NLO calculations are presented.

%

%% file: 02Models.tex


\section{Models with non-minimal Higgs sectors} \label{sec: models}
In this section, we briefly review the HSM, the 2HDM and the IDM.
Before moving on to the discussion on each extended Higgs model, we review concepts of general constraints on parameter spaces that are independent of models.
We then define the extended Higgs models in order.

\subsection{Constraints on extended Higgs models}
First of all, the size of Higgs quartic couplings is constrained by the perturbative unitarity bound, which was originally introduced to obtain the upper limit on the mass of the Higgs boson in the SM \cite{Lee:1977yc, Lee:1977eg}.
Using the equivalence theorem~\cite{Cornwall:1974km}, this bound requires that the magnitude of partial wave amplitudes for the elastic scatterings of two-body to two-body scalar boson processes, including the Nambu--Goldstone (NG) bosons, does not exceed a certain value.
Each eigenvalue of the $s$-wave amplitude $a_{0}^{i}$ should satisfy
\begin{align}
\abs{a_{0}^{i}}\leq\xi,
\end{align}
where $\xi=1$ \cite{Lee:1977yc, Lee:1977eg} or $1/2$.
We take $\xi=1/2$ in this paper.

Next, the vacuum stability bound provides an independent constraint on scalar quartic couplings.
This bound requires that the Higgs potential is bounded from below in any direction with large field values.
This condition is trivially satisfied in the SM by taking the Higgs quartic coupling to be positive.
However, this bound requires a set of inequalities in terms of Higgs quartic couplings in extended Higgs models \cite{Deshpande:1977rw}.

Furthermore, in extended Higgs models, wrong local vacua can appear in addition
to the true vacuum giving the correct value of the Fermi constant $G_{F}$.
We have to avoid parameter regions where the depth of such wrong vacua becomes deeper than that of the true one.
The condition to avoid the wrong vacua can be written by combinations of dimensionful and dimensionless parameters in the Higgs potential, and it provides an independent constraint from the above two constraints.

Apart from these theoretical constraints, we need to take into account bounds from experimental data.
At the LEP/SLC experiments, various EW observables have been precisely measured such as the masses and widths of the weak gauge bosons.
These precision measurements can be used to constrain the size of new physics effects which can enter into the two-point functions for weak gauge bosons.
Such indirect effects, so-called oblique corrections, are conveniently parameterized by the $S, T$ and $U$ parameters \cite{Peskin:1990zt, Peskin:1991sw}, which are expressed in terms of two-point functions of the weak bosons.
From the global fit of EW parameters \cite{Zyla:2020zbs}, new physics effects on the $S$ and $T$ parameters under $U = 0$ are constrained by
\begin{align}
S = 0.05\pm0.09\qc
T = 0.08\pm0.07,
\end{align}
with the correlation factor of $+0.91$ and the reference values of the masses of the SM Higgs boson and the top quark being $m_{h}^{\mathrm{ref}} = 126$ GeV and $m_{t}^{\mathrm{ref}} = 173$ GeV, respectively. 

Flavor experiments also provide important constraints on the parameter space in extended Higgs models, particularly in multi-doublet models.
We will discuss these constraints in more detail in Sec.~\ref{subsec: 2HDM} about the 2HDM.
Additional scalars have been directly searched at the LHC \cite{Aad:2020zxo, Aad:2019zwb, Aaboud:2018mjh, Aaboud:2018knk, Aaboud:2017gsl, Aad:2020fpj, Aad:2020tps, Aad:2020ncx, Aad:2021xzu, Sirunyan:2018taj, Sirunyan:2019wph, Sirunyan:2017uhk, Sirunyan:2017isc, Sirunyan:2019pqw, Sirunyan:2019xjg, Sirunyan:2020hwv, Sirunyan:2019hkq}, and constraints are obtained for parameters of extended Higgs models.
In addition, Higgs coupling measurements also give constraints especially on the mixing parameters in the HSM and the 2HDM \cite{Aad:2019mbh, CMS:2020gsy}.
The application of these constraints to each extended Higgs sector will be discussed in the following subsections.

\subsection{Higgs singlet model} \label{subsec: HSM}
In the HSM, we have one isospin doublet scalar field $\Phi$ with the hypercharge $Y=1/2$ and one real singlet field $S$ with $Y=0$.
We parametrize these scalar fields as
\begin{align}
\Phi = \mqty(G^{+} \\ \frac{1}{\sqrt{2}}(v+\phi+iG^{0}))\qc
S = v_{S}+s,
\end{align}
where $v$ is the vacuum expectation value (VEV) of the doublet field which is related to the Fermi constant by $v=(\sqrt{2}G_{F})^{-1/2}\simeq 246\ \mathrm{GeV}$, while $v_{S}$ is the VEV of the singlet field.
The component fields $G^{\pm}$ and $G^{0}$ in the doublet field correspond to the NG bosons.

The most general Higgs potential is given by
\begin{align}
V_{\mathrm{HSM}}(\Phi, S) = m_{\Phi}^{2}\abs{\Phi}^{2}+\lambda\abs{\Phi}^{4}+\mu_{\Phi S}\abs{\Phi}^{2}S
+\lambda_{\Phi S}\abs{\Phi}^{2}S^{2}+t_{S}S+m_{S}^{2}S^{2}+\mu_{S}S^{3}+\lambda_{S}S^{4},
\end{align}
where all the parameters are real.
We can take any value of $v_{S}$ without changing physical results \cite{Chen:2014ask}, and we fix $v_{S}=0$ in the following discussion.

In the HSM, we have two physical neutral Higgs bosons.
Their mass eigenstates are defined by introducing the mixing angle $\alpha$ as
\begin{align}
\mqty(s \\ \phi) = R(\alpha)\mqty(H \\ h) \qq{with}
R(\theta) = \mqty(c_{\theta} & -s_{\theta} \\ s_{\theta} & c_{\theta}),
\end{align}
with the shorthand notation for the trigonometric functions as $s_{\theta}\equiv \sin{\theta}$ and $c_{\theta}\equiv \cos{\theta}$.
We define the domain of $\alpha$ as $-\pi/2 \leq \alpha \leq \pi/2$.
We identify $h$ as the discovered Higgs boson with a mass of 125 GeV.
After solving the tadpole conditions, the squared masses of neutral Higgs bosons are expressed as
\begin{align}
m_{H}^{2} &= M_{11}^{2}c_{\alpha}^{2}+M_{22}^{2}s_{\alpha}^{2}+M_{12}^{2}s_{2\alpha},  \label{eq: m_bH_HSM}\\
m_{h}^{2} &= M_{11}^{2}s_{\alpha}^{2}+M_{22}^{2}c_{\alpha}^{2}-M_{12}^{2}s_{2\alpha}, \label{eq: m_h_HSM} \\
\tan{2\alpha} &= \frac{2M_{12}^{2}}{M_{11}^{2}-M_{22}^{2}}, \label{eq: tan2a_HSM}
\end{align}
where the squared mass matrix elements $M_{ij}^{2}\ (i,j=1,2)$ in $(s,\ \phi)^{T}$ basis are given by
\begin{align}
M_{11}^{2}=M^{2}+\lambda_{\Phi S}v^{2}\qc
M_{22}^{2}=2\lambda v^{2}\qc
M_{12}^{2}=\mu_{\Phi S}v,
\end{align}
with $M^{2}\equiv 2m_{S}^{2}$.
The parameters $m_{\Phi}^{2}$ and $t_{S}$ are eliminated by using the stationary conditions for $\phi$ and $s$.
We can replace the parameters $\lambda, m_{S}^{2}$ and $\mu_{\Phi S}$ with $m_{H}^{2}, m_{h}^{2}$ and $\alpha$ by using Eqs.~\eqref{eq: m_bH_HSM}-\eqref{eq: tan2a_HSM}.
We choose the following five parameters to be the free input parameters in the HSM:
\begin{align}
m_{H}\qc \lambda_{\Phi S}\qc \mu_{S}\qc \lambda_{S}\qc c_{\alpha}, \label{eq: inputs_HSM}
\end{align}
and the two parameters $m_{h}$ and $v$ are fixed by experiments.
If the Higgs potential respects an exact discrete $Z_{2}$ symmetry, the $t_{S}, \mu_{S}$ and $\mu_{\Phi S}$ terms are forbidden.
This corresponds to the case with $\alpha\to 0$ and $\mu_{S}\to 0$.

The kinetic terms of scalar fields are given by
\begin{align}
\mathcal{L}_{\mathrm{kin}}^{\mathrm{HSM}} = \abs{D_{\mu}\Phi}^{2}+\frac{1}{2}\qty(\partial_{\mu}S)^{2},
\end{align}
where $D_{\mu}$ is the covariant derivative for the Higgs doublet.
The gauge-gauge-scalar type interaction terms are given by
\begin{align}
\mathcal{L}_{\mathrm{kin}}^{\mathrm{HSM}} \supset
g m_{W}(c_{\alpha}W_{\mu}^{+}W^{-\mu}h+s_{\alpha}W_{\mu}^{+}W^{-\mu}H)
+\frac{g_{Z}m_{Z}}{2}(c_{\alpha}Z_{\mu}Z^{\mu}h+s_{\alpha}Z_{\mu}Z^{\mu}H),
\end{align}
where $g$ is the weak gauge coupling and $g_{Z}=g/c_{W}$ with $\theta_{W}$ being the weak mixing angle.

The Yukawa interaction terms are same as those in the SM and given by
\begin{align}
\mathcal{L}_{\mathrm{Y}}^{\mathrm{HSM}} =
-Y_{u}\overline{Q}_{L}\widetilde{\Phi}u_{R}-Y_{d}\overline{Q}_{L}\Phi d_{R}
-Y_{e}\overline{L}_{L}\Phi e_{R}+\mathrm{h.c.},
\end{align}
where $\widetilde{\Phi}=i\sigma_{2}\Phi^{*}$.
In the above equation, $Q_{L}$ and $L_{L}$ are left-handed quark and lepton doublets, respectively, while $u_{R}, d_{R}$ and $e_{R}$ are right-handed up-type quark, down-type quark and charged lepton singlets, respectively.
The interaction terms for $h$ and $H$ with fermions are given by
\begin{align}
\mathcal{L}_{\mathrm{Y}}^{\mathrm{HSM}} \supset
-\sum_{f=u,d,e}\frac{m_{f}}{v}(c_{\alpha}\bar{f}fh+s_{\alpha}\bar{f}fH).
\end{align}
We note that the SM-like Higgs bosons couplings with the SM particles are universally suppressed by $c_{\alpha}$ as compared to those SM values.

The parameters in the Higgs potential are constrained by perturbative unitarity, vacuum stability and the conditions to avoid wrong vacua.
For the perturbative unitarity bound, there are four independent eigenvalues given in Refs.~\cite{Cynolter:2004cq, Kanemura:2016lkz}.
The necessary and sufficient conditions to satisfy vacuum stability are given by \cite{Pruna:2013bma}
\begin{align}
\lambda_{\Phi}>0\qc
\lambda_{S}>0\qc
2\sqrt{\lambda_{\Phi}\lambda_{S}}+\lambda_{\Phi S}>0. \label{eq: vs_HSM}
\end{align}
For the conditions to avoid wrong vacua can be found in Refs.~\cite{Espinosa:2011ax, Chen:2014ask, Lewis:2017dme}.

The one-loop corrected two-point functions for weak bosons are found in Ref.~\cite{Lopez-Val:2014jva}.
Imposing the constraint from the $S$ and $T$ parameters, we can obtain the upper limit on $m_{H}$ depending on the value of $c_{\alpha}$.
Constraints on the mass of the additional Higgs boson and the mixing angle from the LHC data have been studied in Refs.~\cite{Robens:2015gla, Robens:2016xkb}.

\subsection{Two Higgs doublet model} \label{subsec: 2HDM}
In the 2HDM, we have two isospin doublet scalar fields $\Phi_{i}$ with the hypercharge $Y=1/2$.
We parametrize these doublets as
\begin{align}
\Phi_{i} = \mqty(\omega^{+}_{i} \\ \frac{1}{\sqrt{2}}(v_{i}+h_{i}+iz_{i}))\qc
(i=1,2),
\end{align}
where $v_{1}$ and $v_{2}$ are the VEVs of two doublets with $v=\sqrt{v_{1}^{2}+v_{2}^{2}}$.

In the most general 2HDM, flavor-changing neutral currents (FCNCs) appear at tree level, and it is severely constrained by experiments.
In order to avoid such FCNCs, we introduce a discrete $Z_{2}$ symmetry, where two doublets transform as $\Phi_{1}\to \Phi_{1}$ and $\Phi_{2}\to -\Phi_{2}$ \cite{Glashow:1976nt, Paschos:1976ay}.
One can introduce the soft breaking term of the $Z_{2}$ symmetry in the Higgs potential without spoiling the desirable property of the flavor sector.

The most general Higgs potential under the softly-broken $Z_{2}$ symmetry is given by
\begin{align}
V_{\mathrm{2HDM}}(\Phi_{1}, \Phi_{2}) &= 
m_{1}^{2}\abs{\Phi_{1}}^{2}+m_{2}^{2}\abs{\Phi_{2}}^{2}-m_{3}^{2}\qty(\Phi_{1}^{\dagger}\Phi_{2}+\mathrm{h.c.}) \notag \\
&\quad
+\frac{1}{2}\lambda_{1}\abs{\Phi_{1}}^{4}+\frac{1}{2}\lambda_{2}\abs{\Phi_{2}}^{4}
+\lambda_{3}\abs{\Phi_{1}}^{2}\abs{\Phi_{2}}^{2}+\lambda_{4}\abs{\Phi_{1}^{\dagger}\Phi_{2}}^{2} \notag \\
&\quad
+\frac{1}{2}\lambda_{5}\qty[\qty(\Phi_{1}^{\dagger}\Phi_{2})^{2}+\mathrm{h.c.}].
\end{align}
Although $m_{3}^{2}$ and $\lambda_{5}$ are generally complex, we take them to be real and consider the CP-conserving case for simplicity.
The mass eigenstates of the Higgs fields are defined as
\begin{align}
\mqty(\omega_{1}^{\pm} \\ \omega_{2}^{\pm}) = R(\beta)\mqty(G^{\pm} \\ H^{\pm})\qc
\mqty(z_{1} \\ z_{2}) = R(\beta)\mqty(G^{0} \\ A)\qc
\mqty(h_{1} \\ h_{2}) = R(\alpha)\mqty(H \\ h),
\end{align}
where $\tan{\beta}=v_{2}/v_{1}$, and $H^{\pm}$ and $A$ are the charged and CP-odd Higgs bosons respectively, while $H$ and $h$ are the CP-even Higgs bosons.
We define the domain of $\beta$ to be $0\leq\beta\leq\pi/2$.
We identify $h$ as the discovered Higgs boson with a mass of 125 GeV.
After solving two tadpole conditions for $h_{1}$ and $h_{2}$, the squared masses of the charged and CP-odd Higgs bosons are given by
\begin{align}
m_{H^{\pm}}^{2}=M^{2}-\frac{1}{2}(\lambda_{4}+\lambda_{5})v^{2}\qc
m_{A}^{2}=M^{2}-\lambda_{5}v^{2},
\end{align}
where $M^{2}=m_{3}^{2}/(s_{\beta}c_{\beta})$ which describes the softly-breaking scale of the $Z_{2}$ symmetry.
The squared masses of the neutral Higgs bosons and the mixing angle $\beta-\alpha$ are given by
\begin{align}
m_{H}^{2} &= M_{11}^{2}c_{\beta-\alpha}^{2}+M_{22}^{2}s_{\beta-\alpha}^{2}-M_{12}^{2}s_{2(\beta-\alpha)},  \label{eq: m_bH_2HDM}\\
m_{h}^{2} &= M_{11}^{2}s_{\beta-\alpha}^{2}+M_{22}^{2}c_{\beta-\alpha}^{2}+M_{12}^{2}s_{2(\beta-\alpha)}, \label{eq: m_h_2HDM} \\
\tan{2(\beta-\alpha)} &= -\frac{2M_{12}^{2}}{M_{11}^{2}-M_{22}^{2}}, \label{eq: tan2a}
\end{align}
where $M_{ij}^{2}\ (i,j=1,2)$ are the squared mass matrix elements for the CP-even scalar states in the Higgs basis \cite{Davidson:2005cw} $(h_{1},h_{2})R(\beta)$:
\begin{align}
M_{11}^{2}&=(\lambda_{1}c_{\beta}^{4}+\lambda_{2}s_{\beta}^{4})v^{2}+\frac{1}{2}\lambda_{345}v^{2}s_{2\beta}^{2}, \\
M_{22}^{2}&=M^{2}+\frac{1}{4}(\lambda_{1}+\lambda_{2}-2\lambda_{345})v^{2}s_{2\beta}^{2}, \\
M_{12}^{2}&=-\frac{1}{2}(\lambda_{1}c_{\beta}^{2}-\lambda_{2}s_{\beta}^{2}-\lambda_{345}c_{2\beta})v^{2}s_{2\beta},
\end{align}
with $\lambda_{345}\equiv\lambda_{3}+\lambda_{4}+\lambda_{5}$.
We define the domain of $\beta-\alpha$ to be $0\leq\beta-\alpha\leq\pi$ so that $s_{\beta-\alpha}$ is always positive and $c_{\beta-\alpha}$ has the opposite sign from $M_{12}^{2}$ \cite{Bernon:2015qea}.
The eight parameters in the Higgs potential are expressed by the following six input parameters:
\begin{align}
m_{H}\qc m_{A}\qc m_{H^{\pm}}\qc M^{2}\qc \tan{\beta}\qc s_{\beta-\alpha}, \label{eq: inputs_2HDM}
\end{align}
and the two parameters $m_{h}$ and $v$ are fixed by experiments.
In addition, we have a degree of freedom of the sign of $c_{\beta-\alpha}$.

The kinetic terms of the Higgs doublets are given by
\begin{align}
\mathcal{L}_{\mathrm{kin}}^{\mathrm{2HDM}} = \abs{D_{\mu}\Phi_{1}}^{2}+\abs{D_{\mu}\Phi_{2}}^{2}.
\end{align}
In the mass eigenbasis of the Higgs bosons, the gauge-gauge-scalar type interaction terms are given by
\begin{align}
\mathcal{L}_{\mathrm{kin}}^{\mathrm{2HDM}} \supset
g m_{W}(s_{\beta-\alpha}W_{\mu}^{+}W^{-\mu}h+c_{\beta-\alpha}W_{\mu}^{+}W^{-\mu}H)
+\frac{g_{Z}m_{Z}}{2}(s_{\beta-\alpha}Z_{\mu}Z^{\mu}h+c_{\beta-\alpha}Z_{\mu}Z^{\mu}H).
\end{align}

\begin{table}[t]
\centering
  \begin{tabular}{l|ccccccc|ccc} \hline
  \multicolumn{1}{l|}{ } & \multicolumn{7}{c|}{$Z_{2}$ charge} & \multicolumn{3}{c}{Mixing factor}\\ \cline{2-11}
  & $\Phi_{1}$ & $\Phi_{2}$ & $Q_{L}$ & $L_{L}$ & $u_{R}$ & $d_{R}$ & $e_{R}$ & $\zeta_{u}$ & $\zeta_{d}$ & $\zeta_{e}$\\ \hline
    Type-I & $+$ & $-$ & $+$ & $+$ & $-$ & $-$ & $-$ & $\cot{\beta}$ & $\cot{\beta}$ & $\cot{\beta}$ \\
    Type-II & $+$ & $-$ & $+$ & $+$ & $-$ & $+$ & $+$ & $\cot{\beta}$ & $-\tan{\beta}$ & $-\tan{\beta}$ \\
    Type-X (lepton specific) & $+$ & $-$ & $+$ & $+$ & $-$ & $-$ & $+$ & $\cot{\beta}$ & $\cot{\beta}$ & $-\tan{\beta}$ \\ 
    Type-Y (flipped) & $+$ & $-$ & $+$ & $+$ & $-$ & $+$ & $-$ & $\cot{\beta}$ & $-\tan{\beta}$ & $\cot{\beta}$\\ \hline
    \end{tabular}
   \caption{Charge assignment of the softly-broken $Z_{2}$ symmetry and the mixing factors in Yukawa interactions.}
   \label{tab: Z2}
\end{table}

The Yukawa interaction terms under the $Z_{2}$ symmetry are given by
\begin{align}
\mathcal{L}_{\mathrm{Y}}^{\mathrm{2HDM}} =
-Y_{u}\overline{Q}_{L}\widetilde{\Phi}_{u}u_{R}-Y_{d}\overline{Q}_{L}\Phi_{d} d_{R}
-Y_{e}\overline{L}_{L}\Phi_{e} e_{R}+\mathrm{h.c.},
\end{align}
where $\Phi_{u,d,e}$ are either $\Phi_{1}$ or $\Phi_{2}$.
As in Table~\ref{tab: Z2}, there are four types of Yukawa interactions according to the $Z_{2}$ charge assignment \cite{Barger:1989fj, Aoki:2009ha}.
The interaction terms for the physical Higgs bosons with the fermions are given by
\begin{align}
\mathcal{L}_{\mathrm{Y}}^{\mathrm{2HDM}} &\supset
-\sum_{f=u,d,e}\frac{m_{f}}{v}\qty[(s_{\beta-\alpha}+\zeta_{f}c_{\beta-\alpha})\bar{f}fh
+(c_{\beta-\alpha}-\zeta_{f}s_{\beta-\alpha})\bar{f}fH
-2iI_{f}\zeta_{f}\bar{f}\gamma_{5}fA] \notag \\
&\quad
+\frac{\sqrt{2}}{v}\qty[V_{ud}\bar{u}(m_{u}\zeta_{u}P_{L}-m_{d}\zeta_{d}P_{R})dH^{+}-m_{e}\zeta_{e}\bar{\nu}P_{R}eH^{+}+\mathrm{h.c.}],
\end{align}
with $I_{f}=1/2\ (-1/2)$ for $f=u\ (d, e)$ and $V_{ud}$ is the Cabbibo--Kobayashi--Maskawa matrix element.

The parameters in the Higgs potential are constrained by perturbative unitarity, vacuum stability and the condition to avoid wrong vacua.
For the perturbative unitarity bound, there are twelve independent eigenvalues of the $s$-wave amplitude matrix \cite{Kanemura:1993hm, Akeroyd:2000wc, Ginzburg:2005dt, Kanemura:2015ska}.
The vacuum stability bound is sufficiently and necessarily satisfied by imposing the following conditions \cite{Deshpande:1977rw, Klimenko:1984qx, Sher:1988mj, Nie:1998yn, Kanemura:1999xf}
\begin{align}
\lambda_{1}>0\qc
\lambda_{2}>0\qc
\sqrt{\lambda_{1}\lambda_{2}}+\lambda_{3}+\mathrm{MIN}(0,\ \lambda_{4}+\lambda_{5},\ \lambda_{4}-\lambda_{5}) > 0. \label{eq: vacuum_stability_2HDM}
\end{align}
In addition, the wrong vacua can be avoided by taking $M^{2} \geq 0$ \cite{Barroso:2013awa}.
We thus only take the positive value of $M^{2}$ in the following discussion.

The expressions of the two-point functions for the weak bosons in the 2HDM are found in Refs.~\cite{Toussaint:1978zm, Bertolini:1985ia, Peskin:2001rw, Grimus:2008nb, Kanemura:2011sj}.
Imposing the constraint of the $S$ and $T$ parameters, we can find that the charged Higgs bosons and one of the neutral Higgs bosons should be approximately degenerate in their mass.
This results from the constraint of the $T$ parameter, and it can be satisfied if the Higgs potential respects the custodial symmetry \cite{Pomarol:1993mu, Gerard:2007kn}.
Constraints on the parameters in 2HDMs from the LHC data have been discussed in Refs.~\cite{Bernon:2015qea, Chang:2015goa, Dorsch:2016tab, Wang:2017xml, Arbey:2017gmh, Aiko:2020ksl}.

In the 2HDM, constraints from flavor experiments are important to be taken into account.
These bounds particularly provide the lower limit on the mass of the charged Higgs boson $m_{H^{\pm}}$ depending on the type of Yukawa interaction and $\tan{\beta}$.
For example, from the $B_{s}\to X_{s}\gamma$ data, $m_{H^{\pm}}$ has to be greater than about $800$ GeV at 95\% confidence level (C.L.) in the Type-II and Type-Y 2HDMs with $\tan{\beta}\gtrsim 2$ \cite{Misiak:2020vlo}, while $\mathcal{O}(100)$ GeV of $m_{H^{\pm}}$ is allowed in the Type-I and TypeX 2HDMs with $\tan{\beta}\gtrsim 2$ \cite{Misiak:2017bgg}.
Constraints on $m_{H^{\pm}}$ and $\tan{\beta}$ from various flavor observables are also shown in Ref.~\cite{Haller:2018nnx} in the four types of 2HDMs.

\subsection{Inert doublet model}
The contents of the Higgs bosons in the IDM are the same as those in the 2HDM.
In the IDM, we assume an exact $Z_{2}$ symmetry and prohibit the $m_{3}^{2}$ term in the Higgs potential which softly breaks the $Z_{2}$ symmetry in the 2HDM.
We also assume that the second Higgs doublet $\Phi_{2}$ does not develop the VEV to avoid the spontaneous breaking down of the $Z_{2}$ symmetry.

We parametrize the doublets as
\begin{align}
\Phi_{1} = \mqty(G^{+} \\ \frac{1}{\sqrt{2}}(v+h+iG^{0}))\qc
\Phi_{2} = \mqty(H^{+} \\ \frac{1}{\sqrt{2}}(H+iA)).
\end{align}
The squared masses of the Higgs bosons are given by
\begin{align}
m_{h}^{2} &= \lambda_{1}v^{2}, \\
m_{H}^{2} &= M^{2}+\frac{1}{2}(\lambda_{3}+\lambda_{4}+\lambda_{5})v^{2}, \\
m_{A}^{2} &= M^{2}+\frac{1}{2}(\lambda_{3}+\lambda_{4}-\lambda_{5})v^{2}, \\
m_{H^{\pm}}^{2} &= M^{2}+\frac{1}{2}\lambda_{3}v^{2},
\end{align}
where $M^{2}\equiv m_{2}^{2}$.
We note that in addition to the absence of the $m_{3}^{2}$ term, there is no tadpole condition for $H$.
Therefore, the mass formulae for the scalar bosons are different from those in the 2HDM.
We choose the following five parameters to be the free input parameters in the IDM:
\begin{align}
m_{H}\qc m_{A}\qc m_{H^{\pm}}\qc M^{2}\qc \lambda_{2}, \label{eq: inputs_IDM}
\end{align}
and the two parameters $m_{h}$ and $v$ are fixed by experiments.

The parameters in the Higgs potential are constrained by perturbative unitarity, vacuum stability and the condition to guarantee the inert vacuum.
The same conditions for perturbative unitarity and vacuum stability in the 2HDM can be
applied to the IDM, because these bounds are given in terms of the scalar quartic couplings.
In addition, there is the condition to guarantee the inert vacuum with $(\expval{\Phi_{1}^{0}},\ \expval{\Phi_{2}^{0}})=(v/\sqrt{2},0)$ \cite{Ginzburg:2010wa},
\begin{align}
\frac{m_{1}^{2}}{\sqrt{\lambda_{1}}}<\frac{M^{2}}{\sqrt{\lambda_{2}}}. \label{eq: inert_vacuum_cond}
\end{align}
Since the tadpole condition makes $m_{1}^{2}$ negative, and the vacuum stability condition constraints $\lambda_{1}$ and $\lambda_{2}$ to be positive, the condition given in Eq.~\eqref{eq: inert_vacuum_cond} is satisfied by taking $M^{2} > 0$.
We refer to this condition as the one to avoid wrong vacua, according to the other two models discussed above.

For the constraints of the $S$ and $T$ parameters, we can use the same expressions as those in the 2HDM with $s_{\beta-\alpha}=1$.
As similar to the case in the 2HDMs, the charged Higgs bosons and one of the neutral Higgs bosons should be approximately degenerate in their mass in order to satisfy the constraint from the $T$ parameter.
In the IDM, constraints on the masses of the additional Higgs bosons from collider experiments are relatively weak since the additional Higgs bosons do not couple to the SM fermions.
Constraints from the LEP and the LHC have been studied in Ref.~\cite{Lundstrom:2008ai} and Refs.~\cite{Belanger:2015kga, Belyaev:2016lok}, respectively.
Dark matter constraints from relic density and direct detection also limit the parameter space; see, e.g., Refs.~\cite{Belyaev:2016lok, Ilnicka:2018def} for details.

\begin{table}[t]
\centering
  \begin{tabular}{l|ccc} \hline
                       & HSM & 2HDMs & IDM \\ \hline
  $\kappa_{V}$ & $c_{\alpha}$ & $s_{\beta-\alpha}$ & $1$ \\
  $\kappa_{f}$  & $c_{\alpha}$ & $s_{\beta-\alpha}+\xi_{f}c_{\beta-\alpha}$ & $1$ \\ \hline
  \end{tabular}
  \caption{Scaling factors for the SM-like Higgs boson couplings in the extended Higgs models.}
\label{tab: kappa}
\end{table}
Finally, we summarize the scaling factors for the SM-like Higgs boson couplings to the weak bosons $\kappa_{V}$ and the fermions $\kappa_{f}$ in Table~\ref{tab: kappa}.

%

%% file: 03Process.tex


\section{Electroweak corrections to the process $e^{+}e^{-}\to hZ$} \label{sec: process}
In this section, we define the notation for the process $e^{+}e^{-}\to hZ$ and discuss the helicity amplitudes based on the form-factor decomposition.
We list relevant renormalized quantities for this process and give the formulae of form factors including the one-loop corrections.
The differential cross section with arbitrary sets of electron and $Z$ boson polarization is also presented.
For the numerical evaluation, we use the SM input parameters given in Appendix~\ref{sec: SM_inputs}.

\subsection{Helicity amplitudes and cross section}
\begin{figure}[t]
\centering
\includegraphics[scale=0.45]{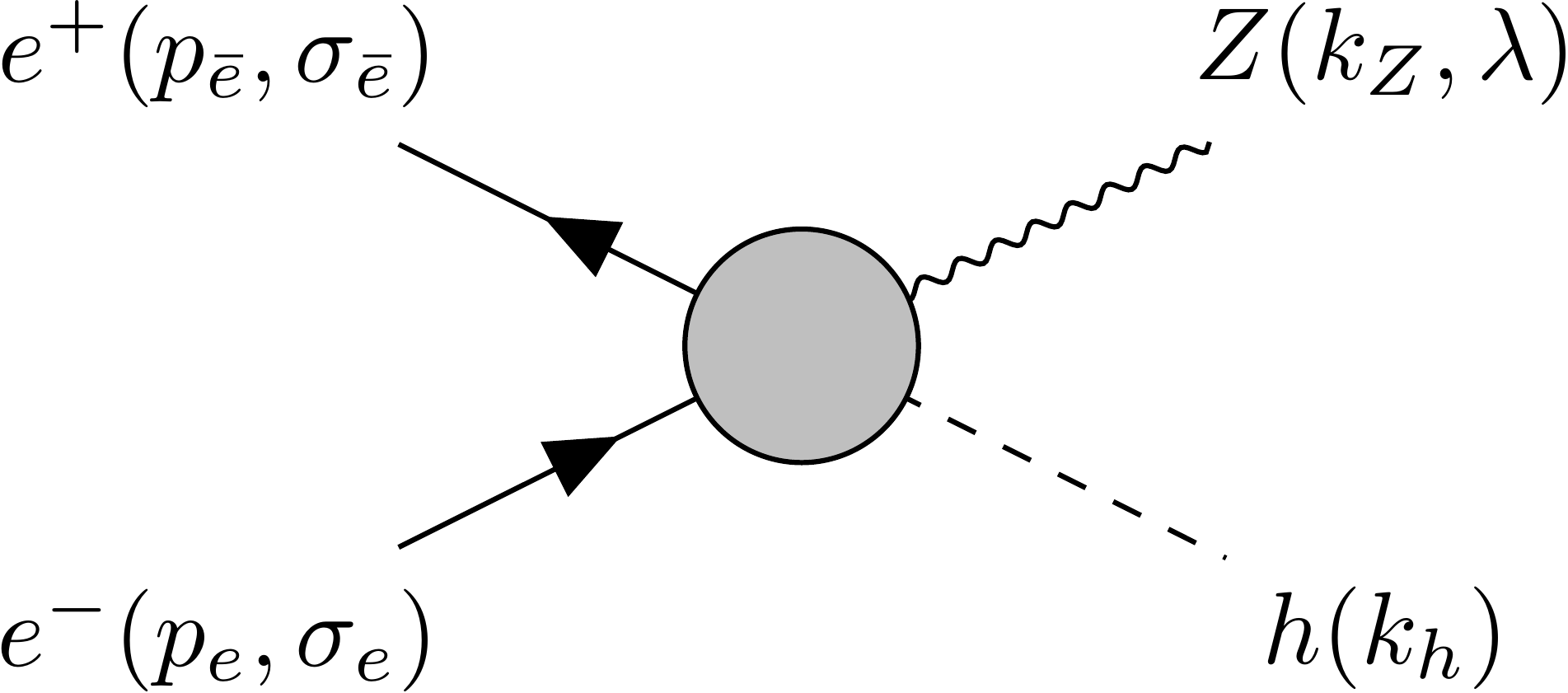}
\caption{The process $e^{+}e^{-}\to hZ$ with momentum and helicity assignments. The momenta $p_{e}$ and $p_{\bar{e}}$ are incoming, while $k_{h}$ and $k_{Z}$ are outgoing.}
\label{fig: diagram_eehZ}
\end{figure}

The process
\begin{align}
e^{-}(p_{e}, \sigma_{e})+e^{+}(p_{\bar{e}}, \sigma_{\bar{e}}) \to h(k_{h}) + Z(k_{Z}, \lambda)
\end{align}
is depicted in Fig.~\ref{fig: diagram_eehZ}.
The momenta and helicities of the incoming electron and positron are denoted by $(p_{e}, \sigma_{e})$ and $(p_{\bar{e}}, \sigma_{\bar{e}})$, respectively.
Correspondingly, $(k_{Z}, \lambda)$ is used for the outgoing $Z$ boson, and $k_{h}$ is the momentum of the outgoing Higgs boson.
The signs `$+$' and `$-$' of the variables $\sigma_{e}$ and $\sigma_{\bar{e}}$  refer to helicities $+1/2$ and $-1/2$, respectively.
The helicity $\lambda$ takes `$\pm$' or `$0$'.
In the following discussion, we neglect the mass of the electron whenever it is possible.

The Mandelstam variables are denoted by
\begin{align}
s &= (p_{e}+p_{\bar{e}})^{2} = (k_{Z} + k_{h})^{2}, \\
t &= (p_{e} - k_{Z})^{2} = (p_{\bar{e}} - k_{h})^{2}, \\
u &= (p_{e} - k_{h})^{2} = (p_{\bar{e}} - k_{Z})^{2},
\end{align}
and they satisfy $s+t+u = m_{Z}^{2}+m_{h}^{2}$.
In the CM frame of the $e^{+}e^{-}$ collision, the momenta of each external particle are
\begin{align}
p_{e}^{\mu} &= \frac{\sqrt{s}}{2}(1,0,0,1), \\
p_{\bar{e}}^{\mu} &= \frac{\sqrt{s}}{2}(1,0,0,-1), \\
k_{Z}^{\mu} &=\frac{\sqrt{s}}{2}\qty(1+\frac{m_{Z}^{2}-m_{h}^{2}}{s}, \beta\sin{\theta}, 0, \beta\cos{\theta}), \\
k_{h}^{\mu} &= \frac{\sqrt{s}}{2}\qty(1-\frac{m_{Z}^{2}-m_{h}^{2}}{s}, -\beta\sin{\theta}, 0, -\beta\cos{\theta}), \label{eq: momenta}
\end{align}
where $\beta$ is defined by
\begin{align}
\beta = \frac{\abs{\vb{k}_{Z}}}{E}=\frac{1}{s}\sqrt{\qty[s-(m_{Z}+m_{h})^{2}]\qty[s-(m_{Z}-m_{h})^{2}]},
\end{align}
with the beam energy $E=\sqrt{s}/2$.
We use the scattering angle $\theta$ between $e^{-}$ and $Z$ boson; $\hat{\vb{p}}_{e}\vdot \hat{\vb{k}}_{Z} =\cos{\theta}$ where the hat indicates the unit vector.
The scattering angle $\theta$ is related to $t$ and $u$ via
\begin{align}
t &= \frac{1}{2}(m_{Z}^{2}+m_{h}^{2}-s)+\frac{s}{2}\beta\cos{\theta}, \\
u &= \frac{1}{2}(m_{Z}^{2}+m_{h}^{2}-s)-\frac{s}{2}\beta\cos{\theta}.
\end{align}

The helicity amplitudes for $e^{+}e^{-}\to hZ$ vanish for $\sigma_{e}=\sigma_{\bar{e}}$ in the limit $m_{e}\to 0$ due to the chirality conservation.
Therefore, we use $\sigma = \sigma_{e} = -\sigma_{\bar{e}}$ for the non-vanishing amplitudes.
The helicity amplitude $\mathcal{M}_{\sigma \lambda}(s, t)$ can be decomposed into a set of basic matrix elements $\mathcal{M}_{i, \sigma \lambda}$ and corresponding form factors $F_{i, \sigma}(s,t)$ as \cite{Denner:1992bc}
\begin{align}
\mathcal{M}_{\sigma \lambda}(s, t)
= \sum_{i=1}^{3} F_{i, \sigma}(s,t)\mathcal{M}_{i, \sigma \lambda}(s, t). \label{eq: hel_amp}
\end{align}
The basic matrix elements are given by
\begin{align}
\mathcal{M}_{i, \sigma \lambda} = j_{\sigma, \mu}(p_{e},p_{\bar{e}})T^{\mu\nu}_{i}(s, t)\varepsilon^{*}_{\nu}(k_{Z}, \lambda),
\end{align}
where $\varepsilon^{*\mu}(k_{Z}, \lambda)$ is the polarization vector for $Z$ boson, and $j_{\sigma}^{\mu}(p_{e},p_{\bar{e}})$ is the fermion current of the initial electron and positron,
\begin{align}
\varepsilon^{*\mu}(k_{Z}, \pm) &= \frac{1}{\sqrt{2}}[0, \mp \cos{\theta}, i, \pm \sin{\theta}], \\
\varepsilon^{*\mu}(k_{Z}, 0) &= \frac{\sqrt{s}}{2m_{Z}}[\beta, \alpha \sin{\theta}, 0, \alpha \cos{\theta}], \label{eq: longitudinal_polarization_vector} \\
j_{\sigma}^{\mu}(p_{e},p_{\bar{e}}) &= \bar{v}(p_{\bar{e}})\gamma^{\mu}P_{\sigma}u(p_{e})
= \sqrt{s}[0, 1, \sigma i, 0],
\end{align}
with the chirality projection operator $P_{\sigma} = (1+\sigma\gamma_{5})/2$ and $\alpha = 1+(m_{Z}^{2}-m_{h}^{2})/s$.
The three basis tensor $T^{\mu\nu}_{i}$ are defined by
\begin{align}
T_{1}^{\mu\nu} = g^{\mu\nu}, \quad
T_{2}^{\mu\nu} = k_{Z}^{\mu}(p_{e}+p_{\bar{e}})^{\nu}, \quad
T_{3}^{\mu\nu} = k_{Z}^{\mu}(p_{e}-p_{\bar{e}})^{\nu}.
\end{align}

In the CM frame, the first elements of basic matrix ${\cal M}_{1, \sigma \lambda}$ become
\begin{align}
\mathcal{M}_{1,\, \sigma \pm} &= \sigma \sqrt{\frac{s}{2}}(1\pm\sigma \cos{\theta}) = \sigma \sqrt{2s}\, d_{\sigma, \pm}^{1}(\theta), \\
\mathcal{M}_{1,\, \sigma 0} &= -\frac{s\alpha}{2m_{Z}}\sin{\theta}
=\sigma\frac{s\alpha}{\sqrt{2}m_{Z}}\, d_{\sigma, 0}^{1}(\theta),
\end{align}
where $d_{m', m}^{j} (\theta)$ is the Wigner's $d$ function.
The second elements ${\cal M}_{2, \sigma \lambda}$ are
\begin{align}
\mathcal{M}_{2, \sigma \pm}&= 0, \\
\mathcal{M}_{2, \sigma 0}&= -\frac{s^{2}\beta^{2}}{4m_{Z}}\sin{\theta}
=\sigma \frac{s^{2}\beta^{2}}{2\sqrt{2}m_{Z}}\, d_{\sigma, 0}^{1}(\theta).
\end{align}
The third elements ${\cal M}_{3, \sigma \lambda}$ are
\begin{align}
\mathcal{M}_{3,\, \sigma \pm} &= \pm\frac{s\beta}{2}\sqrt{\frac{s}{2}}\sin^{2}{\theta}
=\pm s\beta \sqrt{\frac{s}{3}}\, d_{2,0}^{2}(\theta), \\
\mathcal{M}_{3,\, \sigma 0}&= -\frac{s^{2}\alpha\beta}{4m_{Z}}\cos{\theta}\sin{\theta}
=\frac{s^{2}\alpha\beta}{4m_{Z}^{2}}\qty[d_{2,1}^{2}(\theta)-d_{2,-1}^{2}(\theta)].
\end{align}
The six physical helicity amplitudes are given in terms of the form factors $F_{i, \sigma}$ by
\begin{align}
\mathcal{M}_{\sigma \pm}(s, t) &=
\sqrt{\frac{s}{2}}\qty[F_{1, \sigma}(s, t)\pm\frac{s\beta}{2}(\sigma\mp\cos{\theta})F_{3, \sigma}(s, t)](\sigma\pm\cos{\theta}), \\
\mathcal{M}_{\sigma 0}(s, t) &=
-\frac{s}{2m_{Z}}\qty[\alpha F_{1, \sigma}(s, t)
+\frac{s\beta^{2}}{2}F_{2, \sigma}(s, t)
-\frac{s\alpha\beta}{2}\cos{\theta}F_{3, \sigma}(s, t)]\sin{\theta}.
\end{align}

We denote the tree- and one-loop contributions to the helicity amplitude as
\begin{align}
\mathcal{M}^{(0)}_{\sigma \lambda}(s, t) &= \sum_{i=1}^{3}F_{i, \sigma}^{(0)}(s,t)\mathcal{M}_{i, \sigma \lambda}, \\
\mathcal{M}^{(1)}_{\sigma \lambda}(s, t) &= \sum_{i=1}^{3}F_{i, \sigma}^{(1)}(s,t)\mathcal{M}_{i, \sigma \lambda}.
\end{align}
The helicity-dependent differential cross section at NLO in EW is given by
\begin{align}
{\dv{\sigma}{\Omega}}(\sigma, \lambda; s, t)
&= 
\frac{\beta}{64\pi^{2} s}\qty{
|\mathcal{M}^{(0)}_{\sigma \lambda}(s, t)|^{2}
+2\Re[\mathcal{M}^{(1)}_{\sigma \lambda}(s, t)\mathcal{M}^{(0)*}_{\sigma \lambda}(s, t)]} \label{eq: diff_xsec}.
\end{align}
The helicity-dependent cross section $\sigma(\sigma, \lambda; s)$ can be obtained by integrating Eq.~\eqref{eq: diff_xsec} over the solid angle.

In a realistic setup, one needs to introduce the degree of polarization of initial electron $P_{e}$ and positron $P_{\bar{e}}$.
We use the convention where a purely left-handed (right-handed) electron corresponds to $P_{e}=-1\ (+1)$.
The polarized differential cross section is given by
\begin{align}
{\dv{\sigma}{\Omega}}(P_{e}, P_{\bar{e}}, \lambda; s, t)
&=\sum_{\sigma=\pm}\frac{1}{4}(1+\sigma P_{e})(1-\sigma P_{\bar{e}})
{\dv{\sigma}{\Omega}}(\sigma, \lambda; s, t). \label{eq: pol_diff_xsec}
\end{align}
The unpolarized cross section $\sigma(\lambda; s)$ corresponds to $P_{e}=P_{\bar{e}}=0$.
$(P_{e}, P_{\bar{e}})=(\mp 0.8, \pm 0.3)$ is planned polarization at the ILC \cite{Fujii:2017vwa}.

The polarized cross section can be rewritten in terms of the helicity-dependent cross section as \cite{Fujii:2018mli}
\begin{align}
\sigma(P_{e}, P_{\bar{e}}, \lambda; s)
&= \frac{1}{4}(1-P_{e})(1+P_{\bar{e}})\sigma(-, \lambda; s)+\frac{1}{4}(1+P_{e})(1-P_{\bar{e}})\sigma(+, \lambda; s) \notag \\
&= 2\sigma(\lambda; s)(1-P_{e}P_{\bar{e}})(1-P_{\mathrm{eff}} A_{\mathrm{LR}}), \label{eq: pol_xsec}
\end{align}
where the effective polarization $P_{\mathrm{eff}}$ and the left-right asymmetry $A_{\mathrm{LR}}$ are defined as
\begin{align}
P_{\mathrm{eff}} &= \frac{P_{e}-P_{\bar{e}}}{1-P_{e}P_{\bar{e}}}, \\
A_{\mathrm{LR}} &= \frac{\sigma(-,\lambda; s)-\sigma(+,\lambda; s)}{\sigma(-,\lambda; s)+\sigma(+,\lambda; s)}.
\end{align}
By using Eq.~\eqref{eq: pol_xsec}, one can easily evaluate the effect of the beam polarization from the helicity-dependent cross sections. 
Therefore, in the following discussion, we focus on the unpolarized and helicity-dependent cross section to exhibit analytical behaviors.

\subsection{Tree-level contribution to the helicity amplitudes}
\begin{figure}[t]
\centering
\includegraphics[scale=0.4]{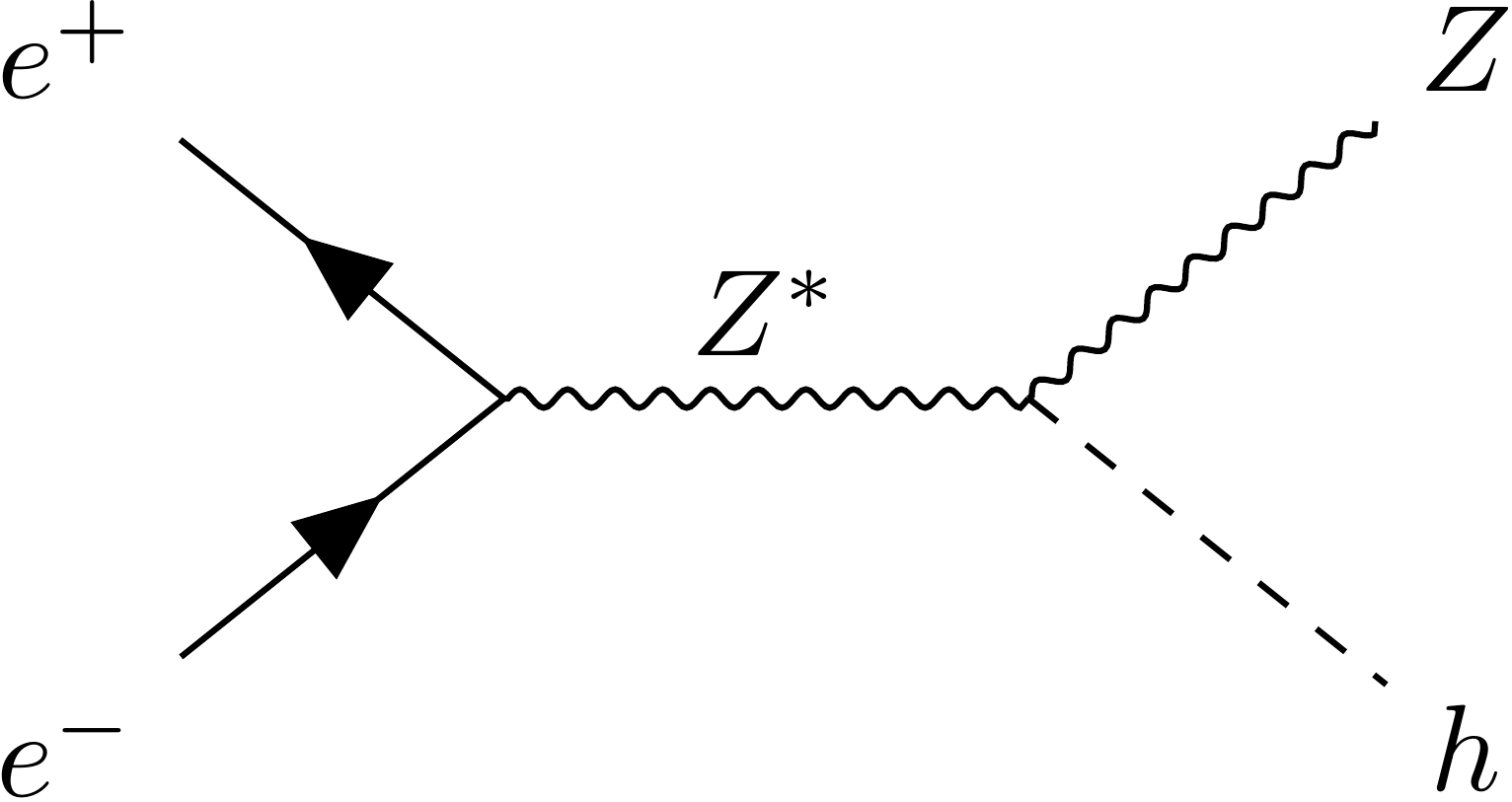}
\caption{Tree-level diagram for the process $e^{+}e^{-}\to hZ$.}
\label{fig: tree}
\end{figure}
At LO, only one diagram of Fig.~\ref{fig: tree} is relevant since the electron-Higgs coupling is proportional to $m_{e}$, and it is negligible.
The contribution of the tree-level diagram to the form factors is expressed as
\begin{align}
F_{i, \sigma}^{(0)} = \frac{g_{Z}\Gamma_{hZZ}^{1, (0)}}{s-m_{Z}^{2}}g_{\sigma}f_{i}^{(0)},
\end{align}
with the tree level $hZZ$ coupling $\Gamma_{hZZ}^{1, (0)}=2\kappa_{Z}m_{Z}^{2}/v$.
The scaling factor $\kappa_{Z}$ is given in Table~\ref{tab: kappa}.
If there is no mixing between CP-even scalars, the cross section in the extended Higgs models is the same as that in the SM at LO.
The couplings $g_{\pm}$ are defined by
\begin{align}
g_{+} = s_{W}^{2}\qc
g_{-} =  -\frac{1}{2}+s_{W}^{2}.
\end{align}
The coefficients $f_{i}^{(0)}$ are
\begin{align}
f_{1}^{(0)}=1,\quad
f_{2}^{(0)}=f_{3}^{(0)}=0.
\end{align}
The lowest order differential cross section is given by
\begin{align}
{\dv{\sigma_{\mathrm{LO}}}{\Omega}}(\sigma, \lambda; s, t)
&=\frac{\beta}{64\pi^{2}s}\big|F_{1,\sigma}^{(0)}\big|^{2}
\begin{dcases}
{2s\abs{d_{\sigma, \pm}^{1} (\theta)}^{2} \quad (\lambda = \pm)}, \\
{\frac{s^{2}\alpha^{2}}{2m_{Z}^{2}}\abs{d_{\sigma, 0}^{1} (\theta)}^{2} \quad (\lambda = 0)}.
\end{dcases}
\end{align}

\begin{figure}[t]
\begin{minipage}{0.45\hsize}
\centering
\includegraphics[scale=0.55]{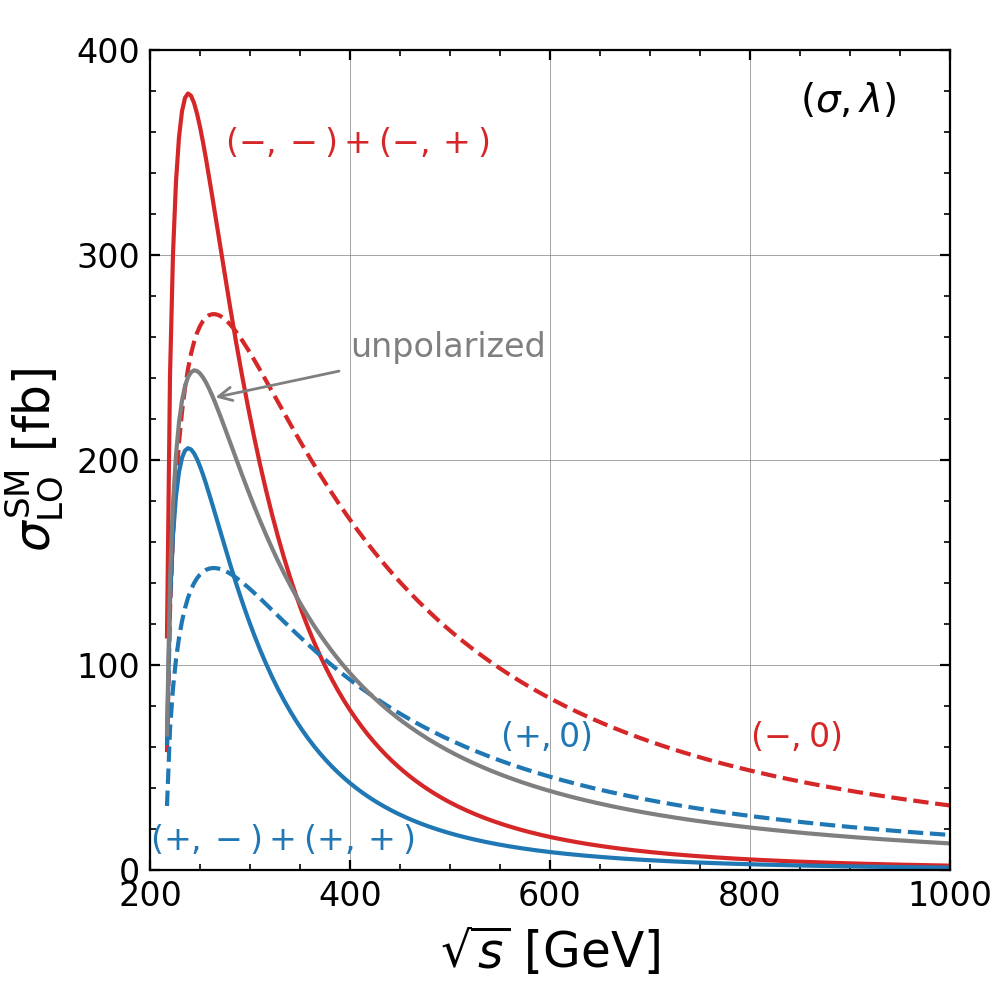}
\end{minipage}
\begin{minipage}{0.45\hsize}
\centering
\includegraphics[scale=0.55]{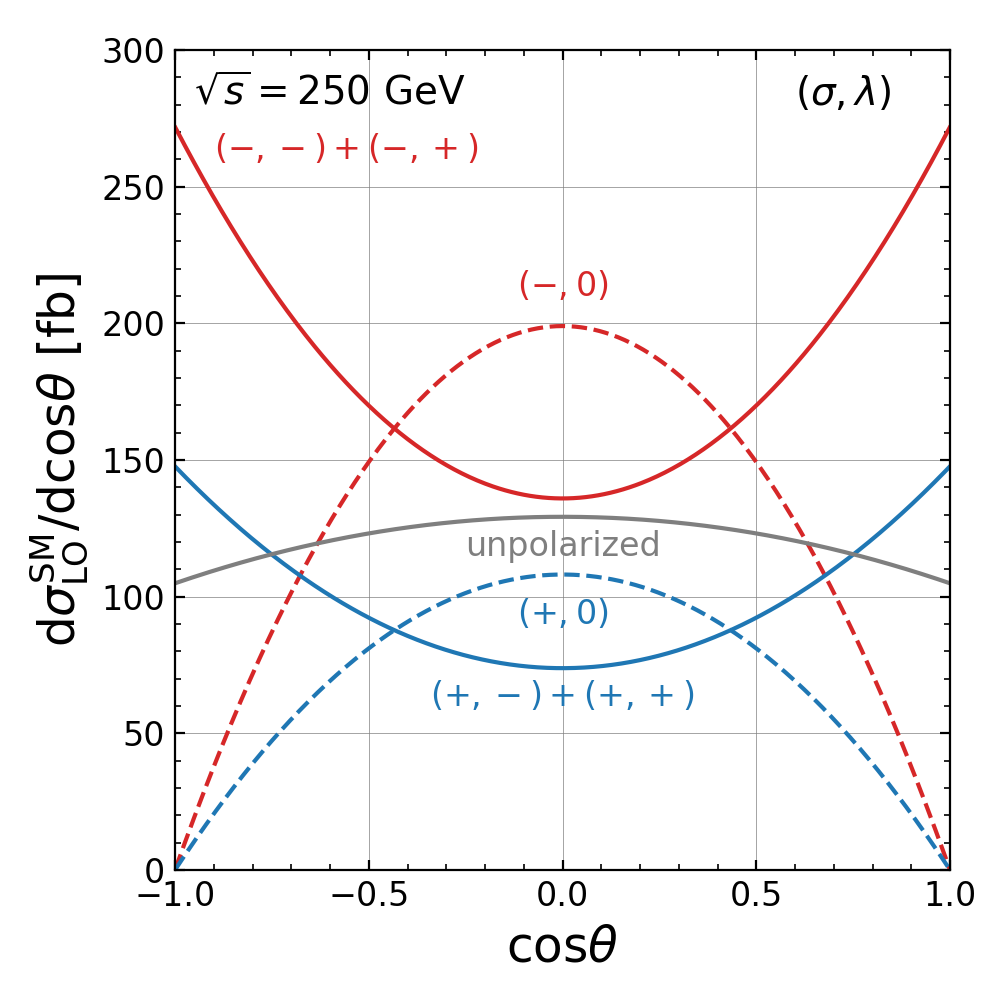}
\end{minipage}
\caption{(Left) Helicity-dependent cross sections of $e^{+}e^{-}\to hZ$ at LO in the SM as a function of CM energy. The red lines show the results for the left-handed electron and the right-handed positron, while the blue lines show those for the right-handed electron and the left-handed positron. The solid (dashed) lines show the results for the transversely (longitudinally) polarized $Z$ bosons. The black solid line corresponds to that for the unpolarized cross section where the polarization of $Z$ boson is also summed.
(Right) Helicity-dependent differential cross sections of $e^{+}e^{-}\to hZ$ at LO in the SM as a function of $\cos{\theta}$ at $\sqrt{s}=250$ GeV. The line colors and styles are the same as those of the left figure.}
\label{fig: xsec_SM_LO}
\end{figure}

In the left panel of Fig.~\ref{fig: xsec_SM_LO}, we show the helicity-dependent cross sections at LO as a function of the CM energy.
In the numerical evaluation, we take $G_{F}$ as an input given in Appendix~\ref{sec: SM_inputs} and use the tree-level relation $v=(\sqrt{2}G_{F})^{-1/2}$.
The solid and dashed lines correspond to the transversely ($\lambda=\pm$) and the longitudinally ($\lambda=0$) polarized $Z$ bosons, respectively.
The cross sections peak just above the threshold $\sqrt{s}= m_{Z}+m_{h}$ and monotonically decrease at higher energies.
For energies well above the threshold, the longitudinally polarized $Z$ boson dominates the cross section.
This is due to the factor $s/m_{Z}^{2}$ originated from the longitudinal polarization vector defined in Eq.~\eqref{eq: longitudinal_polarization_vector}.
The cross section for the left-handed electron is larger than that for the right-handed electron because the left-handed electron more strongly couples to the $Z$ boson than the right-handed electron $(g_{-}/g_{+})^{2}\simeq 1.8$.
At $\sqrt{s}=250$ GeV, the unpolarized cross section is 242 fb, and the polarized cross section is 379 fb with $(P_{e}, P_{\bar{e}})=(-0.8, 0.3)$.
Therefore, the beam polarization significantly changes the size of the cross section.
We note that the predicted value of the LO cross section highly depends on the input schemes.

The angular distributions of the LO cross section at $\sqrt{s}=250$ GeV are given in the right panel of Fig.~\ref{fig: xsec_SM_LO}.
They are determined by $d_{\sigma, \lambda}^{1}(\theta)$.
The cross section for the transversely polarized $Z$ boson is proportional to $1+\cos^{2}{\theta}$, and it takes maximal value in the forward-backward direction.
On the other hand, that for the longitudinally polarized $Z$ boson is proportional to $1-\cos^{2}{\theta}$, and it vanishes in the forward-backward direction and takes maximal value at $\cos{\theta}=0$.


\subsection{One-loop contributions to the form factors}
\begin{figure}[t]
\centering
\begin{minipage}{0.3\hsize}
\centering
\includegraphics[scale=0.35]{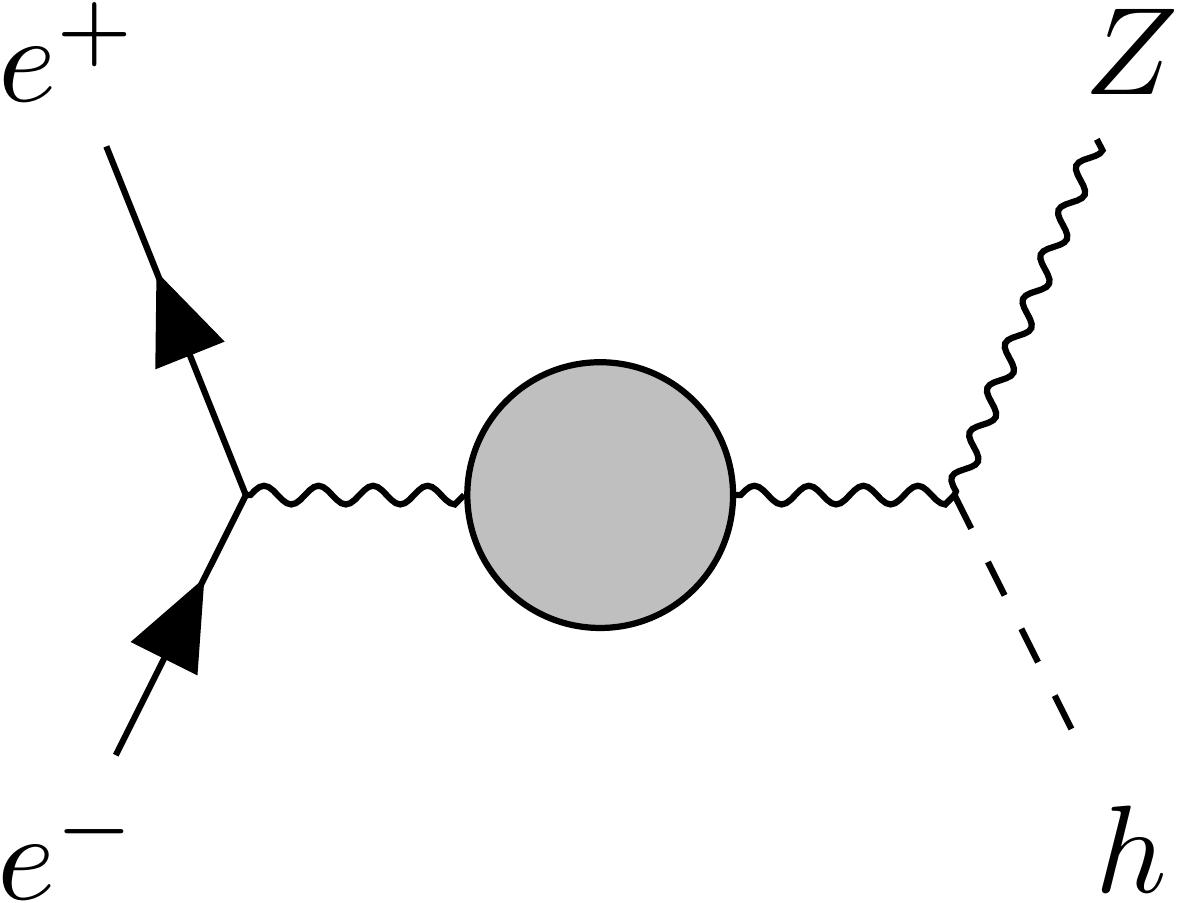}
\subcaption{}
\end{minipage}
\begin{minipage}{0.3\hsize}
\centering
\includegraphics[scale=0.35]{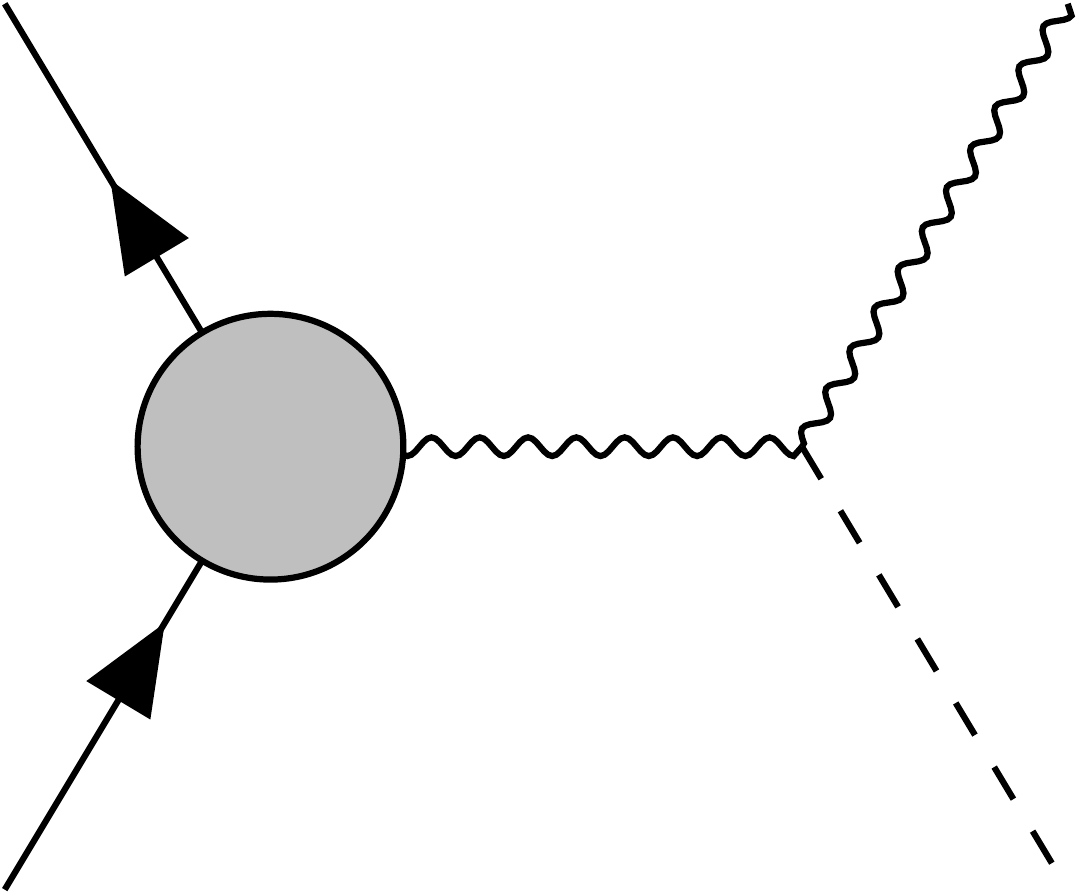}
\subcaption{}
\end{minipage}
\begin{minipage}{0.3\hsize}
\centering
\includegraphics[scale=0.35]{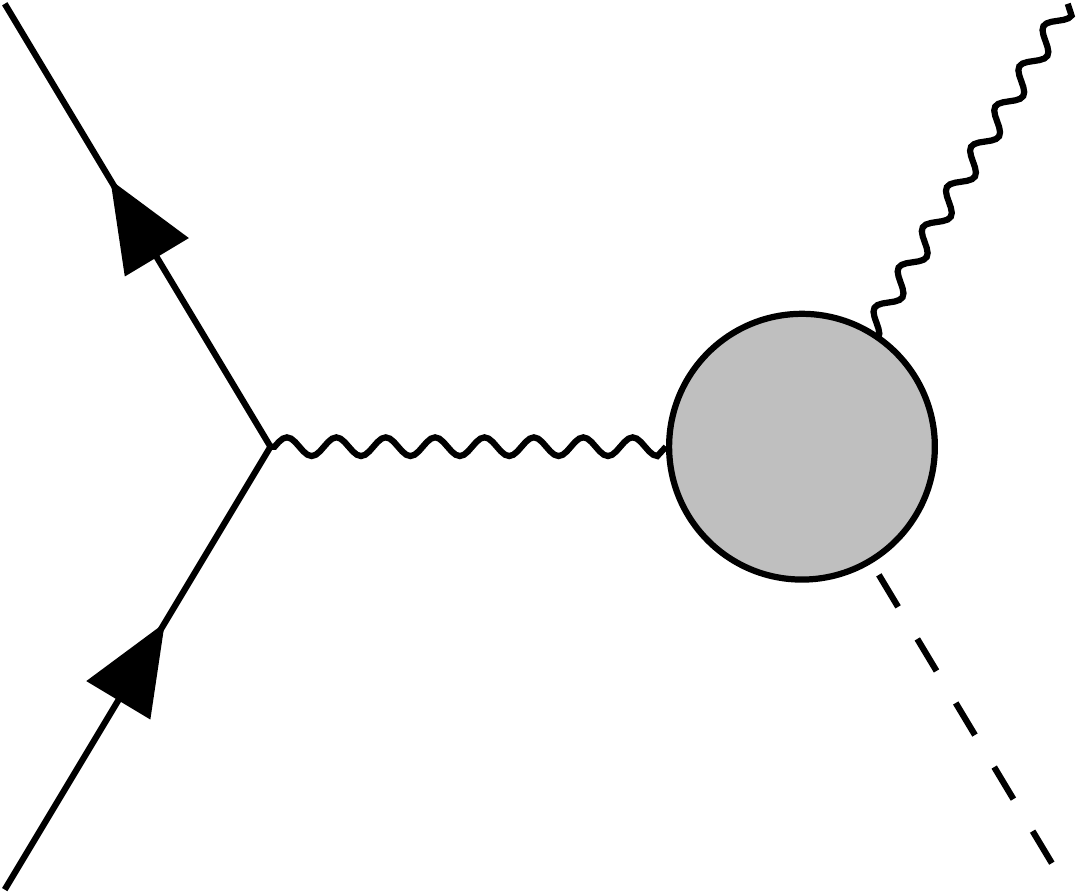}
\subcaption{}
\end{minipage} \\
\begin{minipage}{0.3\hsize}
\centering
\includegraphics[scale=0.35]{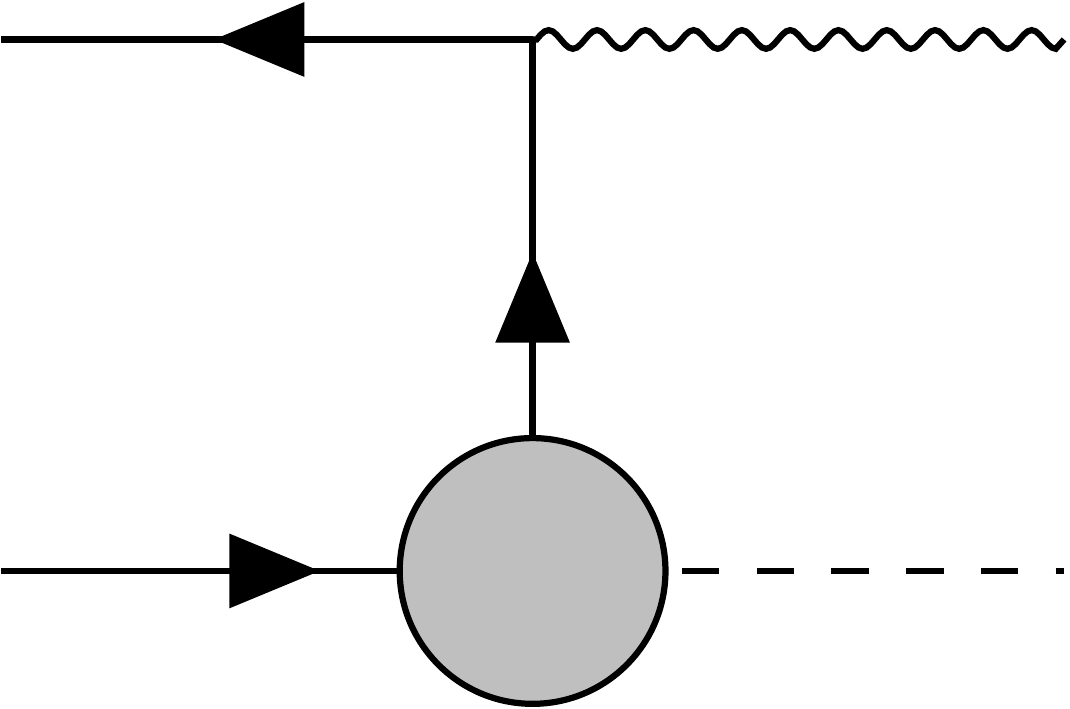}
\subcaption{}
\end{minipage}
\begin{minipage}{0.3\hsize}
\centering
\includegraphics[scale=0.35]{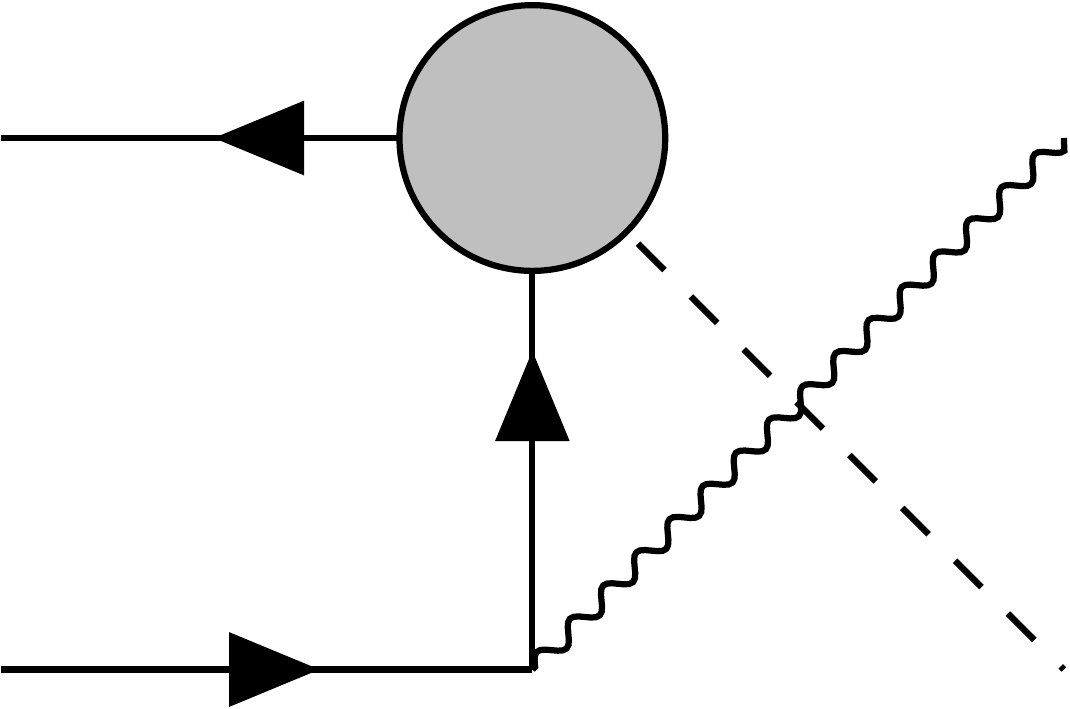}
\subcaption{}
\end{minipage}
\begin{minipage}{0.3\hsize}
\centering
\includegraphics[scale=0.4]{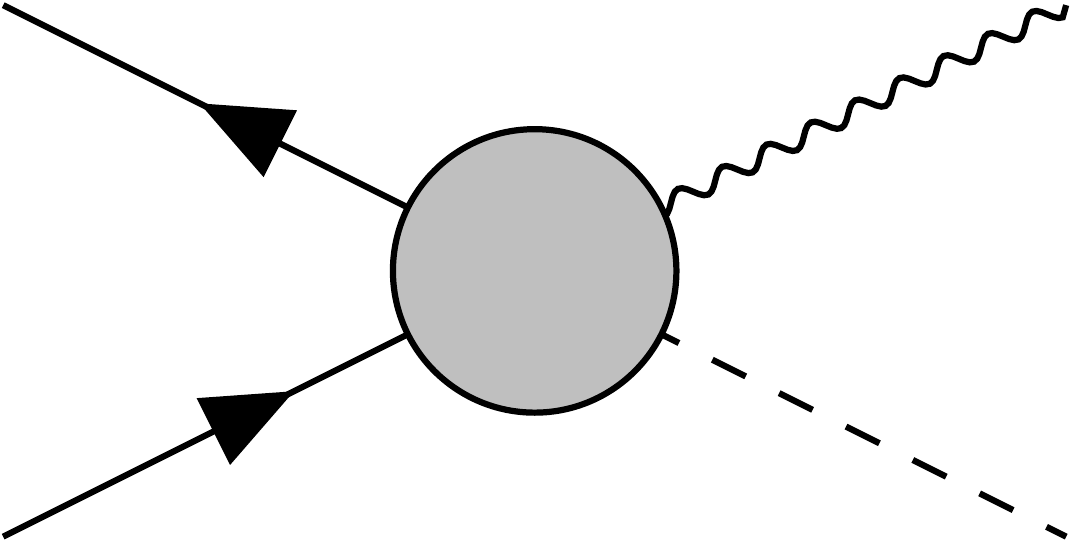}
\subcaption{}
\end{minipage}
\caption{NLO corrections for $e^{+}e^{-}\to hZ$; (a) Gauge boson self-energy, (b) $Ze\bar{e}$ vertex, (c) $hZZ$ and $hZ\gamma$ vertices, (d-e) $he\bar{e}$ vertex, (f) Box diagrams.}
\label{fig: NLO_diagrams}
\end{figure}

As shown in Fig.~\ref{fig: NLO_diagrams}, the one-loop contributions to the form factors $F_{i, \sigma}^{(1)}$ consist of (a) the $Z$ boson self-energy and the $Z\gamma$ mixing, (b) the $Ze\bar{e}$ vertex correction, (c) the $hZZ$ and $hZ\gamma$ vertex corrections, (d-e) the $he\bar{e}$ vertex correction and (f) the box diagrams.
In addition, the renormalization factors of the weak gauge bosons are not to be unity in our renormalization scheme \cite{Bohm:1986rj, Hollik:1988ii}.
Therefore, we have the term $-\Re{\Pi'_{ZZ}(m_{Z}^{2})}/2$ from the wave function renormalization of the on-shell $Z$ boson.
Furthermore, the EW correction to the Fermi decay constant $\Delta r$ appears when one replaces the VEV in the tree-level amplitude with $G_{F}$, since the tree-level relation between these two parameters is no longer valid at the one-loop level.
This replacement corresponds to the resummation of universal higher-order leading corrections such as large logarithms from light fermion masses \cite{Kniehl:1991hk}.
The one-loop contributions to the form factors $F_{i, \sigma}^{(1)}$ are given by
\begin{align}
F_{i, \sigma}^{(1)} &= F_{i, \sigma}^{ZZ}+F_{i, \sigma}^{Z\gamma}+F_{i, \sigma}^{Ze\bar{e}}
+F_{i, \sigma}^{hZZ}+F_{i, \sigma}^{hZ\gamma}+F_{i, \sigma}^{he\bar{e}}
+F_{i, \sigma}^{\mathrm{Box}} \notag \\
&\quad
+F_{i, \sigma}^{\Pi'_{ZZ}}+F_{i, \sigma}^{\Delta r}, \label{eq: NLO_form_fact}
\end{align}
where the terms in the first line correspond to the contributions from the diagrams in Fig.~\ref{fig: NLO_diagrams}, while the terms in the second line come from the renormalization procedure. 
For the computation of these EW corrections, we adopt the modified on-shell renormalization scheme defined in Ref.~\cite{Kanemura:2017wtm}.
In the on-shell renormalization scheme, all the counterterms in the amplitude of $e^{+}e^{-}\to hZ$ are determined in terms of the one-particle irreducible (1PI) diagrams for one- and two-point functions of Higgs bosons, gauge bosons and fermions by imposing a set of the renormalization conditions.
Adding these counterterms, one can obtain the ultra-violet (UV) finite one-loop corrected vertices.

In the wide range of extended Higgs models, there are mixings among Higgs bosons, and the gauge dependence appears in the renormalization of these mixing angles.
We apply the pinch technique to remove the gauge dependence in the renormalized vertex functions \cite{Bojarski:2015kra, Krause:2016oke, Kanemura:2017wtm}.

Apart from the UV divergences, there are infrared (IR) divergences when we calculate virtual photon loop contributions.
In the calculation of individual photon loop contributions, we regularize them with a finite photon mass $\mu$.
The photon mass dependences in the one-loop calculation are exactly canceled by adding contributions of real photon emissions.
The analytic expression of the real photon contribution with the soft-photon approximation is given by~\cite{Fleischer:1982af, Kniehl:1991hk, Denner:1992bc},
\begin{align}
\dd \sigma_{\mathrm{soft}}
=
\dd \sigma_{\mathrm{LO}}
\qty{-\frac{\alpha}{\pi}\qty[\ln{\frac{4\Delta E^{2}}{\mu^{2}}}\qty(1+\ln{\frac{m_{e}^{2}}{s}})+\frac{1}{2}\ln^{2}{\frac{m_{e}^{2}}{s}}+\ln{\frac{m_{e}^{2}}{s}}+\frac{\pi^{2}}{3}]},
\end{align}
with the photon energy cutoff $\Delta E$.
The dependence of $\Delta E$ vanishes in the inclusive cross section where one also includes the contribution of hard photon emissions \cite{Kniehl:1991hk}.
The inclusive cross section still depends on $\ln(m_{e}^{2}/s)$, and this logarithmic term potentially takes a large value.
This dependence can also be eliminated by introducing the electron structure functions as discussed in Ref.~\cite{Xie:2018yiv}.
However, the treatment of hard photon emission highly depends on the experimental setup.
The hard photon changes the kinematics of the process, and these effects would be eliminated by applying appropriate experimental cuts.
In addition, if one considers the scenario with $\kappa_{Z}\simeq 1$, these effects in extended Higgs models are almost the same as in the SM.
Therefore, we do not consider the electromagnetic effects when we focus on the difference between the predictions in the extended models and those in the SM.

In the one-loop calculation, we choose the fine structure constant $\alpha_{\mathrm{em}}$, the Fermi constant $G_{F}$ and the $Z$ boson mass $m_{Z}$ as the input EW parameters.
In addition to these EW parameters, we also use the shift of the fine structure constant $\Delta \alpha_{\mathrm{em}}$, the strong coupling constant $\alpha_{s}$ and the masses of the fermions and the discovered Higgs boson as the input parameters.
The values of these SM input parameters are given in Appendix~\ref{sec: SM_inputs}.
We use the input parameters given in Eqs.~\eqref{eq: inputs_HSM}, \eqref{eq: inputs_2HDM} and \eqref{eq: inputs_IDM} in the HSM, the 2HDMs and the IDM, respectively.

\subsubsection{Renormalized vertices}
In the following calculation, the $Zf\bar{f},\, hZZ,\, hZ\gamma$ and $hf\bar{f}$ vertices are relevant, where $hZ\gamma$ vertex is one-loop induced.
Each of these vertices can be decomposed into several form factors depending on their Lorentz structure.

The renormalized $Zf\bar{f}$ vertex can be decomposed in the massless limit of external fermions as
\begin{align}
\widehat{\Gamma}_{Zf\bar{f}}^{\mu}(p_{f}, p_{\bar{f}}, p_{Z})
=g_{Z}\gamma^{\mu}\qty[
\widehat{\Gamma}_{Zf\bar{f}}^{V}-\gamma_{5}\widehat{\Gamma}_{Zf\bar{f}}^{A}](p_{f}^{2}, p_{\bar{f}}^{2}, p_{Z}^{2}),
\end{align}
where $p_{f}\, (p_{\bar{f}})$ is the incoming four-momentum of the fermion (anti-fermion), and $p_{Z}$ is the outgoing four-momentum of the $Z$ boson.
We can further decompose these vertices into the tree-level and one-loop level contributions
\begin{align}
\widehat{\Gamma}_{Zf\bar{f}}^{i} = \Gamma_{Zf\bar{f}}^{i, (0)}+\Gamma_{Zf\bar{f}}^{i, (1)},
\qq{with}
\Gamma_{Zf\bar{f}}^{i, (1)} = \Gamma_{Zf\bar{f}}^{i, \mathrm{1PI}}+\delta \Gamma_{Zf\bar{f}}^{i}\qc (i=V, A).
\end{align}
The tree-level contribution is given by
\begin{align}
\Gamma_{Zf\bar{f}}^{V, (0)} = \frac{I_{f}}{2}-Q_{f}s_{W}^{2}\qc
\Gamma_{Zf\bar{f}}^{A, (0)} = \frac{I_{f}}{2}.
\end{align}
In the massless limit of external fermions, expressions of these vertices in the HSM, the 2HDMs and the IDM are the same as those in the SM.
Analytic expressions for the 1PI diagrams and counterterms are presented in Appendix B in Ref.~\cite{Kanemura:2019kjg}.

The renormalized $hZZ$ vertex can be decomposed as
\begin{align}
\widehat{\Gamma}_{hZZ}^{\mu\nu}(p_{1}, p_{2}, p_{h})
=\qty[g^{\mu\nu}\widehat{\Gamma}_{hZZ}^{1}
+\frac{p_{1}^{\nu}p_{2}^{\mu}}{m_{Z}^{2}}\widehat{\Gamma}_{hZZ}^{2}
+i\epsilon^{\mu\nu\rho\sigma}\frac{p_{1\rho}p_{2\sigma}}{m_{Z}^{2}}\widehat{\Gamma}_{hZZ}^{3}](p_{1}^{2}, p_{2}^{2}, p_{h}^{2}),
\end{align}
where $p_{1}$ and $p_{2}$ are incoming four-momenta of the $Z$ bosons, and $p_{h}$ is the outgoing four-momentum of the SM-like Higgs boson.
We can further decompose these vertices into the tree-level and one-loop level contributions
\begin{align}
\widehat{\Gamma}_{hZZ}^{i} = \Gamma_{hZZ}^{i, (0)}+\Gamma_{hZZ}^{i, (1)},
\qq{with}
\Gamma_{hZZ}^{i, (1)} = \Gamma_{hZZ}^{i, \mathrm{1PI}}+\delta \Gamma_{hZZ}^{i}\qc (i=1, 2, 3). \label{eq: ren_hZZ}
\end{align}
The tree-level contribution is given by
\begin{align}
\Gamma_{hZZ}^{1, (0)} = 2\kappa_{Z}m_{Z}^{2}\qty(\sqrt{2}G_{F})^{1/2}\qc
\Gamma_{hZZ}^{2, (0)} = \Gamma_{hZZ}^{3, (0)} = 0.
\end{align}
The form factor $\widehat{\Gamma}_{hZZ}^{3}$ is non-zero only when the SM-like Higgs boson is a CP-mixed state.
Therefore this form factor vanishes in the model with CP conservation in the Higgs sector.
Analytic expressions for the 1PI diagrams and counterterms are presented in Ref.~\cite{Kanemura:2016lkz} for the HSM, Ref.~\cite{Kanemura:2015mxa} for the 2HDMs and Ref.~\cite{Kanemura:2016sos} for the IDM.

Similarly, we define the loop-induced $hZ\gamma$ vertex as
\begin{align}
\widehat{\Gamma}_{hZ\gamma}^{\mu\nu}(p_{Z}, p_{\gamma}, p_{h})
=
\qty[g^{\mu\nu}\widehat{\Gamma}_{hZ\gamma}^{1}
+\frac{p_{Z}^{\nu}p_{\gamma}^{\mu}}{p_{h}^{2}}\widehat{\Gamma}_{hZ\gamma}^{2}
+i\epsilon^{\mu\nu\rho\sigma}\frac{p_{Z\rho}p_{\gamma\sigma}}{p_{h}^{2}}\widehat{\Gamma}_{hZ\gamma}^{3}](p_{Z}^{2}, p_{\gamma}^{2}, p_{h}^{2}).
\end{align}
The form factor $\widehat{\Gamma}_{hZ\gamma}^{3}$ vanishes in the model with CP conservation in the Higgs sector.
Analytic expressions for the 1PI diagrams and counterterms are presented in Refs.~\cite{Kanemura:2018esc, Kanemura:2019kjg}.

The renormalized $hf\bar{f}$ vertex can be decomposed as
\begin{align}
\widehat{\Gamma}_{hf\bar{f}}^{\mu}(p_{f}, p_{\bar{f}}, p_{h})&=
\widehat{\Gamma}^{S}_{hf\bar{f}}+\gamma_{5}\widehat{\Gamma}^{P}_{hf\bar{f}}
+\slashed{p}_{f}\widehat{\Gamma}^{V_{1}}_{hf\bar{f}}
+\slashed{p}_{\bar{f}}\widehat{\Gamma}^{V_{2}}_{hf\bar{f}} \notag \\
&\quad
+\slashed{p}_{f}\gamma_{5}\widehat{\Gamma}^{A_{1}}_{hf\bar{f}}
+\slashed{p}_{\bar{f}}\gamma_{5}\widehat{\Gamma}^{A_{2}}_{hf\bar{f}}
+\slashed{p}_{f}\slashed{p}_{\bar{f}}\widehat{\Gamma}^{T}_{hf\bar{f}}
+\slashed{p}_{f}\slashed{p}_{\bar{f}}\gamma_{5}\widehat{\Gamma}^{PT}_{hf\bar{f}},
\end{align}
where $p_{f}\, (p_{\bar{f}})$ is the incoming four-momentum of the fermion (anti-fermion), and $p_{h}$ is the outgoing four-momentum of the SM-like Higgs boson.
Analytic expressions for the renormalized vertices are presented in Ref.~\cite{Kanemura:2016lkz} for the HSM, Ref.~\cite{Kanemura:2015mxa} for the 2HDMs and Ref.~\cite{Kanemura:2016sos} for the IDM.

\subsubsection{Expression of form factors including one-loop corrections}
We list the one-loop contributions to the form factors in terms of the renormalized quantities.
The one-loop propagator corrections appear in the sum in Eq.~\eqref{eq: NLO_form_fact} as the term
\begin{align}
F_{i, \sigma}^{\mathrm{SE}} = \frac{g_{Z}\Gamma_{hZZ}^{1, (0)}}{s-m_{Z}^{2}}\qty[
-g_{\sigma}\frac{\widehat{\Pi}_{ZZ}^{T}(s)}{s-m_{Z}^{2}}
-Q_{e}s_{W}c_{W}\frac{\widehat{\Pi}_{Z\gamma}^{T}(s)}{s}]f_{i}^{(0)},
\end{align}
with the renormalized self-energies $\widehat{\Pi}_{VV'}^{T}(s)$ of the neutral vector bosons.
The renormalized $Ze\bar{e}$ corrections appear as
\begin{align}
F_{i, \sigma}^{Ze\bar{e}} = \frac{g_{Z}\Gamma_{hZZ}^{1, (0)}}{s-m_{Z}^{2}}\qty[
\widehat{\Gamma}_{Zf\bar{f}}^{V}(m_{e}^{2}, m_{e}^{2}, s)
-\sigma \widehat{\Gamma}_{Zf\bar{f}}^{A}(m_{e}^{2}, m_{e}^{2}, s)]f_{i}^{(0)}.
\end{align}
The renormalized $hZV\ (V=Z,\ \gamma)$ corrections appear as
\begin{align}
F_{i, \sigma}^{hZV} = \frac{g_{Z}}{s-m_{Z}^{2}}g_{\sigma}f_{i}^{Z(1)}(s)
+\frac{g_{Z}}{s}Q_{e}s_{W}c_{W}f_{i}^{\gamma(1)}(s),
\end{align}
with
\begin{align}
f_{1}^{Z(1)}(s) &= \widehat{\Gamma}_{hZZ}^{1}(m_{Z}^{2}, s, m_{h}^{2}), \\
f_{2}^{Z(1)}(s) &= -\frac{1}{m_{Z}^{2}}\widehat{\Gamma}_{hZZ}^{2}(m_{Z}^{2}, s, m_{h}^{2}), \\
f_{1}^{\gamma(1)}(s) &= \widehat{\Gamma}_{hZ\gamma}^{1}(m_{Z}^{2}, s, m_{h}^{2}), \\
f_{2}^{\gamma(1)}(s) &= -\frac{1}{m_{h}^{2}}\widehat{\Gamma}_{hZ\gamma}^{2}(m_{Z}^{2}, s, m_{h}^{2}).
\end{align}
The renormalized $he\bar{e}$ corrections appear as
\begin{align}
F_{i, \sigma}^{he\bar{e}} = -g_{Z}g_{\sigma}\qty{
\qty[\widehat{\Gamma}_{he\bar{e}}^{V_{1}}+\sigma \widehat{\Gamma}_{he\bar{e}}^{A_{1}}](t, 0, m_{h}^{2})
-\qty[\widehat{\Gamma}_{he\bar{e}}^{V_{2}}+\sigma \widehat{\Gamma}_{he\bar{e}}^{A_{2}}](0, u, m_{h}^{2})
}f_{i}^{(0)}.
\end{align}
The $1/t$ and $1/u$ terms originated from the fermion propagator are canceled by the vertex corrections \cite{Denner:1992bc}.
However, the renormalized vertices depend on $t$ and $u$, and they cause non-trivial $\cos{\theta}$ dependence.

There are the five $W$ boson mediated and one $Z$ boson mediated box diagrams in the massless limit of the electron.
The amplitudes of the $W$ boson mediated diagrams can be written as
\begin{align}
{\cal M}^{k}_{\sigma \lambda}
&=
\frac{\kappa_{V}}{16\pi^{2}}\delta_{\sigma -}
\sum_{i=1}^{3}C^{k}F_{i}^{k}(s,t){\cal M}_{i, \sigma\lambda}\qc (k=1,2,\cdots,5), \label{eq: W_box}
\end{align}
with
\begin{align}
F_{1}^{k}(s,t) &= F^{k}(s,t), \\
F_{2}^{k}(s,t) &= \frac{1}{2}\qty[F^{k}_{e}(s,t)+F^{k}_{\bar{e}}(s,t)], \\
F_{3}^{k}(s,t) &= \frac{1}{2}\qty[F^{k}_{e}(s,t)-F^{k}_{\bar{e}}(s,t)].
\end{align}
The amplitude of the $Z$ boson mediated diagram has a different structure from others, and it can be written as
\begin{align}
{\cal M}^{6}_{\sigma \lambda}
=
\frac{\kappa_{V}}{16\pi^{2}}\sum_{i=1}^{3}g_{\sigma}^{3}C^{6}F_{i}^{6}(s,t){\cal M}_{i, \sigma \lambda}, \label{eq: Z_box}
\end{align}
with
\begin{align}
F_{1}^{6}(s,t) &= F^{6}(s,t), \\
F_{2}^{6}(s,t) &= \frac{1}{2}\qty[F^{6}_{e}(s,t)+F^{6}_{\bar{e}}(s,t)], \\
F_{3}^{6}(s,t) &= \frac{1}{2}\qty[F^{6}_{e}(s,t)-F^{6}_{\bar{e}}(s,t)].
\end{align}
The expressions of $C^{k}$ and $F_{i}^{k}(s,t)$ are given in Appendix~\ref{sec: boxs}.
We define $B_{i, \sigma}(s, t)$ by
\begin{align}
B_{i, \sigma}(s, t) = \frac{\kappa_{V}}{16\pi^{2}}\qty[\delta_{\sigma -}\sum_{k=1}^{5}C^{k}F_{i}^{k}(s,t)
+g_{\sigma}^{3}C^{6}F_{i}^{6}(s,t)]. \label{eq: form_factors}
\end{align}

Finally, the form factors at one-loop level are given by
\begin{align}
F_{i, \sigma}^{(1)} &=
\frac{g_{Z}}{s-m_{Z}^{2}}\qty{
\Gamma_{hZZ}^{1, (0)}\qty[-g_{\sigma}\frac{\widehat{\Pi}_{ZZ}^{T}(s)}{s-m_{Z}^{2}}
+\widehat{\Gamma}_{Ze\bar{e}}^{\sigma}(m_{e}^{2}, m_{e}^{2}, s)]f_{i}^{(0)}
+g_{\sigma}f_{i}^{Z(1)}(s)} \notag \\
&\quad
+\frac{eQ_{e}}{s}\qty[
-\Gamma_{hZZ}^{1, (0)}\frac{\widehat{\Pi}_{Z\gamma}^{T}(s)}{s-m_{Z}^{2}}f_{i}^{(0)}
+f_{i}^{\gamma(1)}(s)] \notag \\
&\quad
-g_{Z}g_{\sigma}\qty{
\qty[\widehat{\Gamma}_{he\bar{e}}^{V_{1}}+\sigma \widehat{\Gamma}_{he\bar{e}}^{A_{1}}](t, 0, m_{h}^{2})
-\qty[\widehat{\Gamma}_{he\bar{e}}^{V_{2}}+\sigma \widehat{\Gamma}_{he\bar{e}}^{A_{2}}](0, u, m_{h}^{2})}f_{i}^{(0)} \notag \\
&\quad
+ B_{i, \sigma}(s, t)
+\frac{g_{Z}\Gamma_{hZZ}^{1, (0)}}{s-m_{Z}^{2}}g_{\sigma}
\qty[-\frac{1}{2}\Re{\Pi'_{ZZ}(m_{Z}^{2})}-\Delta r]f_{i}^{(0)}.
\end{align}

%

%% file: 04Numerical_results.tex


\section{Numerical results} \label{sec: Numerical_results}
In this section, we begin with an analysis on the behavior of the NLO weak corrections to the cross section of the $e^{+}e^{-}\to hZ$ process in the SM.
We then analyze the deviations from the SM values at NLO in the HSM, the 2HDM and the IDM.

We evaluate the form factors $F_{i, \sigma}^{(1)}$ by using the \texttt{H-COUP} program \cite{Kanemura:2017gbi, Kanemura:2019slf}, where $G_{F}$ is taken as an input.
In order to compare our results in the SM with the previous works~\cite{Denner:1992bc, Belanger:2003sd, Sun:2016bel, Gong:2016jys}, we extend the \texttt{H-COUP} program and take $m_{W}$ as an input instead of $G_{F}$.
With this extension, we have confirmed that our results are in agreement with the previous results.
In the following, we show the results obtained in the scheme where $\alpha_{\mathrm{em}}(0), G_{F}$ and $m_{Z}$ are input parameters.

In order to study the theoretical behavior of the one-loop corrections, we parametrize the differential cross section as
\begin{align}
\dd \sigma
= \dd \sigma_{\mathrm{LO}}(1+\delta_{\mathrm{weak}}+\delta_{\mathrm{em}}),
\end{align}
where $\delta_{\mathrm{weak}}$ and $\delta_{\mathrm{em}}$ denote the relative weak and electromagnetic corrections, respectively.

As we will see below, the NP effects mainly come from $\widehat{\Gamma}_{hZZ}^{1}$ vertex, and that appear independently of $\sigma$ and $\lambda$.
Therefore, we show the results of the unpolarized cross section where the polarization of the $Z$ boson is also summed.
In order to analyze the NP effects in each renormalized quantity, we introduce $\overline{\Delta}_{X}^{\mathrm{EW}}$,
\begin{align}
\overline{\Delta}_{X}^{\mathrm{EW}} = \Delta_{X, \mathrm{NP}}^{\mathrm{EW}}- \Delta_{X, \mathrm{SM}}^{\mathrm{EW}}\qc (X = ZZ,\, Z\gamma,\, Ze\bar{e},\, hZZ,\, hZ\gamma,\, he\bar{e},\, \mathrm{Box},\, \Pi'_{ZZ},\, \Delta r), \label{eq: Del_EW}
\end{align}
with $\Delta_{X}^{\mathrm{EW}}= \sigma_{X}/\sigma_{\mathrm{LO}}$.
We evaluate $\sigma_{X}$ by substituting
\begin{align}
\mathcal{M}_{\sigma\lambda}^{(1)X}(s,t) = \sum_{i=1}^{3}F_{i,\sigma}^{X}\mathcal{M}_{i,\sigma\lambda}
\end{align}
into $\mathcal{M}_{\sigma\lambda}^{(1)}(s,t)$ in Eq.~\eqref{eq: diff_xsec}, where $F_{i,\sigma}^{X}$ is defined in Eq.~\eqref{eq: NLO_form_fact}.
We also evaluate the ratios of the total cross sections to exhibit the deviations from the predictions in the SM,
\begin{align}
\Delta R^{hZ} = \frac{\sigma_{\mathrm{NP}}}{\sigma_{\mathrm{SM}}}-1.
\label{eq: dRhZ}
\end{align}

\subsection{Standard Model}
\begin{figure}[t]
\begin{minipage}{0.45\hsize}
\centering
\includegraphics[scale=0.5]{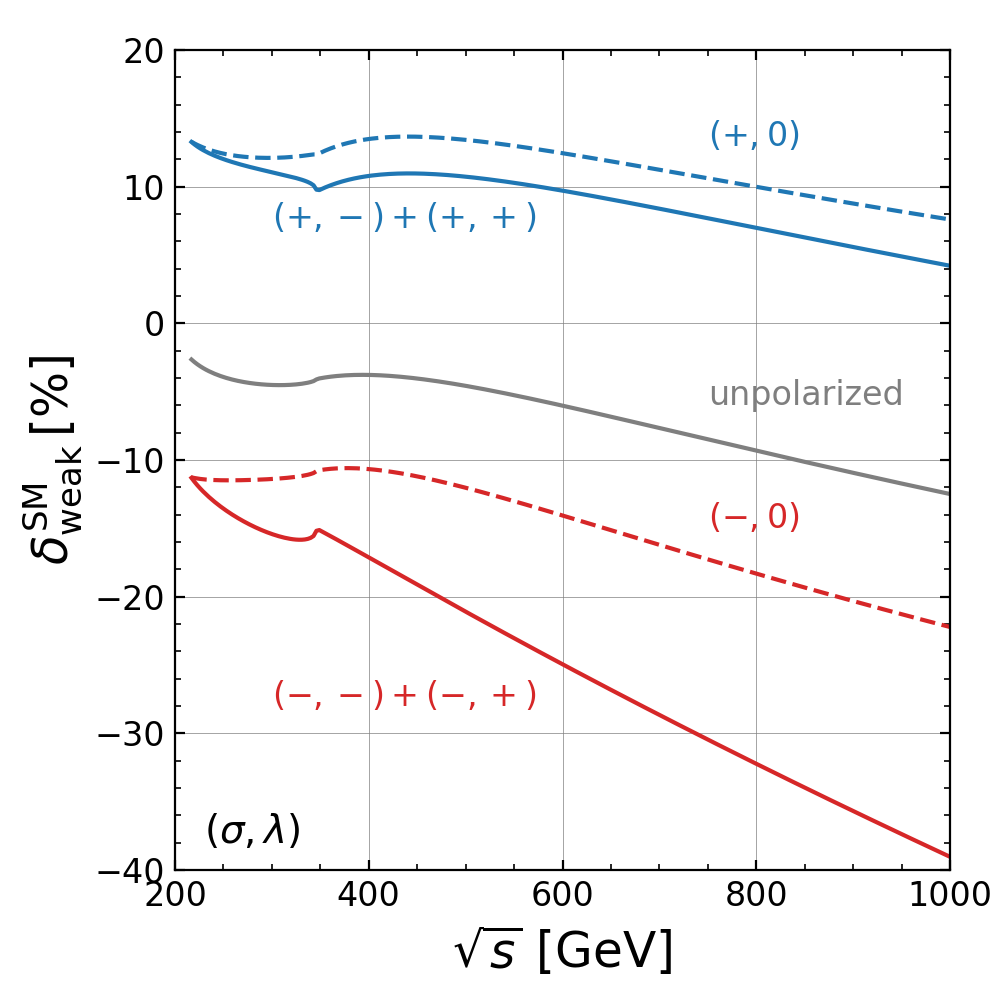}
\end{minipage}
\begin{minipage}{0.45\hsize}
\centering
\includegraphics[scale=0.5]{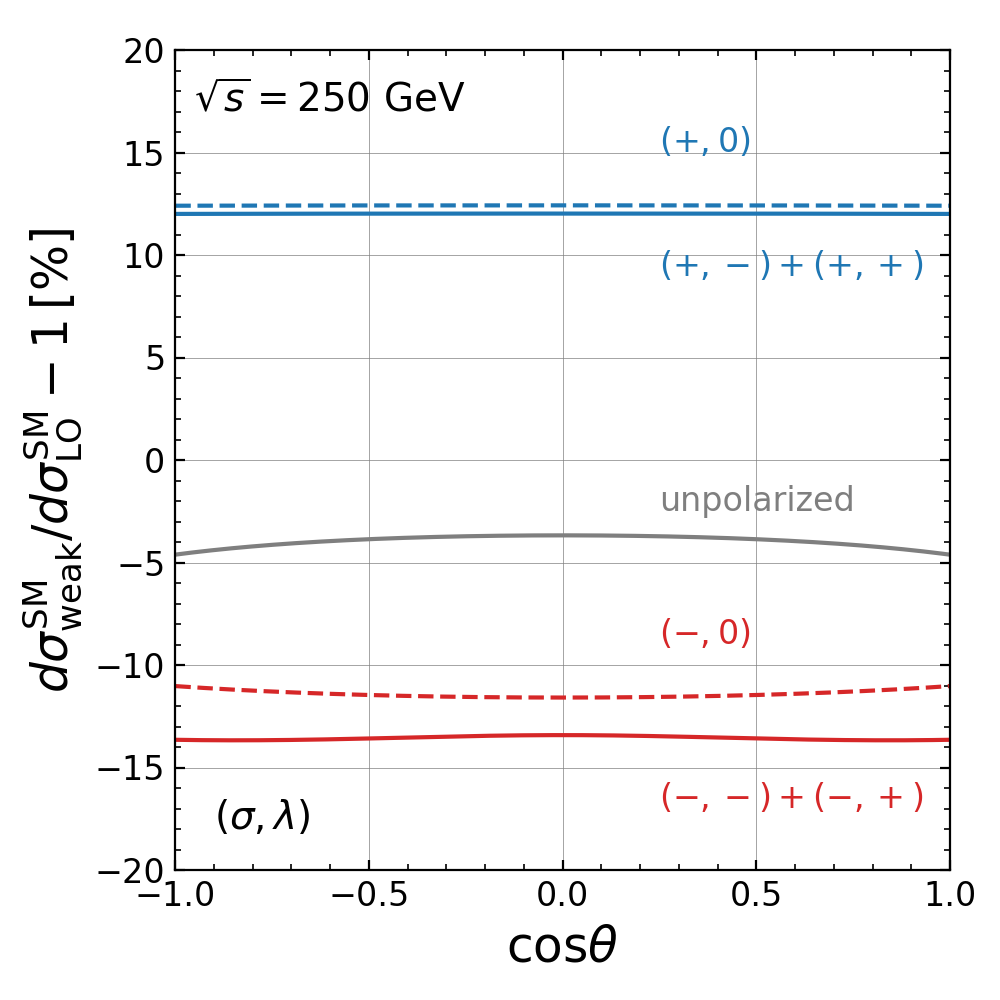}
\end{minipage}
\caption{(Left) Weak corrections to the helicity-dependent cross sections in the SM.
The red lines show the results for the left-handed electron and the right-handed positron, while the blue lines show those for the right-handed electron and the left-handed positron.
The solid (dashed) lines show the results for the transversely (longitudinally) polarized $Z$ bosons. The black solid line corresponds to that for the unpolarized cross section where the polarization of $Z$ boson is also summed.
(Right) Weak corrections to the helicity-dependent differential cross sections in the SM at $\sqrt{s}=250$ GeV. The line colors and styles are the same as those of the left figure.}
\label{fig: del_weak_SM}
\end{figure}

In the left panel of Fig.~\ref{fig: del_weak_SM}, we show the weak one-loop corrections to the helicity-dependent cross sections as a function of the CM energy.
The weak corrections to the cross section for the right-handed electron are positive, and they increase the cross section by about $10\%$.
On the other hand, those for the left-handed electron are negative, and the size of these corrections strongly depends on the CM energy.
The reason for this difference comes from negative contributions from the box diagrams.
Among the six box diagrams, the five $W$ boson mediated diagrams only contribute to the helicity amplitudes for the left-handed electron.
They give negative contributions to the helicity-dependent cross sections for the left-handed electron.
In addition, their effects become relevant at higher energies and give large negative corrections.
The peak around $\sqrt{s}\simeq 350$ GeV corresponds to the threshold at $2m_{t}$ in the top-loop contributions.

In the right panel of Fig.~\ref{fig: del_weak_SM}, we show the weak one-loop corrections to the differential cross sections as a function of the CM energy.
From Eq.~\eqref{eq: form_factors}, we can see that only the $he\bar{e}$ vertex and box corrections cause different $\cos{\theta}$ dependence from those at LO.
At $\sqrt{s}=250$ GeV, this effect is not so large, and the angular distribution of the $Z$ boson is almost determined by the $d_{\sigma, \lambda}^{1} (\theta)$ functions.
At higher energies, the angular distribution of the $Z$ boson is significantly modified through the box contributions \cite{Denner:1992bc}.
However, the size of the cross sections decreases in such a higher energy region.

\subsection{Higgs singlet model}
\begin{figure}[t]
\begin{minipage}{0.45\hsize}
\centering
\includegraphics[scale=0.5]{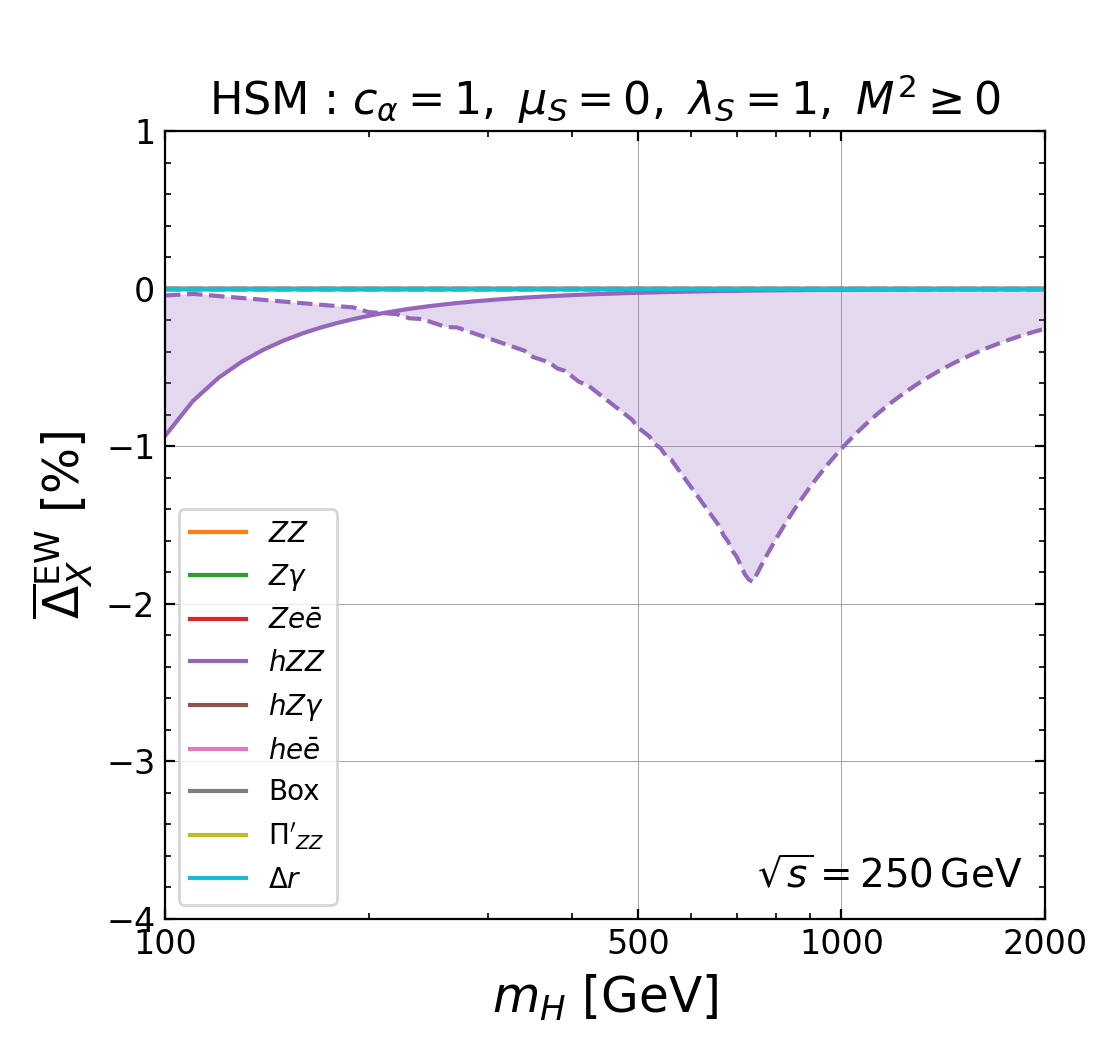}
\end{minipage}
\begin{minipage}{0.45\hsize}
\centering
\includegraphics[scale=0.5]{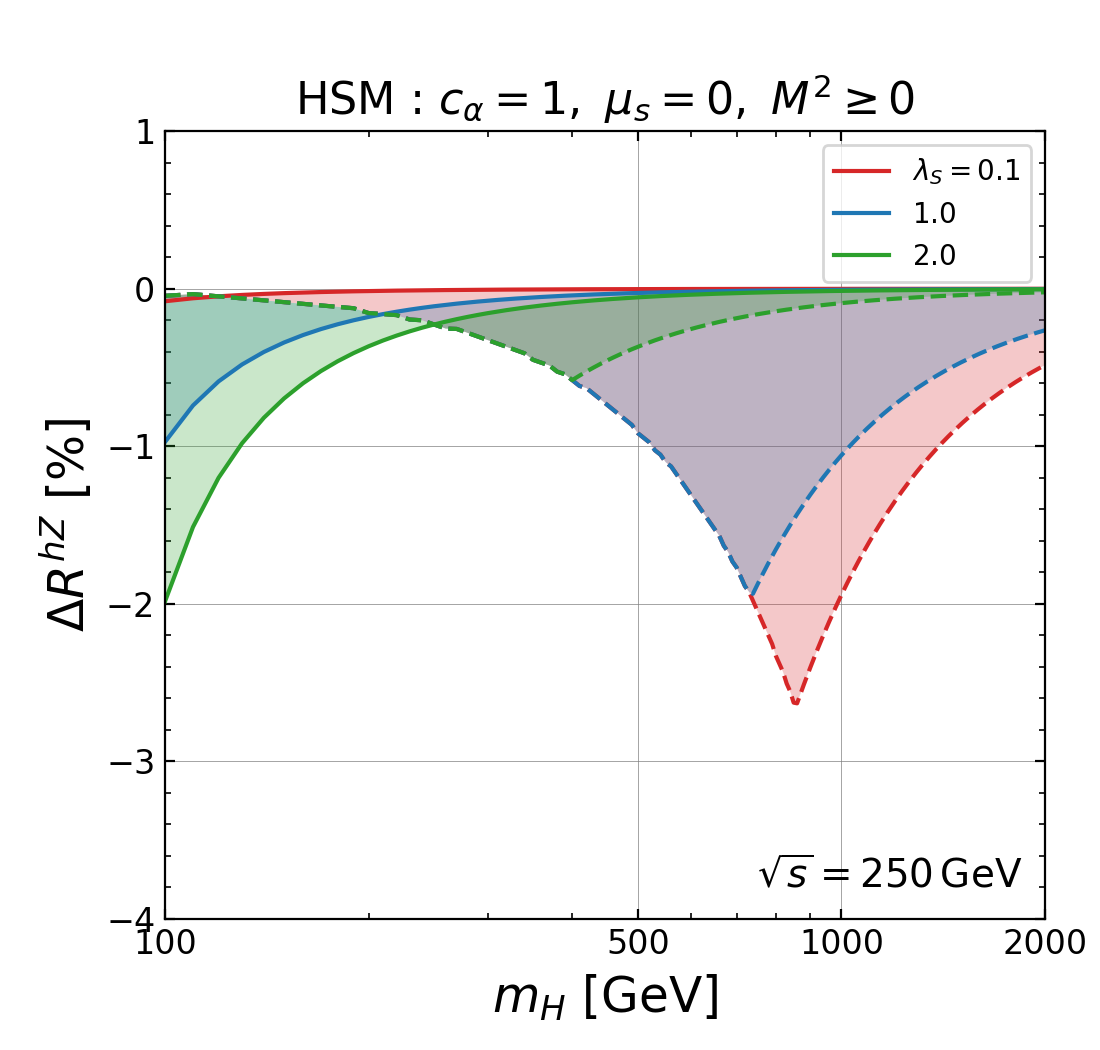}
\end{minipage}
\caption{NP effects in the EW corrections as a function of the mass of the additional Higgs boson in the $Z_{2}$ symmetric HSM at $\sqrt{s}=250$ GeV. The left panel shows $\overline{\Delta}_{X}^{\mathrm{EW}}$ with $\lambda_{S}=1$. The right panel shows $\Delta R^{hZ}$ with $\lambda_{S}=0.1$ (red), $1$ (blue) and $2$ (green). The solid and dashed curves denote the case with the maximal and minimal values of $M^{2}$, respectively. Perturbative unitarity, vacuum stability, avoiding wrong vacua, and the constraints of the EW $S$ and $T$ parameters are imposed.}
\label{fig: Z2-sym_HSM}
\end{figure}

First, we consider the $Z_{2}$ symmetric scenario in the HSM where $c_{\alpha}=1$ and $\mu_{S}=0$.
There remain three input parameters, $m_{H},\, \lambda_{\Phi S}$ and $\lambda_{S}$.
In the following analysis, we impose perturbative unitarity, vacuum stability, avoiding wrong vacua, and the constraints of the EW $S$ and $T$ parameters.
In order to analyze the theoretical behavior,  we here do not impose the constraints from the direct searches of the additional Higgs boson and the Higgs coupling measurements.
In addition, we impose $M^{2}>0$ as in the case of the IDM and the 2HDMs.

In the left panel of Fig.~\ref{fig: Z2-sym_HSM}, we show $\overline{\Delta}_{X}^{\mathrm{EW}}$ defined in Eq.~\eqref{eq: Del_EW} as a function of the mass of the additional Higgs boson in the $Z_{2}$ symmetric HSM at $\sqrt{s}=250$ GeV.
We here take $\lambda_{S}=1$ and scan $\lambda_{\Phi S}$ for $\abs{\lambda_{\Phi S}}< 4\pi$.
In the $Z_{2}$ symmetric HSM, only $\overline{\Delta}_{hZZ}^{\mathrm{EW}}$ takes non-zero value. 
Furthermore, among the components of the renormalized $hZZ$ vertex in Eq.~\eqref{eq: ren_hZZ}, the wave function renormalization factor of the SM-like Higgs boson $\delta Z_{h}$ only gives the NP effects. 
$\delta Z_{h}$ is defined by the two-point function of the SM-like Higgs boson $\Pi_{hh}^{\mathrm{1PI}}(p^{2})$ as
\begin{align}
\delta Z_{h} = \eval{-\dv{p^{2}}\Pi_{hh}^{\mathrm{1PI}}(p^{2})}_{p^{2}=m_{h}^{2}}.
\end{align}
As the NP contributions, there are two $H$ propagated diagrams in $\Pi_{hh}^{\mathrm{1PI}}(p^{2})$.
One of them is proportional to $\lambda_{hhHH}$, while the other is proportional to $\lambda_{hHH}^{2}$.
However, the former does not contribute to $\delta Z_{h}$ because the loop function $A(m_{H})$ does not depend on the external momentum $p^{2}$.
Therefore, only the latter contributes to the helicity amplitude as the NP effect,
\begin{align}
\delta Z_{h}^{\mathrm{HSM}}-\delta Z_{h}^{\mathrm{SM}}
&= -\frac{\lambda_{hHH}^{2}}{16\pi^{2}} \eval{\dv{p^{2}}B_{0}(p^{2}; m_{H}^{2}, m_{H}^{2})}_{p^{2}=m_{h}^{2}}, \label{eq: ddZh_HSM}
\end{align}
with $\lambda_{hHH} = -\lambda_{\Phi S}v$.
We note that this difference does not directly depend on $\lambda_{S}$, but it indirectly determines the possible size of the NP effects through the perturbative unitarity and vacuum stability bounds.

The magnitude of $\overline{\Delta}_{hZZ}^{\mathrm{EW}}$ becomes larger when the mass of the extra Higgs boson is taken to be larger up to around 700 GeV.
This peak corresponds to the point where the minimum value of $M^{2}=m_{H}^{2}-\lambda_{\Phi S}v^{2}$ changes from zero to non-zero due to the perturbative unitarity bound.
At this point, $\lambda_{\Phi S}$ takes the maximal value, and it triggers a sizable effect.
In the case of larger values of $m_{H}$, the magnitude of $\overline{\Delta}_{hZZ}^{\mathrm{EW}}$ monotonically decreases because perturbative unitarity constrains the size of $\lambda_{\Phi S}$.
In such a large mass region, $m_{H}$ is approximately equal to $M$, and the additional Higgs boson almost decouples following the decoupling theorem \cite{Appelquist:1974tg}.

We can also see the relatively large NP effects when the mass of the additional Higgs boson is below 200 GeV.
In this region, $\lambda_{\Phi S}$ takes a negative value satisfying the vacuum stability bound thanks to the sizable $\lambda_{S}$.
While the sign of $\lambda_{\Phi S}$ is flipped, only $\abs{\lambda_{\Phi S}}^{2}$ appears in Eq.~\eqref{eq: ddZh_HSM}.
Therefore, $\overline{\Delta}_{hZZ}^{\mathrm{EW}}$ is negative independently of the sign of $\lambda_{\Phi S}$.

In the right panel of Fig.~\ref{fig: Z2-sym_HSM}, we show the predictions of $\Delta R^{hZ}$ in Eq.~\eqref{eq: dRhZ} as a function of the mass of the additional Higgs boson in the $Z_{2}$ symmetric HSM at $\sqrt{s}=250$ GeV.
We here take $\lambda_{S}$ to $0.1, 1$ and $2$ and scan $\lambda_{\Phi S}$ for $\abs{\lambda_{\Phi S}}< 4\pi$. 
In the $Z_{2}$ symmetric HSM, the $Z_{2}$ symmetry prohibits the mixing of the CP-even states, and $\Delta R^{hZ} = 0$ at LO.
The behavior of $\Delta R^{hZ}$ is only determined by $\delta Z_{h}$, and almost the same as that of $\overline{\Delta}_{hZZ}^{\mathrm{EW}}$. 
The possible magnitude of $\Delta R^{hZ}$ indirectly depends on the value of $\lambda_{S}$ through the conditions of perturbative unitarity and vacuum stability.
For $m_{H}\ge 200$ GeV, the possible magnitude of $\Delta R^{hZ}$ decreases as $\lambda_{S}$ becomes large.
On the other hand, for $m_{H} < 200$ GeV, the possible magnitude of $\Delta R^{hZ}$ increases as $\lambda_{S}$ becomes large because a large negative value of $\lambda_{\Phi S}$ is allowed under the vacuum stability bound.

We here mention the scenario where $\mu_{S}$ softly breaks the $Z_{2}$ symmetry.
We note that $\Delta R^{hZ}$ does not directly depend on $\mu_{S}$, and the behavior of $\Delta R^{hZ}$ is the same as that in the $Z_{2}$ symmetric HSM.
However, $\mu_{S}$ indirectly affects the possible size of the NP effects through the conditions for avoiding wrong vacua.
For example, the region where $m_{H}\lesssim 300\ \mathrm{GeV}$ is excluded if $\mu_{S}=100\ \mathrm{GeV}$.

\begin{figure}[t]
\begin{minipage}{0.45\hsize}
\centering
\includegraphics[scale=0.5]{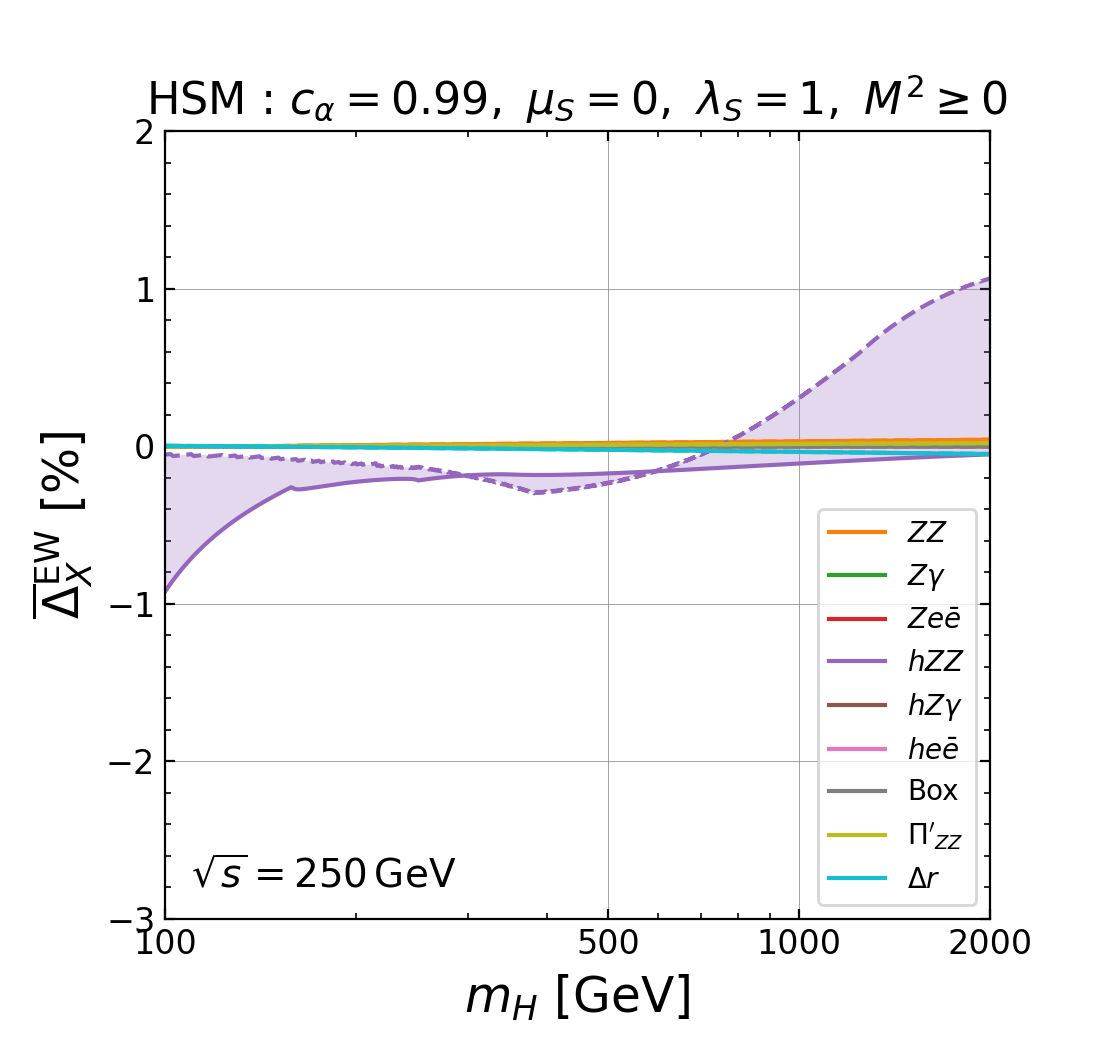}
\end{minipage}
\begin{minipage}{0.45\hsize}
\centering
\includegraphics[scale=0.5]{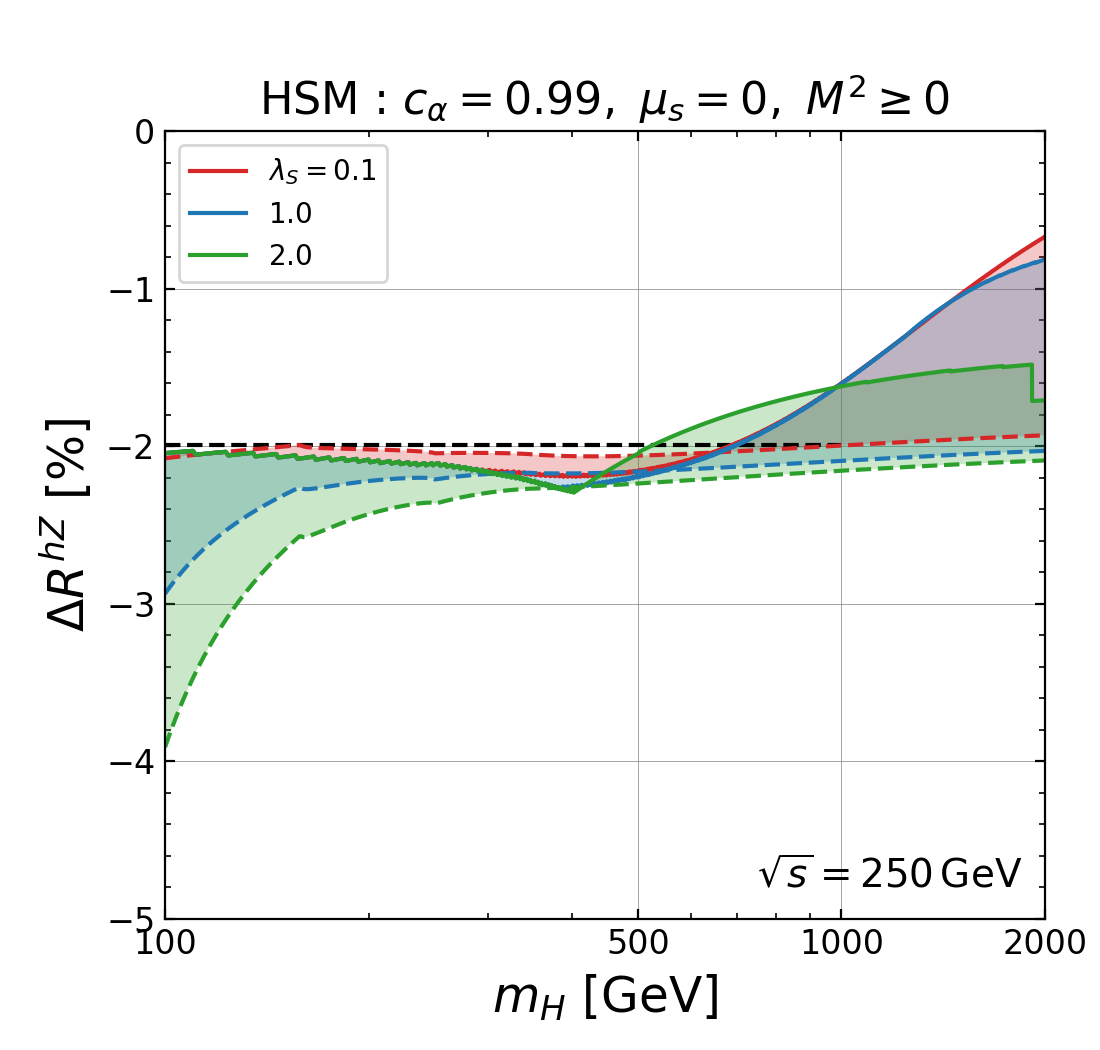}
\end{minipage}
\caption{NP effects in the EW corrections as a function of the mass of the additional Higgs boson in the HSM with $c_{\alpha}=0.99$ and $\mu_{S}=0$ at $\sqrt{s}=250$ GeV. The left panel shows $\overline{\Delta}_{X}^{\mathrm{EW}}$ with $\lambda_{S}=1$. The right panel shows $\Delta R^{hZ}$ with $\lambda_{S}=0.1$ (red), $1$ (blue) and $2$ (green). The solid and dashed curves denote the case with the maximal and minimal values of $M^{2}$, respectively. The black dashed line shows the size of the LO deviation due to the mixing of the CP-even states. Perturbative unitarity, vacuum stability, avoiding wrong vacua, and the constraints of the EW $S$ and $T$ parameters are imposed.}
\label{fig: Z2-broken_HSM}
\end{figure}

Next, we consider the scenario with the mixing of the CP-even states.
In the left panel of Fig.~\ref{fig: Z2-broken_HSM}, we show $\overline{\Delta}_{X}^{\mathrm{EW}}$ as a function of the mass of the additional Higgs boson with $c_{\alpha}=0.99$ and $\mu_{S}=0$ at $\sqrt{s}=250$ GeV.
We here take $\lambda_{S}=1$ and scan $\lambda_{\Phi S}$ for $\abs{\lambda_{\Phi S}}< 4\pi$.
We note that not only the renormalized $hZZ$ vertex but also the other renormalized quantities differ from the SM values unlike the case in the $Z_{2}$ symmetric HSM.
However, the magnitude of $\overline{\Delta}_{hZZ}^{\mathrm{EW}}$ is larger than that of the others.
For $m_{H}\lesssim 700\, \mathrm{GeV}$, $\overline{\Delta}_{hZZ}^{\mathrm{EW}}$ takes a negative value, while it takes a positive value for $m_{H}> 700\ \mathrm{GeV}$.
In order to realize the finite mixing of the CP-even states with a large mass of the additional Higgs boson, the Higgs quartic couplings should take large values, and it triggers a so-called non-decoupling effect.

In the right panel of Fig.~\ref{fig: Z2-broken_HSM}, we show the predictions of $\Delta R^{hZ}$ as a function of the mass of the additional Higgs boson with $c_{\alpha}=0.99$ and $\mu_{S}=0$ at $\sqrt{s}=250$ GeV.
We here take $\lambda_{S}$ to $0.1, 1$ and $2$ and scan $\lambda_{\Phi S}$ for $\abs{\lambda_{\Phi S}}< 4\pi$. 
If there is the mixing of the CP-even states, the LO cross section decreases from its SM value.
When $c_{\alpha}=0.99$, the size of deviation is $\Delta R^{hZ} = -s_{\alpha}^{2} \simeq -0.02$ at LO.
We can see that the magnitude of one-loop effects is compatible with that of the LO contribution, and the NP effects sizably change the predictions for $\Delta R^{hZ}$.
The behavior of $\Delta R^{hZ}$ is almost the same as that of $\overline{\Delta}_{hZZ}^{\mathrm{EW}}$. 
For $m_{H}\lesssim 500\, \mathrm{GeV}$, $\overline{\Delta}_{hZZ}^{\mathrm{EW}}$ increases the magnitude of $\Delta R^{hZ}$, while it decreases the magnitude of $\Delta R^{hZ}$ for $m_{H}> 500\ \mathrm{GeV}$.
For $m_{H}\ge 1000$ GeV, the possible magnitude of $\Delta R^{hZ}$ decreases as $\lambda_{S}$ becomes large.
On the other hand, for $m_{H} < 500$ GeV, it increases as $\lambda_{S}$ becomes large.

\subsection{Inert doublet model}
\begin{figure}[t]
\begin{minipage}{0.45\hsize}
\centering
\includegraphics[scale=0.5]{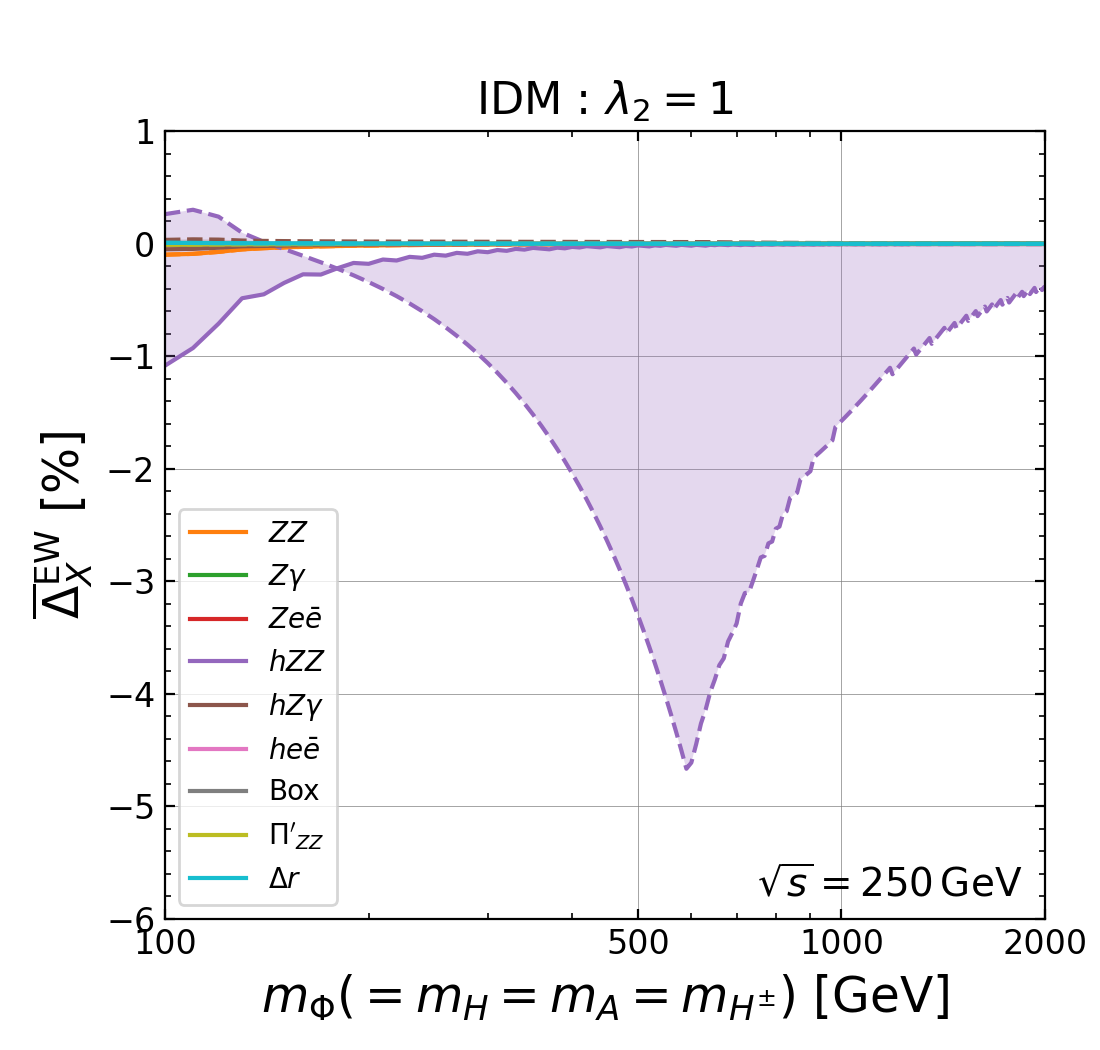}
\end{minipage}
\begin{minipage}{0.45\hsize}
\centering
\includegraphics[scale=0.5]{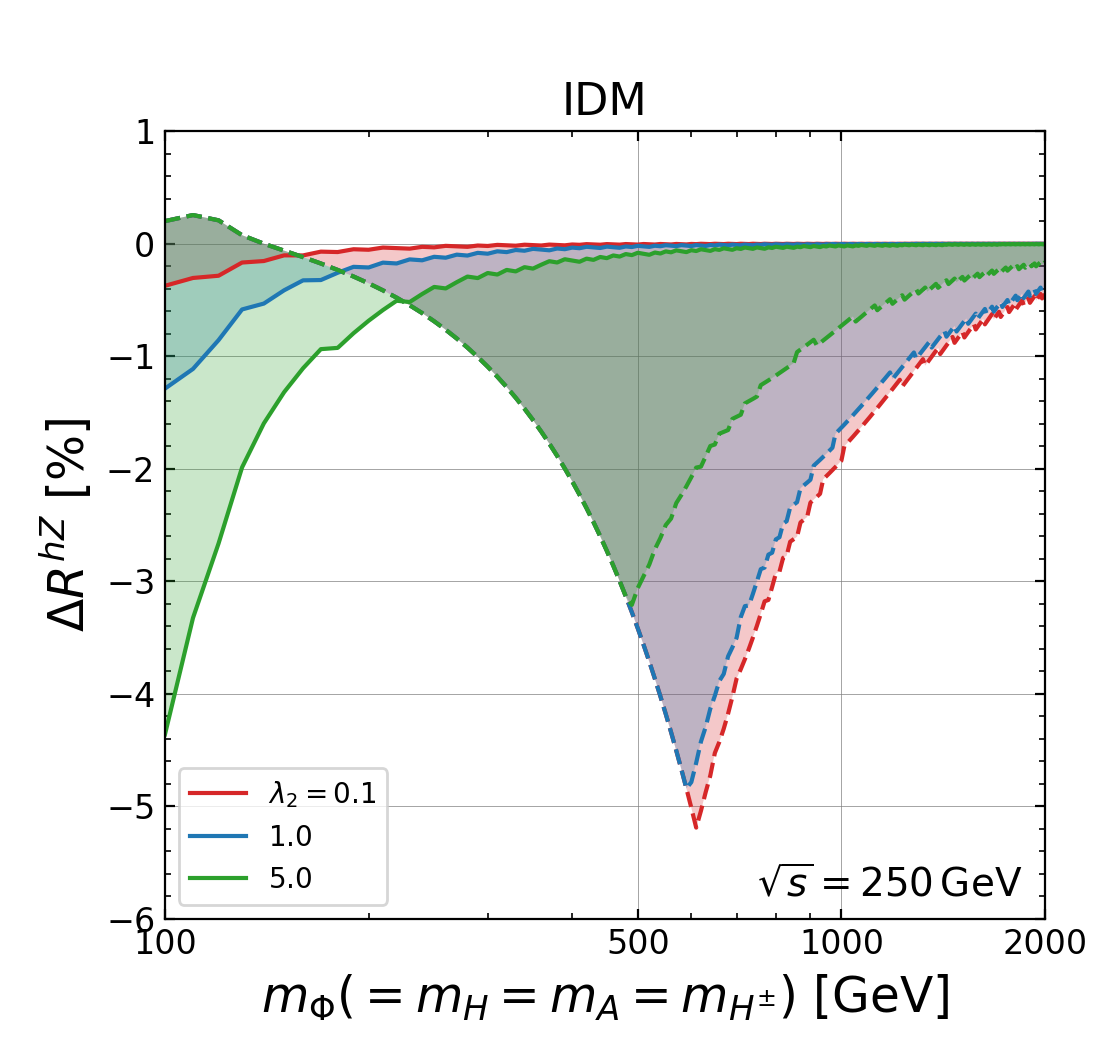}
\end{minipage}
\caption{NP effects in the EW corrections as a function of the masses of the additional Higgs bosons in the IDM at $\sqrt{s}=250$ GeV. We take $m_{H}=m_{A}=m_{H^{\pm}}$. The left panel shows $\overline{\Delta}_{X}^{\mathrm{EW}}$ with $\lambda_{2}=1$. The right panel shows $\Delta R^{hZ}$ with $\lambda_{2}=0.1$ (red), $1$ (blue) and $5$ (green). The solid and dashed curves denote the case with the maximal and minimal value of $M^{2}$, respectively. Perturbative unitarity and vacuum stability bounds and the constraints on the $S$ and $T$ parameters are imposed.}
\label{fig: dR_IDM}
\end{figure}

In the IDM, the $Z_{2}$ symmetry prohibits the mixing of the CP-even states, and $\Delta R^{hZ} = 0$ at LO, similarly to the case in the $Z_{2}$ symmetric HSM.
In the following analysis, we assume that the additional Higgs bosons are degenerate in their mass, $m_{\Phi}\equiv m_{H}=m_{H^{\pm}}=m_{A}$.
There remain two input parameters, $\lambda_{2}$ and $M^{2}$.
We impose perturbative unitarity, vacuum stability, avoiding wrong vacua, and the constraints of the EW $S$ and $T$ parameters.
In order to analyze the theoretical behavior,  we here do not impose the constraints from the direct searches of the additional Higgs boson, the Higgs coupling measurements and the dark matter experiments.

In the left panel of Fig.~\ref{fig: dR_IDM}, we show $\overline{\Delta}_{X}^{\mathrm{EW}}$ as a function of the mass of the additional Higgs bosons in the IDM at $\sqrt{s}=250\, \mathrm{GeV}$.
We here take $\lambda_{2}=1$ and scan $M^{2}$ for $0\leq M^{2}\leq (3\, \mathrm{TeV})^{2}$.
We note that not only the renormalized $hZZ$ vertex but also the other renormalized quantities differ from the SM values unlike in the $Z_{2}$ symmetric HSM.
This is because the additional Higgs bosons are charged under the SM gauge group.
However, the magnitude of $\overline{\Delta}_{hZZ}^{\mathrm{EW}}$ is larger than that of the others in most cases.
In the IDM, there are two main contributions to $\overline{\Delta}_{hZZ}^{\mathrm{EW}}$.
The first one is $\lambda_{h\Phi\Phi}^{2}$ terms originated from $\delta Z_{h}$, similarly to the case in the $Z_{2}$ symmetric HSM.
In addition, there are 1PI diagram contributions proportional to $\lambda_{h\Phi\Phi}$, where the additional Higgs bosons propagate internal lines.
The couplings $\lambda_{h\Phi\Phi}$ are proportional to $(m_{\Phi}^{2}-M^{2})/v$, and large corrections appear when one consider the sizable differences between $m_{\Phi}$ and $M$.
In general, $\delta Z_{h}$ governs the magnitude of $\overline{\Delta}_{hZZ}^{\mathrm{EW}}$, and its behavior is almost the same as that in the $Z_{2}$ symmetric HSM.
The maximal deviation in the IDM is larger than that in the $Z_{2}$ symmetric HSM because we have more than one additional Higgs boson running in the loop in the IDM.
It monotonically decreases when $m_{\Phi}> 600\ \mathrm{GeV}$ following the decoupling theorem.

We can also see the sizable NP effects when $m_{\Phi}$ is below 200 GeV, similarly to the case in the HSM.
In addition, $\overline{\Delta}_{hZZ}^{\mathrm{EW}}$ can be positive if $m_{\Phi}$ is lighter than about 150 GeV.
This is because of the contributions from 1PI diagrams.
While $\delta Z_{h}$ gives negative contributions to $\overline{\Delta}_{hZZ}^{\mathrm{EW}}$, 1PI diagrams for the $hZZ$ vertex give positive contributions.
As we have already mentioned, $\delta Z_{h}$ generally gives larger contributions than 1PI diagram contributions.
However, if $\lambda_{h\Phi\Phi}$ is not so large, 1PI diagram contributions can overcome the contribution of $\delta Z_{h}$, and there are parameter points, where $\overline{\Delta}_{hZZ}^{\mathrm{EW}}$ is positive.

In the right panel of Fig.~\ref{fig: dR_IDM}, we show the predictions of $\Delta R^{hZ}$ as a function of the mass of the additional Higgs bosons in the IDM at $\sqrt{s}=250$ GeV.
We here take $\lambda_{2}$ to $0.1, 1$ and $5$ and scan $M^{2}$ for $0\leq M^{2}\leq (3\, \mathrm{TeV})^{2}$.
The behavior of $\Delta R^{hZ}$ is almost the same as that of $\overline{\Delta}_{hZZ}^{\mathrm{EW}}$. 
The possible magnitude of $\Delta R^{hZ}$ indirectly depends on the value of $\lambda_{2}$ through the conditions of perturbative unitarity and vacuum stability.
For $m_{\Phi}\ge 200$ GeV, the possible magnitude of $\Delta R^{hZ}$ decreases as $\lambda_{2}$ becomes large.
On the other hand, for $m_{\Phi} < 200$ GeV, the possible magnitude of $\Delta R^{hZ}$ increases as $\lambda_{2}$ becomes large, similarly to the case in the $Z_{2}$ symmetric HSM.

\subsection{Two Higgs doublet model}
First, we consider the alignment limit with $s_{\beta-\alpha}=1$, where $\Delta R^{hZ} = 0$ at LO.
We also assume that the additional Higgs bosons are degenerate in mass, $m_{\Phi}\equiv m_{H}=m_{H^{\pm}}=m_{A}$.
There remain two input parameters, $\tan{\beta}$ and $M^{2}$.
We impose perturbative unitarity, vacuum stability, avoiding wrong vacua, and the constraints of the EW $S$ and $T$ parameters.
In order to analyze the theoretical behavior,  we here do not impose the constraints from the direct searches of the additional Higgs boson, the Higgs coupling measurements and the flavor measurements.

We analyze all the four types of 2HDMs, and it turns out that predictions for $\overline{\Delta}_{X}^{\mathrm{EW}}$ and $\Delta R^{hZ}$ are almost the same.
This is because differences among the four types of 2HDMs appear through the down-type quark and lepton Yukawa interactions with the SM-like Higgs boson.
As we will see later, the magnitude of $\overline{\Delta}_{hZZ}^{\mathrm{EW}}$ is larger than that of the others in the 2HDMs, similarly to the case in the HSM and the IDM.
In $\overline{\Delta}_{hZZ}^{\mathrm{EW}}$, the top-quark contributions dominate fermionic contributions, and there is no sizable difference among the four types of 2HDMs.
Therefore, we show the predictions for $\Delta R^{hZ}$ in the Type-I 2HDM as a representative in the following.

\begin{figure}[t]
\begin{minipage}{0.45\hsize}
\centering
\includegraphics[scale=0.5]{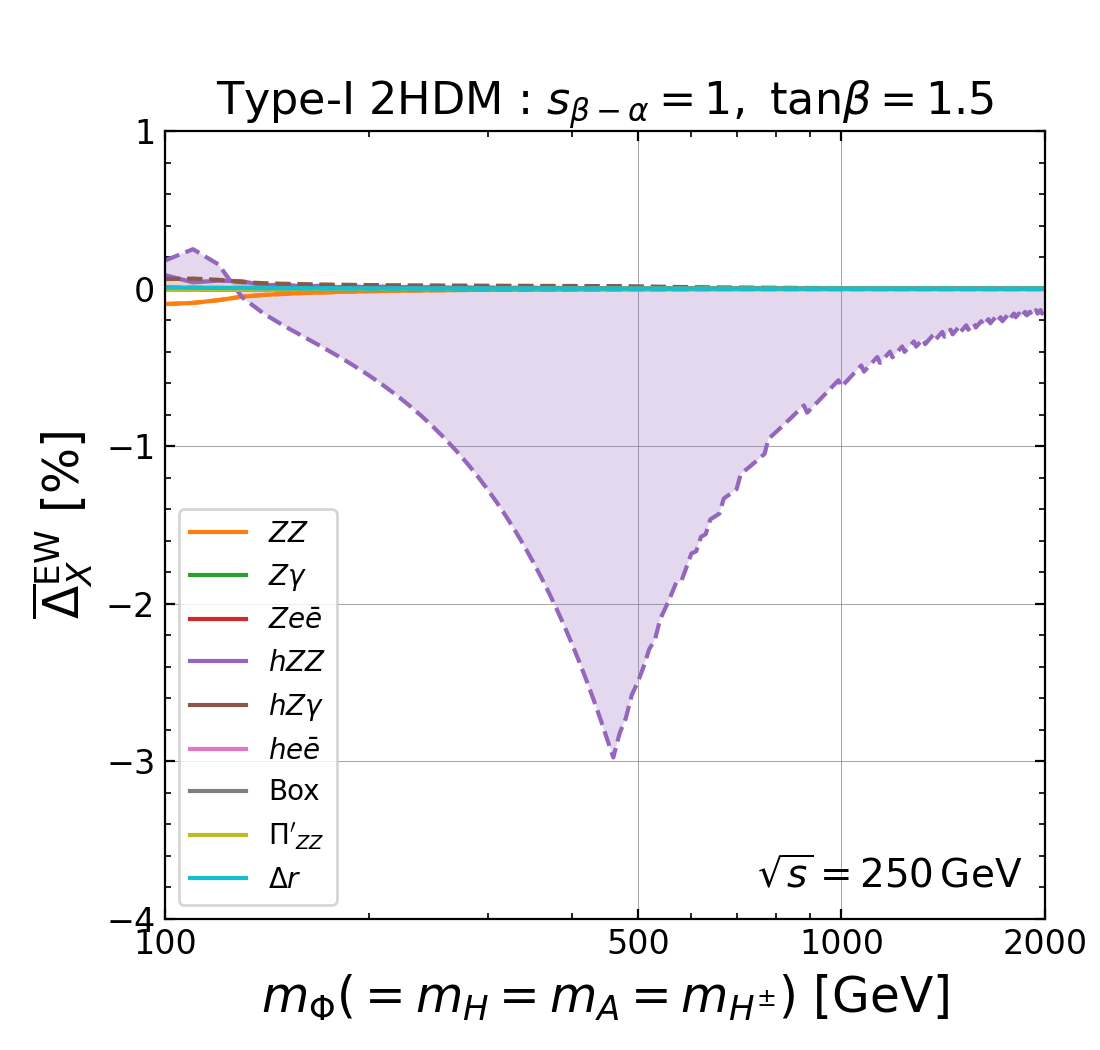}
\end{minipage}
\begin{minipage}{0.45\hsize}
\centering
\includegraphics[scale=0.5]{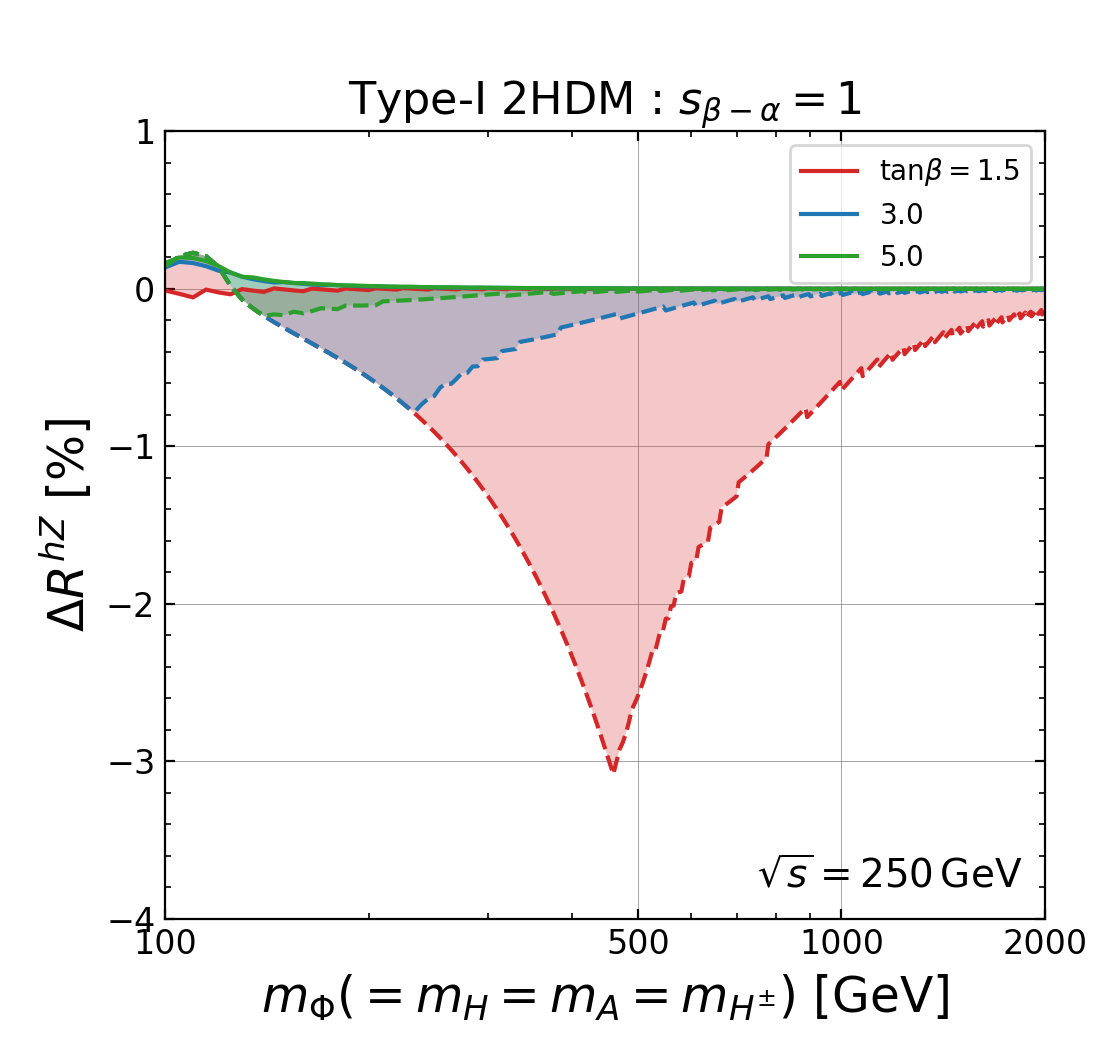}
\end{minipage}
\caption{NP effects in the EW corrections as a function of masses of the additional Higgs bosons in the Type-I 2HDM with $s_{\beta-\alpha}=1$ at $\sqrt{s}=250$ GeV. We take $m_{H}=m_{A}=m_{H^{\pm}}$. The left panel shows $\overline{\Delta}_{X}^{\mathrm{EW}}$ with $\tan{\beta}=1.5$. The right panel shows $\Delta R^{hZ}$ with $\tan{\beta}=1.5$ (red), $3$ (blue) and $5$ (green). The solid and dashed curves denote the case with the maximal and minimal value of $M^{2}$, respectively. Perturbative unitarity and vacuum stability bounds and the constraints on the $S$ and $T$ parameters are imposed.}
\label{fig: dR_2HDM_alignment}
\end{figure}

In the left panel of Fig.~\ref{fig: dR_2HDM_alignment}, we show $\overline{\Delta}_{X}^{\mathrm{EW}}$ as a function of the mass of the additional Higgs bosons in the Type-I 2HDM with $s_{\beta-\alpha}=1$ at $\sqrt{s}=250\, \mathrm{GeV}$.
We here take $\tan{\beta}=1.5$ and scan $M^{2}$ for $0\leq M^{2}\leq (3\, \mathrm{TeV})^{2}$.
We note that not only the renormalized $hZZ$ vertex but also the other renormalized quantities
differ from the SM values because the additional Higgs bosons interact with the gauge bosons, the quarks and the leptons.
Qualitative behaviors of $\overline{\Delta}_{X}^{\mathrm{EW}}$ are almost the same as those in the $Z_{2}$ symmetric HSM and the IDM except for $m_{\Phi}\le 200\, \mathrm{GeV}$.
The magnitude of $\overline{\Delta}_{hZZ}^{\mathrm{EW}}$ is larger than that of the others in most of the parameter space.
It monotonically decreases when $m_{\Phi}> 450\ \mathrm{GeV}$ following the decoupling theorem.

There is no sizable negative correction below $200\, \mathrm{GeV}$ unlike in the $Z_{2}$ symmetric HSM and the IDM.
This is because the Higgs quartic couplings in the 2HDMs are more constrained by vacuum stability than  $\lambda_{S}$ in the HSM and $\lambda_{2}$ in the IDM.
If $m_{\Phi}$ is lighter than about 150 GeV, $\overline{\Delta}_{X}^{\mathrm{EW}}$ can be positive due to the 1PI diagram contributions, similarly to the case in the IDM.

In the right panel of Fig.~\ref{fig: dR_2HDM_alignment}, we show the predictions of $\Delta R^{hZ}$ as a function of the mass of the additional Higgs bosons in the Type-I 2HDM at $\sqrt{s}=250$ GeV.
We here take $\tan{\beta}$ to $1.5, 3$ and $5$ and scan $M^{2}$ for $0\leq M^{2}\leq (3\, \mathrm{TeV})^{2}$.
The behavior of $\Delta R^{hZ}$ is almost the same as that of $\overline{\Delta}_{hZZ}^{\mathrm{EW}}$. 
The possible magnitude of $\Delta R^{hZ}$ decreases as $\tan{\beta}$ becomes large due to the perturbative unitarity and vacuum stability bounds.

\begin{figure}[t]
\begin{minipage}{0.45\hsize}
\centering
\includegraphics[scale=0.5]{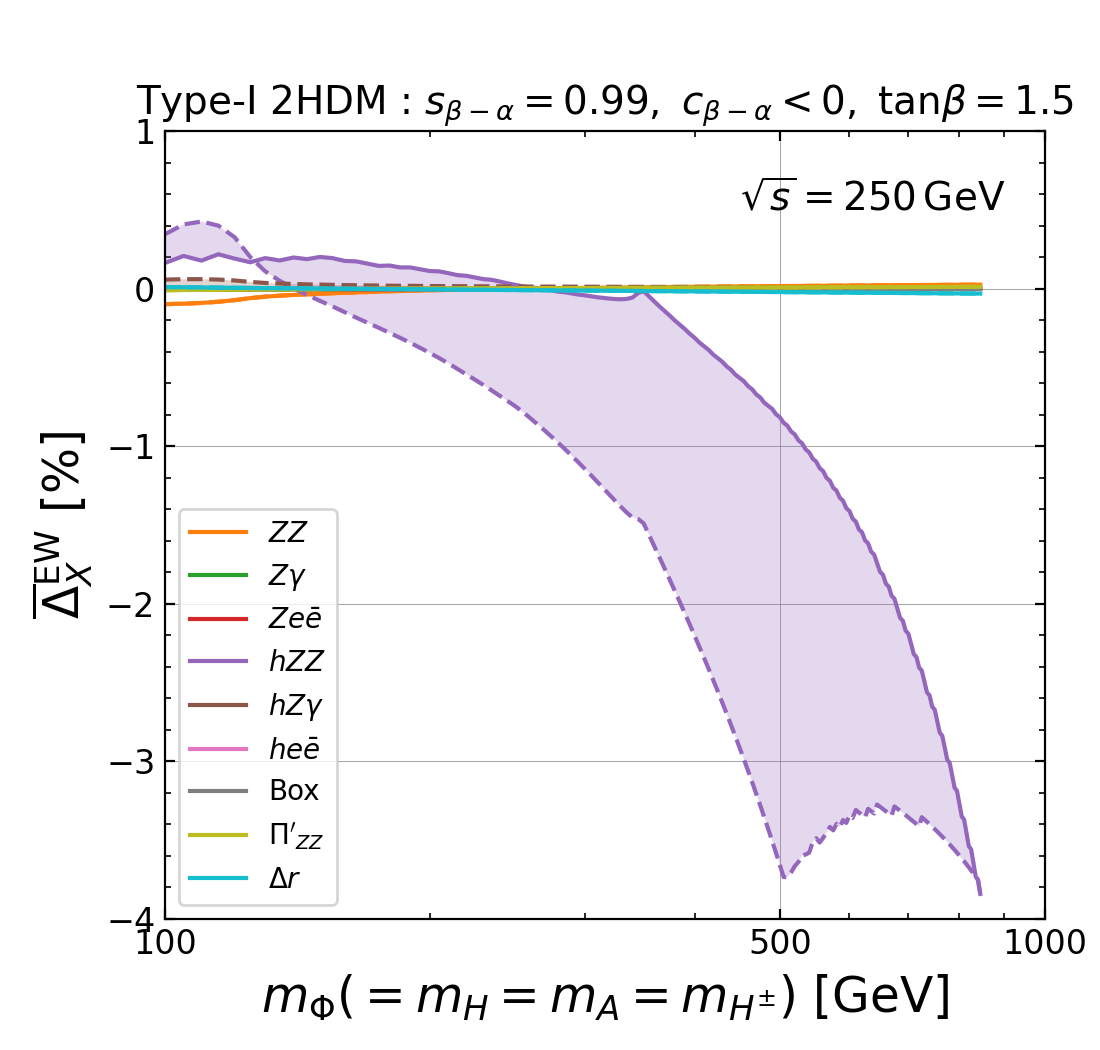}
\end{minipage}
\begin{minipage}{0.45\hsize}
\centering
\includegraphics[scale=0.5]{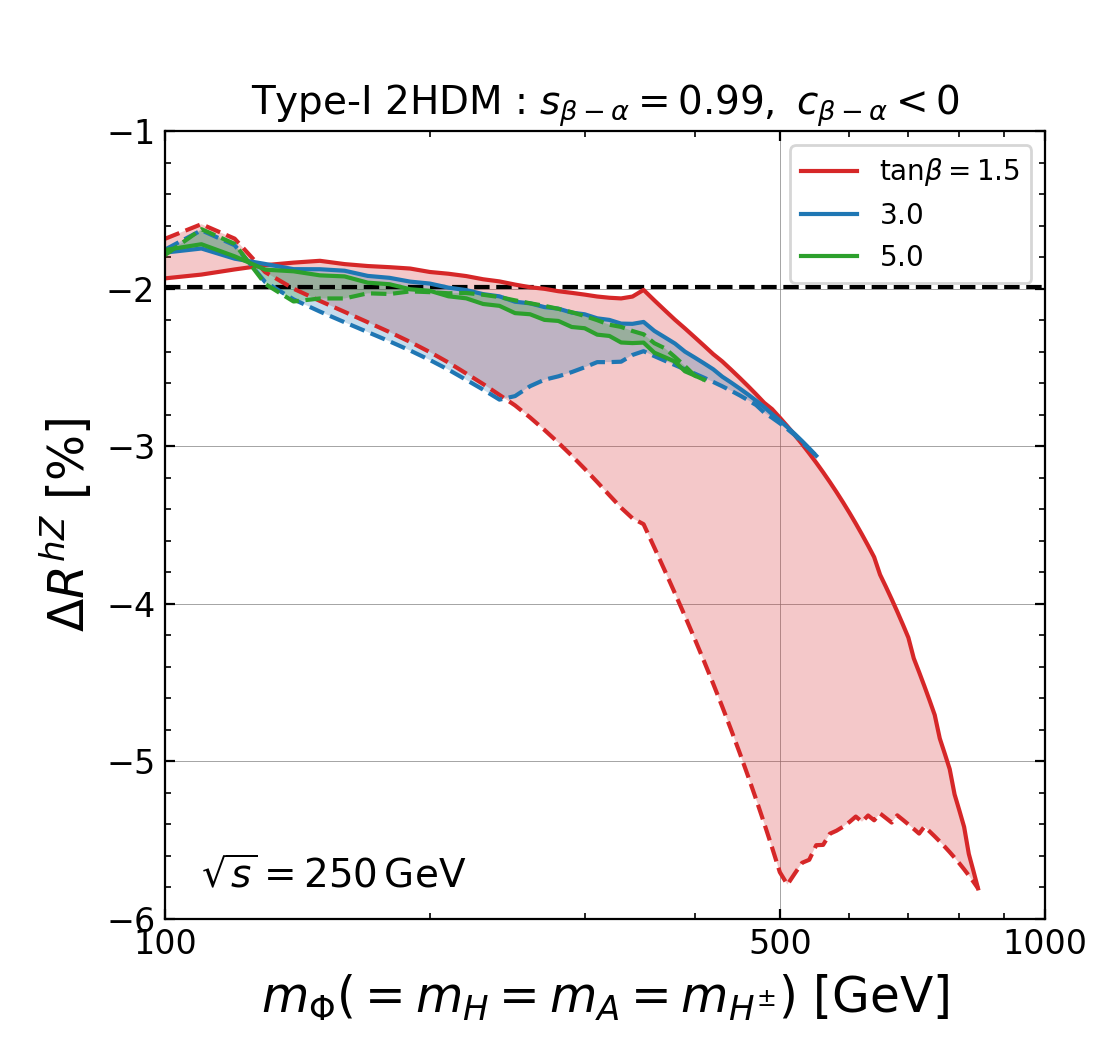}
\end{minipage} \\
\begin{minipage}{0.45\hsize}
\centering
\includegraphics[scale=0.5]{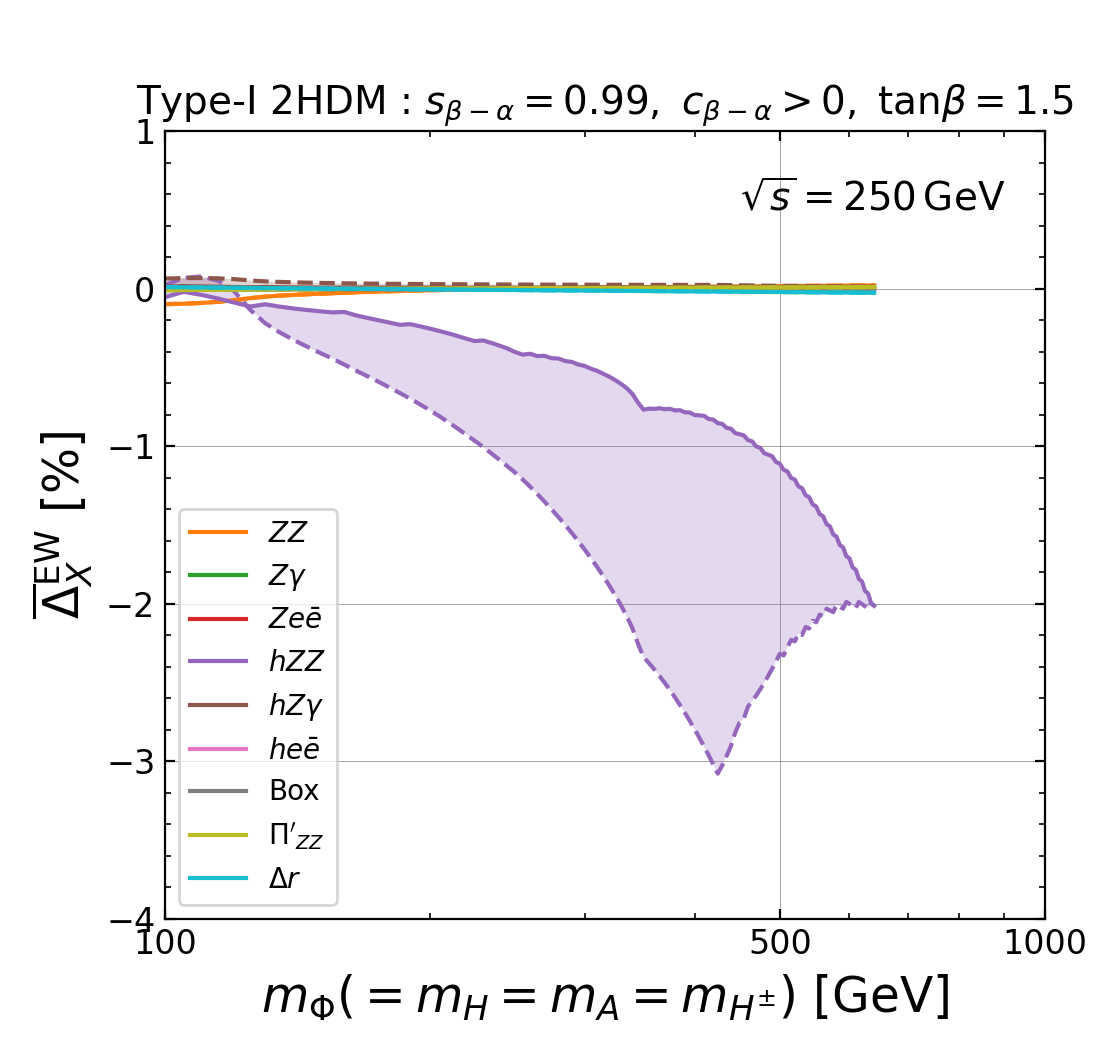}
\end{minipage}
\begin{minipage}{0.45\hsize}
\centering
\includegraphics[scale=0.5]{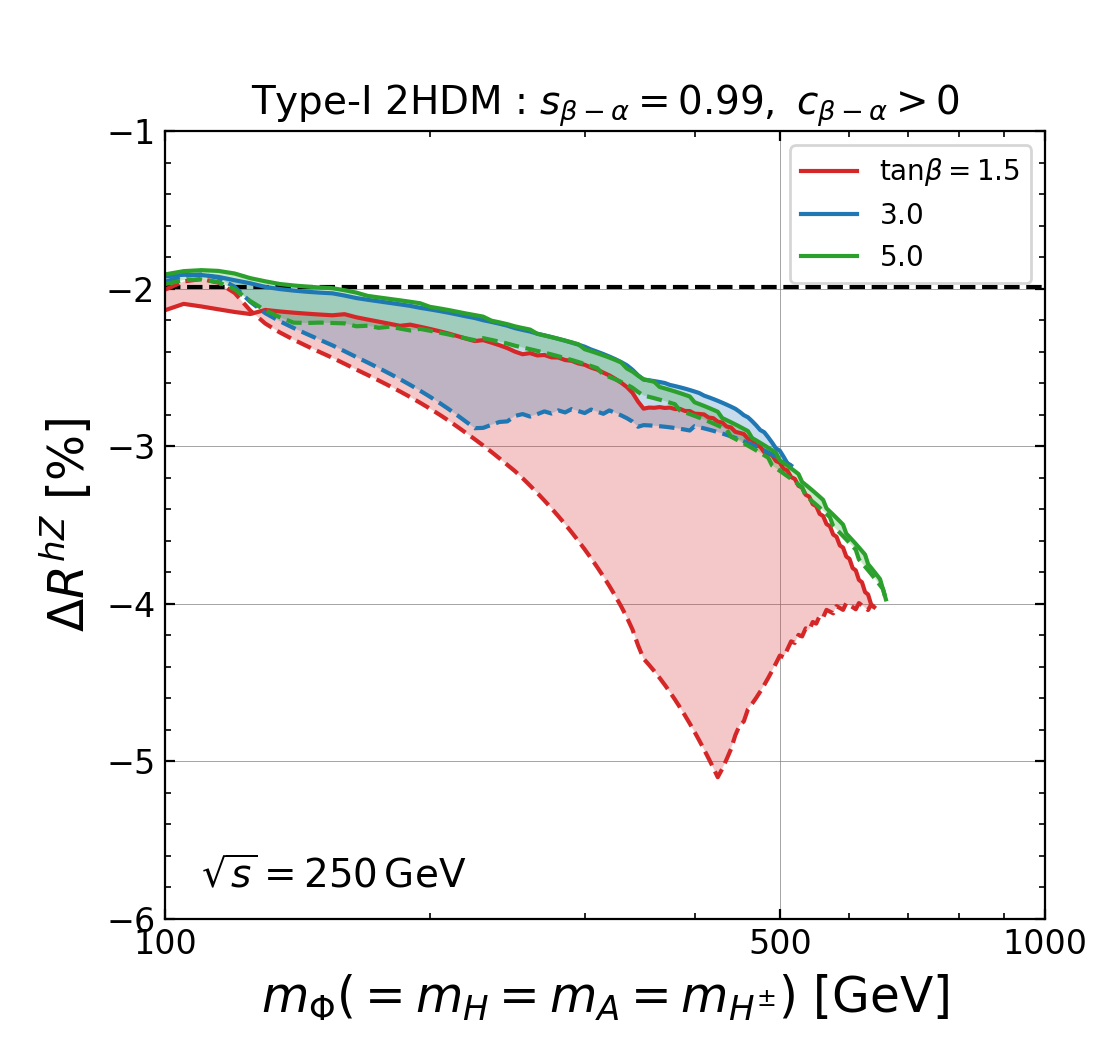}
\end{minipage}
\caption{NP effects in the EW corrections as a function of masses of the additional Higgs bosons in the Type-I 2HDM with $s_{\beta-\alpha}=0.99$ at $\sqrt{s}=250$ GeV. We take $m_{H}=m_{A}=m_{H^{\pm}}$. The top left panel shows $\overline{\Delta}_{X}^{\mathrm{EW}}$ with $\tan{\beta}=1.5$ and $c_{\beta-\alpha}<0$. The top right panel shows $\Delta R^{hZ}$ with $\tan{\beta}=1.5$ (red), $3$ (blue) and $5$ (green) and $c_{\beta-\alpha}<0$. The bottom panels correspond to the case with $c_{\beta-\alpha}>0$. The solid and dashed curves denote the case with the maximal and minimal value of $M^{2}$, respectively. The black dashed line shows the size of the LO deviation due to the mixing. Perturbative unitarity and vacuum stability bounds and the constraints on the $S$ and $T$ parameters are imposed.}
\label{fig: dR_2HDM_nonalignment}
\end{figure}

Next, we consider the scenario with the mixing of the CP-even states.
In the top (bottom) left panel of Fig.~\ref{fig: dR_2HDM_nonalignment}, we show $\overline{\Delta}_{X}^{\mathrm{EW}}$ as a function of the mass of the additional Higgs bosons in the Type-I 2HDM with $s_{\beta-\alpha}=0.99$ and $c_{\beta-\alpha}<0$ ($c_{\beta-\alpha}>0$) at $\sqrt{s}=250\, \mathrm{GeV}$.
We here take $\tan{\beta}=1.5$ and scan $M^{2}$ for $0\leq M^{2}\leq (3\, \mathrm{TeV})^{2}$.
The magnitude of $\overline{\Delta}_{hZZ}^{\mathrm{EW}}$ is larger than that of the others independently of the sign of $c_{\beta-\alpha}$.
In addition, $\overline{\Delta}_{hZZ}^{\mathrm{EW}}$ takes a negative value except for the $m_{H}\lesssim 300\, \mathrm{GeV}$ with $c_{\beta-\alpha}<0$ unlike in the HSM.
We can see the non-decoupling effect in a large mass region of the additional Higgs bosons because they cannot decouple while keeping the finite mixing of the CP-even states, similarly to the case in the HSM.
The maximal value of $m_{\Phi}$ is about 900 GeV for $s_{\beta-\alpha}=0.99$ with $c_{\beta-\alpha}<0$, while it is about 600 GeV with $c_{\beta-\alpha}>0$ for $\tan{\beta}=1.5$.

In the top (bottom) right panel of Fig.~\ref{fig: dR_2HDM_nonalignment}, we show $\Delta R^{hZ}$ as a function of the mass of the additional Higgs bosons in the Type-I 2HDM with $s_{\beta-\alpha}=0.99$ and $c_{\beta-\alpha}<0$ ($c_{\beta-\alpha}>0$) at $\sqrt{s}=250\, \mathrm{GeV}$.
We here take $\tan{\beta}$ to $1.5,\, 3$ and $5$ and scan $M^{2}$ for $0\leq M^{2}\leq (3\, \mathrm{TeV})^{2}$.
The LO cross section decreases from its SM value due to the mixing of the CP-even states.
When $s_{\beta-\alpha} = 0.99$, the size of deviation is $\Delta R^{hZ} = -c_{\beta-\alpha}^{2}\simeq -0.02$ at LO.
We can see that the magnitude of one-loop effects is compatible with that of the LO contribution,  and the NP effects sizably change the predictions for $\Delta R^{hZ}$.
The behavior of $\Delta R^{hZ}$ is almost the same as that of $\overline{\Delta}_{hZZ}^{\mathrm{EW}}$.
In the both signs of $c_{\beta-\alpha}$, $\overline{\Delta}_{hZZ}^{\mathrm{EW}}$ generally increases the magnitude of $\abs{\Delta R^{hZ}}$ except for the region with relatively lighter mass of the additional Higgs bosons.
The possible magnitude of $\Delta R^{hZ}$ decreases as $\tan{\beta}$ becomes large due to the perturbative unitarity and vacuum stability bounds.

Finally, we mention the corrections to the angular distribution of the $Z$ boson.
As we have mentioned, the $he\bar{e}$ vertex and box contributions cause non-trivial $\cos{\theta}$ dependence.
In the limit of the massless electron, only mixing of CP-even states modifies the $he\bar{e}$ vertex and box contributions.
However, as we can see from Figs.~\ref{fig: Z2-broken_HSM} and \ref{fig: dR_2HDM_nonalignment}, these effects are rather small at $\sqrt{s}=250$ GeV.
Therefore, the angular distribution of $Z$ bosons is almost the same as the predictions in the SM.

\subsection{Correlation in the cross section times decay branching ratios}
In this subsection, we analyze the correlation in the cross section times decay branching ratios of the SM-like Higgs boson in the HSM, the IDM and the four types of 2HDMs.

At future collider experiments such as the ILC, the cross section of $e^{+}e^{-}\to hZ$ can be measured without depending on the decay of the SM-like Higgs boson by utilizing the recoil mass technique \cite{Baer:2013cma, Yan:2016xyx}.
This makes it possible to measure the decay branching ratio of the SM-like Higgs boson independently of the cross section.
However, the cross section times decay branching ratios of the SM-like Higgs boson can be measured more precisely.
In Table~\ref{tab: exp_accuracy_ILC250}, we summarize the expected accuracy of the cross section times decay branching ratios of the SM-like Higgs boson at the ILC at $\sqrt{s}=250$ GeV with $250\, \mathrm{fb}^{-1}$ for $(P_{e}, P_{\bar{e}})=(-0.8, +0.3)$.
The values in Table~\ref{tab: exp_accuracy_ILC250} are taken from Table VI in \cite{Baer:2013cma}.

\begin{table}[t]
\centering
  \begin{tabular}{ccccccc} \hline
  $\sigma(e^{+}e^{-}\to Zh)$ & $h\to b\bar{b}$ & $h\to c\bar{c}$ & $h\to \tau\bar{\tau}$ & $h\to WW^{*}$ & $h\to ZZ^{*}$ & $h\to \gamma\gamma$ \\ \hline
  $2\%$ & $1.3\%$ & $8.3\%$ & $3.2\%$ & $4.6\%$ & $18\%$ & $34\%$\\ \hline
  \end{tabular}
  \caption{Expected $1\sigma$ accuracy for the SM-like Higgs boson measurements at the ILC. We quote the values at $\sqrt{s}=250$ GeV with $250\, \mathrm{fb}^{-1}$ for $(P_{e}, P_{\bar{e}})=(-0.8, +0.3)$ in Table~VI in Ref.~\cite{Baer:2013cma}. Except for $\sigma(e^{+}e^{-}\to Zh)$, the numbers correspond to the accuracy of $\sigma(e^{+}e^{-}\to hZ)\times \mathrm{BR}(h\to XY)$.}
\label{tab: exp_accuracy_ILC250}
\end{table}

In the following, we analyze the predictions for $\sigma(e^{+}e^{-}\to h Z)\times \mathrm{BR}(h\to XY)$ at one-loop level, where $X$ and $Y$ denote decay products of the SM-like Higgs boson.
In order to discuss deviations from predictions in the SM, we evaluate the ratio of the total cross section times the decay branching ratios
\begin{align}
\Delta R^{hZ}_{XY} &= \frac{\sigma_{\mathrm{NP}}(e^{+}e^{-}\to hZ)\mathrm{BR}_{\mathrm{NP}}(h\to XY)}{\sigma_{\mathrm{SM}}(e^{+}e^{-}\to hZ)\mathrm{BR}_{\mathrm{SM}}(h\to XY)}-1,
\end{align}
where we assume the beam polarization $(P_{e}, P_{\bar{e}})=(-0.8, +0.3)$.
In the evaluation of the decay branching ratios of the SM-like Higgs boson with the one-loop EW and QCD corrections, we use the \texttt{H-COUP} program \cite{Kanemura:2017gbi, Kanemura:2019slf}.
Although the magnitude of $\Delta R^{hZ}_{XY}$ depends on the treatment of the QED corrections, we do not consider these corrections and discuss the pattern of the deviations in the correlation of $\sigma(e^{+}e^{-}\to h Z)\times \mathrm{BR}(h\to XY)$.
The QED corrections in the cross section universally change the magnitude of $\sigma(e^{+}e^{-}\to hZ)\times \mathrm{BR}(h\to XY)$.
Therefore, the pattern of the deviations is not changed even if we include the QED corrections following a realistic experimental setup.

We scan the input parameters in each model in the following way.
In the HSM, there are five input parameters as given in Eq.~\eqref{eq: inputs_HSM}.
We here use $M^{2}$ as an input parameter instead of $\lambda_{\Phi S}$.
The mass of the additional Higgs boson $H$ is scanned as
\begin{align}
400\, \mathrm{GeV} \le m_{H} < 2000\, \mathrm{GeV},
\end{align}
while $c_{\alpha}$ and $M^{2}$ are scanned as
\begin{align}
0.95 \le c_{\alpha} < 1\qc
0 \le M^{2} < (m_{H}+250\ \mathrm{GeV})^{2}.
\end{align}
We here take $\mu_{s}=0$ and $\lambda_{S}=0.1$ for simplicity.

In the 2HDMs, we have six input parameters given in Eq.~\eqref{eq: inputs_2HDM}.
We assume that the additional Higgs bosons are degenerate in mass as in the previous subsection.
In this scenario, the constraint of the EW $T$ parameter is satisfied independently of the type of the 2HDMs.
The degenerate mass $m_{\Phi}(=m_{H^{\pm}}=m_{H}=m_{A}$) is scanned as
\begin{align}
400\, \mathrm{GeV} &\le m_{\Phi} < 2000\, \mathrm{GeV}\quad \text{for the Type-I and X 2HDMs}, \\
800\, \mathrm{GeV} &\le m_{\Phi} < 2000\, \mathrm{GeV}\quad \text{for the Type-II and Y 2HDMs}.
\end{align}
The lower bound of $m_{\Phi}$ in the Type-I and Type-X 2HDMs comes from the direct search for $A\to \tau\bar{\tau}$ at the LHC \cite{Aiko:2020ksl}.
In the Type-I 2HDM, the parameter regions with $\tan{\beta}\gtrsim 2$ are not excluded.
However, we take the above parameter regions for simplicity.
In the Type-II and Type-Y 2HDMs, the lower bound comes from the flavor experiments, especially from $B_{s}\to X_{s}\gamma$ \cite{Misiak:2020vlo}.
In addition, we scan the other parameters as
\begin{align}
0.98 \leq s_{\beta-\alpha} < 1\qc
2 \leq \tan{\beta} < 10\qc
0\leq M^{2} < (m_{\Phi}+250\ \mathrm{GeV})^{2}.
\end{align}
The lower bound of $\tan{\beta}$ comes from the flavor experiments. We analyze both the positive and negative signs of $c_{\beta-\alpha}$.

In the IDM, we have five input parameters given in Eq.~\eqref{eq: inputs_IDM}.
We take $m_{H}= 63$ GeV which is favored by dark matter constraints. 
In order to satisfy the constraint from the EW $T$ parameter, we assume that $H^{\pm}$ and $A$ are degenerate in mass.
The degenerate mass $m_{H^{\pm}}(=m_{A})$, $M^{2}$ and $\lambda_{2}$ are scanned as
\begin{align}
100\, \mathrm{GeV} &\le m_{H^{\pm}} < 1000\, \mathrm{GeV}, \\
0 &\le M^{2} < (m_{\Phi}+250\, \mathrm{GeV})^{2}, \\
0 &< \lambda_{2} < 4\pi.
\end{align}

Over the above parameter spaces, we impose the constraints discussed in Sec.~\ref{sec: models} such as perturbative unitarity, vacuum stability, avoiding wrong vacua, and the constraints from the EW $S$ and $T$ parameters.
In addition, we take into account the current data of the signal strengths of the discovered Higgs boson at the LHC.
We evaluate the decay rates of the SM-like Higgs boson with higher-order corrections by using the \texttt{H-COUP} program \cite{Kanemura:2017gbi, Kanemura:2019slf}.
We define the scaling factor $\kappa_{XY} = \sqrt{\Gamma_{h\to XY}^{\mathrm{NP}}/\Gamma_{h\to XY}^{\mathrm{SM}}}$ at the one-loop level, and remove the parameter points, where $\kappa_{XY}$ deviates from the observed data at $95\%$ C.L..
In Table~\ref{tab: const_kappa_LHC}, we summarize the current measurements of $\kappa_{XY}$ factors at $1\sigma$ accuracy.
The values in Table~\ref{tab: const_kappa_LHC} are taken from Table XI in Ref.~\cite{Aad:2019mbh}.
We assume that there is no decay mode where the SM-like Higgs boson decays into additional Higgs bosons.

\begin{table}[t]
\centering
  \begin{tabular}{ccccccc} \hline
  $\kappa_{b}$ & $\kappa_{\tau}$ & $\kappa_{\gamma}$ & $\kappa_{g}$ & $\kappa_{Z}$ & $\kappa_{W}$ \\ \hline
  $1.03^{+0.19}_{-0.17}$ & $1.05^{+0.16}_{-0.15}$ & $1.05\pm0.09$ & $0.99^{+0.11}_{-0.10}$ & $1.11\pm0.08$ & $1.05\pm0.09$\\ \hline
  \end{tabular}
  \caption{Current measurements $\kappa_{XY}$ factors at $1\sigma$ accuracy. We quote the values in Table XI in \cite{Aad:2019mbh}. We assume that there is no decay mode where the SM-like Higgs boson decays into additional Higgs bosons.}
\label{tab: const_kappa_LHC}
\end{table}

In the Type-II, X and Y 2HDMs, we have parameter points where Yukawa coupling constants take the negative sign with a large value of $\tan{\beta}$ and $c_{\beta-\alpha}>0$.
These parameter points show the sizable deviation both in the Higgs branching ratio and cross section.
However, we simply omit such particular parameter points in the following analysis in order to extract general features in the 2HDMs.

Before moving on to the numerical results, we discuss the general property of $\Delta R^{hZ}_{XY}$.
The ratio of the total cross section times the decay branching ratio $\Delta R^{hZ}_{XY}$ can be rewritten as
\begin{align}
\Delta R^{hZ}_{XY} &=
\Delta R^{Zh} + \Delta R_{XY} + \Delta R^{Zh}\Delta R_{XY}, \label{eq: dRhZ_XY}
\end{align}
with $\Delta R^{Zh}$ defined in Eq.~\eqref{eq: dRhZ} and $\Delta R_{XY}$ defined as
\begin{align}
\Delta R_{XY} = \frac{\mathrm{BR}_{\mathrm{NP}}(h\to XY)}{\mathrm{BR}_{\mathrm{SM}}(h\to XY)}-1.
\end{align}
The order of loop expansion of $\Delta R^{Zh}\Delta R_{XY}$ is $\mathcal{O}(\hbar^{2})$, and it is sub-leading.
Therefore, the qualitative behavior of $\Delta R^{hZ}_{XY}$ can be understood by independently analyzing $\Delta R^{Zh}$ and $\Delta R_{XY}$.

The behavior of $\Delta R_{XY}$ has been studied in Ref.~\cite{Kanemura:2019kjg} by using the
\texttt{H-COUP} program \cite{Kanemura:2017gbi, Kanemura:2019slf}.
For later convenience, we briefly summarize the behavior of $\Delta R_{XY}$ in the HSM, the IDM and the four types of 2HDMs.
First, in the HSM, the decay branching ratios of $h$ are almost the same as those in the SM predictions, because the partial decay widths are universally suppressed by the radiative corrections and the mixing of the CP-even states.
In our study, both $\Delta R_{\tau\tau}$ and $\Delta R_{bb}$ are at most $0.5 \%$.

The same argument has been claimed for the IDM in Ref.~\cite{Kanemura:2019kjg}.
However, we find that the parameter regions where both $\Delta R_{bb}$ and $\Delta R_{\tau\tau}$ take a few percent deviations.
This difference comes from the large value of $\lambda_{2}$.
In Ref.~\cite{Kanemura:2019kjg}, $\lambda_{2}$ has been fixed to $0.1$.
However, the magnitude of $\lambda_{2}$ indirectly controls the possible size of other Higgs quartic couplings especially through the vacuum stability bound given in Eq.~\eqref{eq: vacuum_stability_2HDM}.
We have obtained the almost same results as those in Ref.~\cite{Kanemura:2019kjg} when we impose $\lambda_{2}\le 0.1$.

In the 2HDMs, the predictions to  $\Delta R_{\tau\tau}$ and $\Delta R_{bb}$ spread out into different directions according to the type of the Yukawa interactions and the sign of $c_{\beta-\alpha}$.
The possible magnitudes of the deviations in the Type-II, X and Y 2HDMs are rather large compared to the Type-I 2HDM, the HSM and the IDM.
They can reach several tens of percent, and especially $\Delta R_{\tau\tau}$ reaches a hundred percent in the Type-X 2HDM.

\begin{figure}[t]
\begin{minipage}{0.45\hsize}
\centering
\includegraphics[scale=0.6]{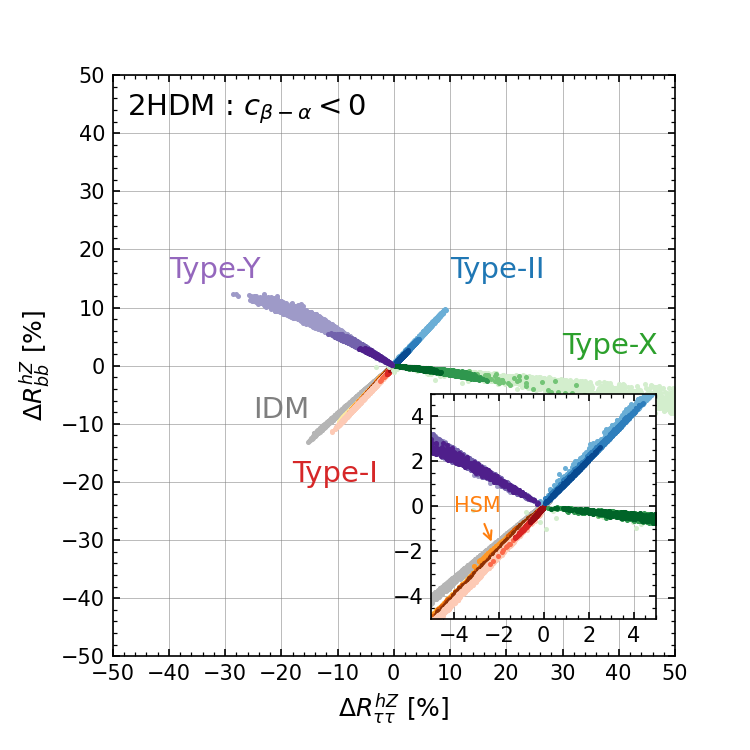}
\end{minipage}
\begin{minipage}{0.45\hsize}
\centering
\includegraphics[scale=0.6]{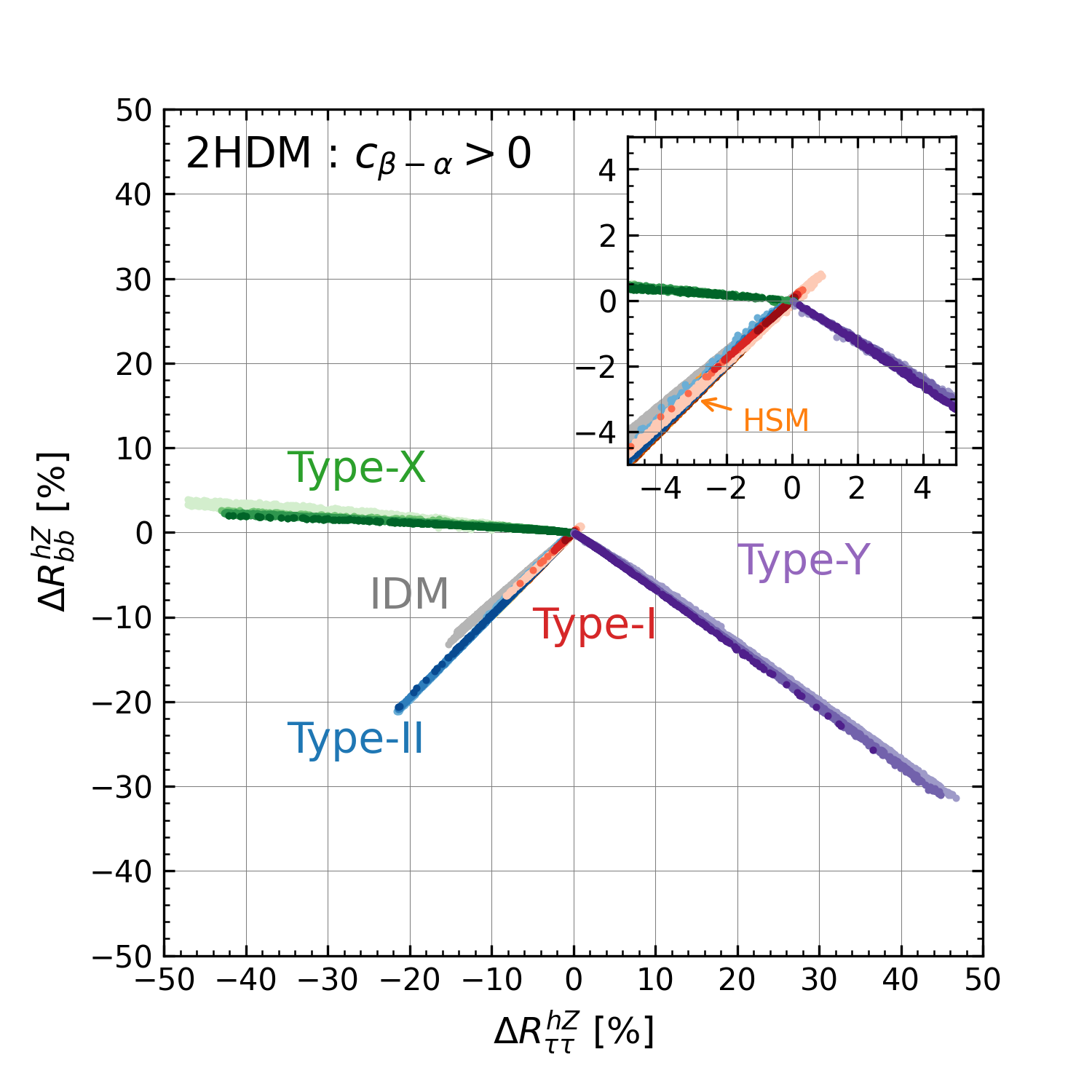}
\end{minipage}
\caption{Correlation between $\Delta R^{hZ}_{\tau\tau}$ and $\Delta R^{hZ}_{bb}$ in the HSM (orange), the IDM (grey) and the Type-I (red), Type-II (blue), Type-X (green), Type-Y (purple) 2HDMs. The left panel shows the results with $c_{\beta-\alpha}<0$ in the 2HDMs, and the right panel shows those with $c_{\beta-\alpha}>0$. The ranges of the parameters are explained in the text. The lighter color corresponds to the lighter mass scale of the additional Higgs bosons, $m_{\Phi}\ge 400,\, 800,\, 1200$ and $1600$ GeV in order.}
\label{fig: dRZh_bb_tata}
\end{figure}

In Fig.~\ref{fig: dRZh_bb_tata}, we show the correlations between $\Delta R^{hZ}_{\tau\tau}$ and $\Delta R^{hZ}_{bb}$ in the HSM, the IDM and the four types of 2HDMs.
We take the color codes where orange, grey, red, blue, green and purple correspond to the HSM, the IDM and the Type-I, II, X, Y 2HDMs, respectively.
The lighter color corresponds to the lighter mass scale of the additional Higgs bosons, $m_{\Phi}\ge 400,\, 800,\, 1200$ and $1600$ GeV in order.
In the left (right) panel, we show the results with $c_{\beta-\alpha}<0$ ($c_{\beta-\alpha}>0$).
The results in the HSM and the IDM are the same both in the left and right panels. 

As discussed in Eq.~\eqref{eq: dRhZ_XY}, $\Delta R^{hZ}_{XY}$ can be rewritten as the sum of $\Delta R^{Zh}$ and $\Delta R_{XY}$, and $\Delta R^{Zh}$ takes a negative value in most cases.
In the Type-II, X and Y 2HDMs, the typical size of $\Delta R_{XY}$ is larger than $\Delta R^{Zh}$.
Therefore, the pattern of the deviation is mainly determined by $\Delta R_{XY}$, and it is consistent with previous analysis in Ref.~\cite{Kanemura:2019kjg}.
In these models, the possible sizes of the deviations are large enough to be detected at the ILC if $m_{\Phi}$ is about one TeV or less.

In the HSM, the Type-I 2HDM and the IDM, we can see sizable deviations both in $\Delta R^{hZ}_{\tau\tau}$ and $\Delta R^{hZ}_{bb}$, and they reach about $10\%$.
Both $\Delta R^{hZ}_{\tau\tau}$ and $\Delta R^{hZ}_{bb}$ take larger values than those in $\Delta R_{\tau\tau}$ and $\Delta R_{bb}$.
This is because the typical size of $\Delta R^{Zh}$ is larger than $\Delta R_{XY}$ in these models, and $\Delta R_{XY}$ also takes a negative value in the HSM, the Type-I 2HDM with $c_{\beta-\alpha}<0$ and the IDM.
In the Type-I 2HDM with $c_{\beta-\alpha}>0$, $\Delta R_{XY}$ takes a positive value.
However, the typical size of $\Delta R_{XY}$ is smaller than $\Delta R^{hZ}$, and $\Delta R^{Zh}_{XY}$ takes a negative value in most of the parameter regions.
In the Type-I 2HDM, $\Delta R^{Zh}_{XY}$ quickly decouples.
This is because a non-zero $c_{\beta-\alpha}$ realizes the maximal deviation in $\Delta R^{hZ}$, especially at LO.
If $m_{\Phi}$ is large, the possible value of $c_{\beta-\alpha}$ is constrained mainly by perturbative unitarity.
On the contrary, in the other types of 2HDMs with $c_{\beta-\alpha}>0$, the decoupling behavior is not clearly seen.
This is because taking the inner parameter $\tan{\beta}$ large keeps the magnitude of the deviation to be large even in the case of large $m_{\Phi}$.
On the other hand, constraints from the Higgs signal strength in the HSM are weaker than those in the 2HDMs, and the deviation in $c_{\alpha}$ realizes the sizable $\Delta R^{Zh}_{XY}$ even if $m_{\Phi}$ is larger than 1 TeV.
In the IDM, the decoupling limit cannot be applied because we fix $m_{H}=63$ GeV.
Therefore, we have a sizable deviation although there is no mixing between the CP-even states.

\begin{figure}[t]
\begin{minipage}{0.45\hsize}
\centering
\includegraphics[scale=0.6]{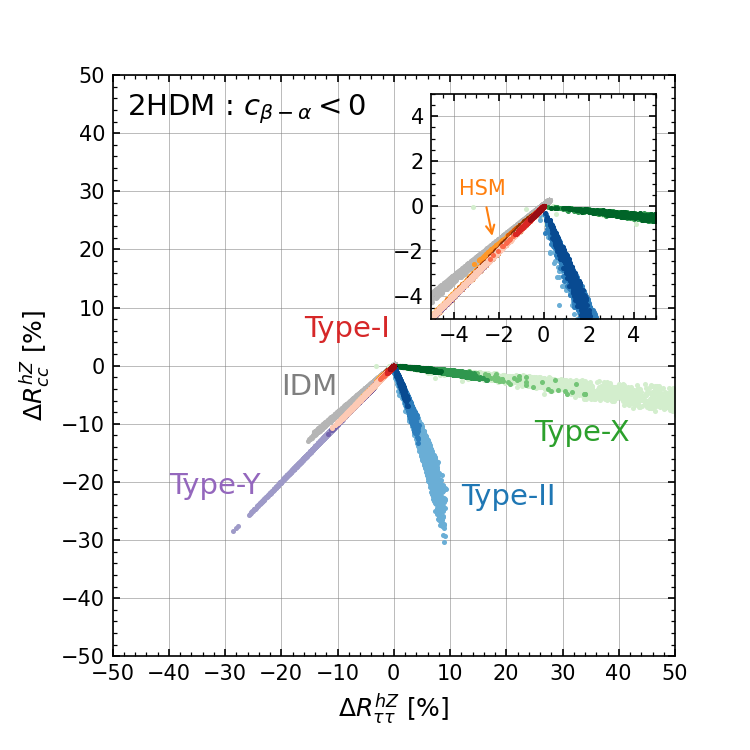}
\end{minipage}
\begin{minipage}{0.45\hsize}
\centering
\includegraphics[scale=0.6]{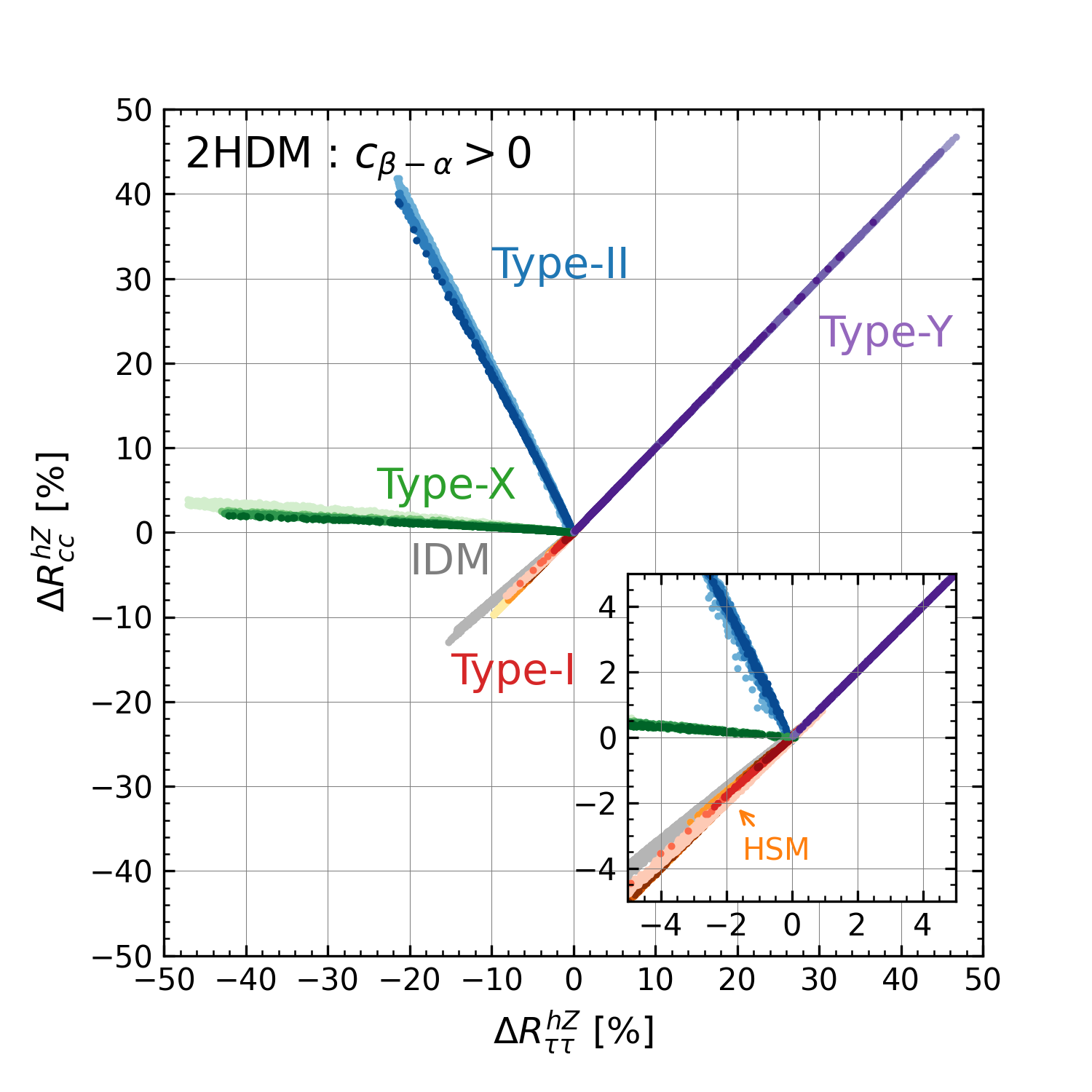}
\end{minipage}
\caption{Correlation between $\Delta R^{hZ}_{\tau\tau}$ and $\Delta R^{hZ}_{cc}$. Color codes and the ranges of the parameters are the same as in those of Fig.~\ref{fig: dRZh_bb_tata}.}
\label{fig: dRZh_cc_tata}
\end{figure}

In Fig.~\ref{fig: dRZh_cc_tata}, we show the correlations between $\Delta R^{hZ}_{\tau\tau}$ and $\Delta R^{hZ}_{cc}$ in the HSM, the IDM and the four types of 2HDMs.
The color codes and gradations are the same as those in Fig.~\ref{fig: dRZh_bb_tata}.
In the left (right) panel, we show the results with $c_{\beta-\alpha}<0$ ($c_{\beta-\alpha}>0$).
The results in the HSM and the IDM are the same both in the left and right panels.

Qualitative behavior of the deviations in each model is the same as in Fig.~\ref{fig: dRZh_bb_tata}.
In the Type-II, X and Y 2HDMs, the typical size of $\Delta R_{XY}$ is larger than $\Delta R^{Zh}$, and the pattern of the deviation is mainly determined by $\Delta R_{XY}$.
On the other hand, we also have sizable deviations in the HSM, the Type-I 2HDM and the IDM, and they reach about $10\%$.
In the Type-I 2HDM with $c_{\beta-\alpha}>0$, $\Delta R_{XY}$ takes a positive value.
However, the typical size of $\Delta R_{XY}$ is smaller than $\Delta R^{hZ}$, and $\Delta R^{Zh}_{XY}$ takes a negative value in most of the parameter regions.

If $c_{\beta-\alpha}$ is negative, the directions of the deviations in $\Delta R^{hZ}_{\tau\tau}$ and $\Delta R^{hZ}_{cc}$ are the same in the HSM, the IDM and the Type-I and Y 2HDMs.
However, this overlap can be partially resolved by looking at the correlation between $\Delta R^{hZ}_{\tau\tau}$ and $\Delta R^{hZ}_{bb}$ where the Type-Y 2HDM shows a different correlation with others.
On the other hand, if $c_{\beta-\alpha}$ is positive, the directions of the deviations in $\Delta R^{hZ}_{\tau\tau}$ and $\Delta R^{hZ}_{bb}$ are the same in the HSM, the IDM and the Type-I and II 2HDMs.
This overlap can also be resolved by looking at the correlation between $\Delta R^{hZ}_{\tau\tau}$ and $\Delta R^{hZ}_{cc}$ where the Type-II 2HDM shows a different correlation with others.
 
\begin{figure}[t]
\begin{minipage}{0.45\hsize}
\centering
\includegraphics[scale=0.6]{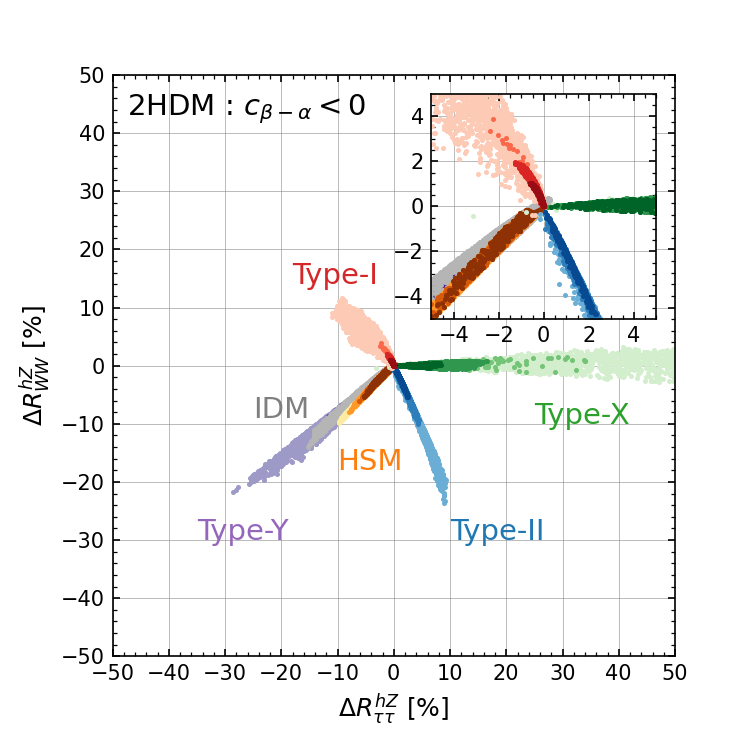}
\end{minipage}
\begin{minipage}{0.45\hsize}
\centering
\includegraphics[scale=0.6]{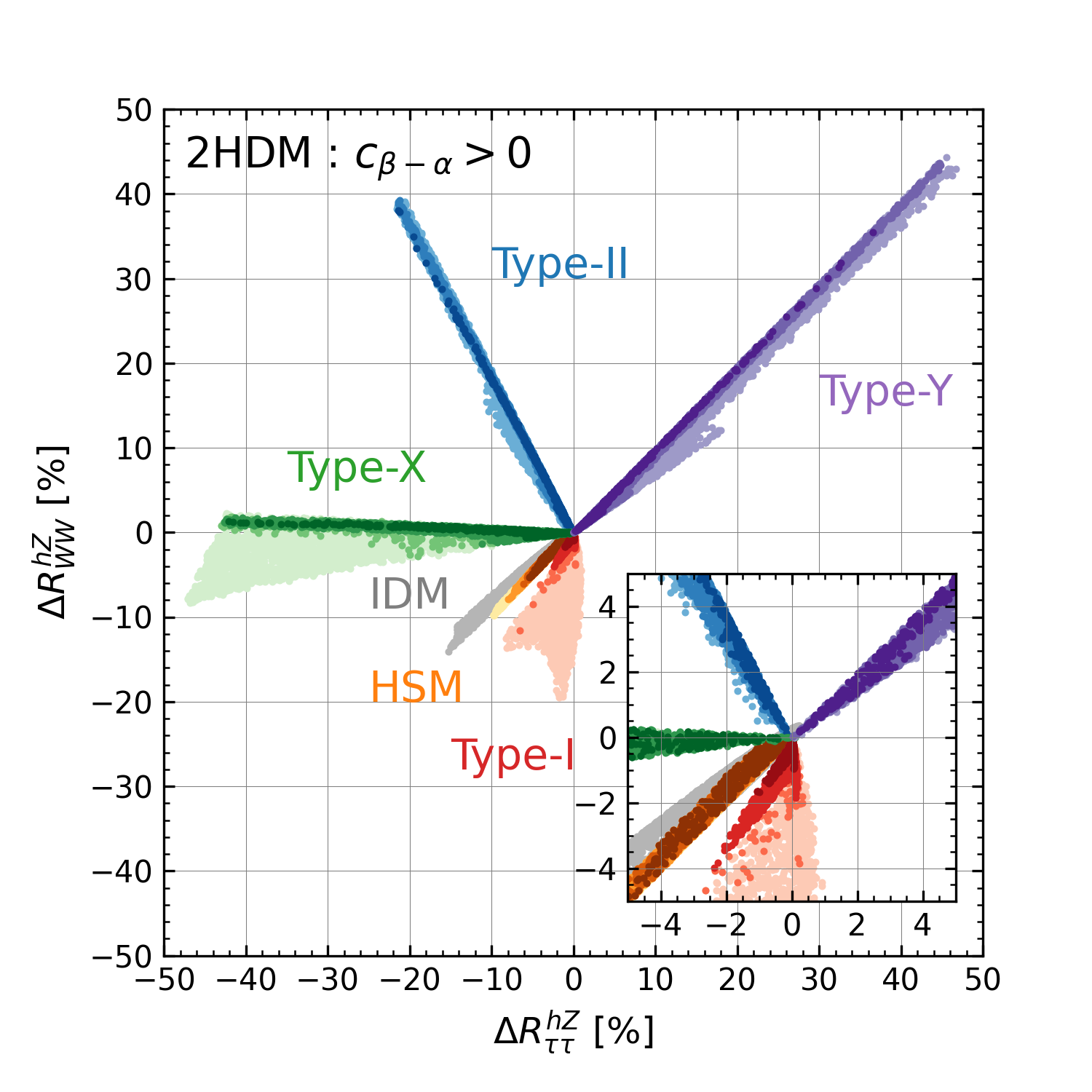}
\end{minipage}
\caption{Correlation between $\Delta R^{hZ}_{\tau\tau}$ and $\Delta R^{hZ}_{WW}$. Color codes and the ranges of the parameters are the same as in those of Fig.~\ref{fig: dRZh_bb_tata}.}
\label{fig: dRZh_WW_tata}
\end{figure}

In order to discriminate the Type-I 2HDM from the HSM and the IDM, we can use the correlation between $\Delta R^{hZ}_{\tau\tau}$ and $\Delta R^{hZ}_{WW}$.
In Fig.~\ref{fig: dRZh_WW_tata}, we show the correlations between $\Delta R^{hZ}_{\tau\tau}$ and $\Delta R^{hZ}_{WW}$ in the HSM, the IDM and the four types of 2HDMs.
The color codes and gradations are the same as those in Fig.~\ref{fig: dRZh_bb_tata}.
In the left (right) panel, we show the results with $c_{\beta-\alpha}<0$ ($c_{\beta-\alpha}>0$).
The results in the HSM and the IDM are the same both in the left and right panels.
Especially in the case of $c_{\beta-\alpha}<0$, $\Delta R^{hZ}_{WW}$ takes a positive value in the Type-I 2HDM, while it takes a negative value in the HSM and the IDM.
Even in the case of $c_{\beta-\alpha}>0$, there is a stronger correlation between $\Delta R^{hZ}_{\tau\tau}$ and $\Delta R^{hZ}_{WW}$ in the HSM and the IDM than those in the Type-I 2HDM.

Finally, we discuss the discrimination between the HSM and the IDM.
The deviation $\Delta R^{hZ}_{\gamma\gamma}$ might be useful to discriminate these models because it is mainly affected by the contribution of the charged Higgs bosons.
The behaviors of $\Delta R^{hZ}_{\tau\tau}$ and $\Delta R^{hZ}_{\gamma\gamma}$ show a different correlation between the HSM and the IDM.
However, the possible size of $\Delta R^{hZ}_{\gamma\gamma}$ is at most $20\%$, and it is rather challenging to discriminate them with $95\%$ C.L. at the ILC.
The large uncertainty in $\Delta R^{hZ}_{\gamma\gamma}$ at the ILC mainly comes from the low statistics, and this would be improved by performing the combined study with the measurements at the ILC and the HL-LHC.

%

%% file: 05Conclusion.tex


\section{Conclusion} \label{sec: conclusion}
We have calculated the cross section for $e^{+}e^{-}\to hZ$ with arbitrary sets of electron and $Z$ boson polarization at the full next-to-leading order in the HSM, the IDM and the four types of 2HDMs.
We have systematically performed complete one-loop calculations to the helicity amplitudes in the on-shell renormalization scheme and present the full analytic results as well as numerical evaluations.
The deviation $\Delta R^{hZ}$ in the total cross section from its SM prediction has been comprehensively analyzed, and the differences among these models have been discussed in detail.
We have found that new physics effects appearing in the renormalized $hZZ$ vertex almost govern the behavior of $\Delta R^{hZ}$.
We have also shown that the predictions for the deviations in the total cross section of $e^{+}e^{-}\to hZ$ times the branching ratios of $h\to XY$.
It has been found that we can discriminate the four types of 2HDMs by analyzing the correlation between $\Delta R^{hZ}_{bb}$ and $\Delta R^{hZ}_{\tau\tau}$ and those between $\Delta R^{hZ}_{cc}$ and $\Delta R^{hZ}_{\tau\tau}$.
Furthermore, the Type-I 2HDM might be specified from the HSM and the IDM by measuring the deviation in $\Delta R^{hZ}_{WW}$.
These signatures can be tested by precision measurements at future Higgs factories such as the ILC.
On the other hand, the discrimination between the HSM and the IDM is rather challenging only by the measurement at the ILC.
However, this problem might be solved by taking into account the deviation in $h\to \gamma\gamma$ signals at the LHC and the HL-LHC.

%

%% file: 06Acknowledgements.tex


\section*{Acknowledgement} \label{sec: acknowledgement}
This work is supported in part by the Grant-in-Aid on Innovative Areas, the Ministry of Education, Culture, Sports, Science and Technology, No.~16H06492 and JSPS KAKENHI Grant No.~20H00160 [S.K.], and JSPS KAKENHI Grant No.~18K03648, 20H05239 and 21H01077 [K.M.].
M. A. was supported in part by the Sasakawa Scientific Research Grant from The Japan Science Society.

%

%% file: A01inputs.tex

\section{Input parameters} \label{sec: SM_inputs}
We work in the scheme where $\alpha_{\mathrm{em}}(0), G_{F}$ and $m_{Z}$ are input paremeters following the \texttt{H-COUP} program.
In Table~\ref{Table: inputs}, we list the SM input parameters.
The values of input parameters are taken from Ref.~\cite{Zyla:2020zbs}.
Other parameters can be evaluated in terms of the above inputs by using tree-level relations.

\begin{table}[t]
\centering
\begin{tabular}{c|c|c} \hline
Input parameter & Symbol & Value \\ \hline
fine-structure constant at the Thomson limit & $\alpha^{-1}_{\mathrm{em}}(0)$ & 137.035999139 \\
Fermi constant & $G_{F}\ \qty[\text{GeV}^{-2}]$ & $1.1663787\times 10^{-5}$ \\
$Z$ boson mass & $m_{Z}\ \qty[\text{GeV}]$ & 91.1876 \\
strong coupling constant at $m_{Z}$ & $\alpha_{s}(m_{Z})$ & 0.1181 \\
Higgs boson mass & $m_{h}\ \qty[\text{GeV}]$ & 125.1 \\
top-quark pole mass & $m_{t}\ \qty[\text{GeV}]$ & 173.1 \\
bottom-quark pole mass & $m_{b}\ \qty[\text{GeV}]$ & 4.78 \\
charm-quark pole mass &  $m_{c}\ \qty[\text{GeV}]$ & 1.67 \\
tauon mass & $m_{\tau}\ \qty[\text{GeV}]$ & 1.77686 \\
muon mass & $m_{\mu}\ \qty[\text{GeV}]$ & 0.105658367 \\ \hline
\end{tabular}
\caption{SM input parameters. The values are taken from Ref.~\cite{Zyla:2020zbs}.}
\label{Table: inputs}
\end{table}

%

%% file: A02Loop_functions.tex


\section{Analytic formulae for the box diagrams} \label{sec: boxs}
We here give the analytic expressions for the box diagram contributions denoted by $F_{i}^{k}(s,t)$ in Eq.~\eqref{eq: W_box} and Eq.~\eqref{eq: Z_box} in terms of the Passarino--Veltman functions defined in Ref.~\cite{Passarino:1978jh}.
We calculate the box diagrams in the 't Hooft--Feynman gauge.

The analytic expressions are given as follows
\allowdisplaybreaks{
\begin{align}
C^{1} &= \frac{1}{2}g^{4}m_{W}c_{W}, \\
F_{e}^{1} &=
4(D_{0}+D_{11}+D_{13}+D_{25})(0, 0, m_{Z}^{2}, m_{h}^{2}, s, u; W, 0, W, W), \\
F_{\overline{e}}^{1} &=
2(D_{13}-D_{12}+2D_{26})(0, 0, m_{Z}^{2}, m_{h}^{2}, s, u; W, 0, W, W), \\
F^{1} &=
-2C_{0}(s, m_{Z}^{2}, m_{h}^{2}; W, W, W)
-[4D_{27}+(s-t+m_{Z}^{2})(D_{0}+D_{11}) \notag \\
&\quad
+uD_{13}+2(u-m_{Z}^{2})D_{12}](0, 0, m_{Z}^{2}, m_{h}^{2}, s, u; W, 0, W, W), \\
C^{2} &= \frac{1}{2}g^{4}m_{W}c_{W}, \\
F^{2}_{e} &=
2(D_{13}-D_{12}+2D_{26})(0, 0, m_{Z}^{2}, m_{h}^{2}, s, t; W, 0, W, W), \\
F^{2}_{\overline{e}} &=
4(D_{0}+D_{11}+D_{13}+D_{25})(0, 0, m_{Z}^{2}, m_{h}^{2}, s, t; W, 0, W, W), \\
F^{2} &=
-2C_{0}(s, m_{Z}^{2}, m_{h}^{2}; W, W, W)
-[4D_{27}+(m_{Z}^{2} + s - u)(D_{0}+D_{11}) \notag \\
&\quad
+2(t - m_{Z}^{2})D_{12}+tD_{13}](0, 0, m_{Z}^{2}, m_{h}^{2}, s, t; W, 0, W, W), \\
C^{3} &= -\frac{1}{4}g^{3}g_{Z}m_{W}, \\
F_{e}^{3} &= 0, \\
F_{\overline{e}}^{3} &=
-4(D_{12}-D_{13})(0, 0, m_{Z}^{2}, m_{h}^{2}, s, u; W, 0, W, G^{\pm}), \\
F^{3} &=
C_{0}(s, m_{Z}^{2}, m_{h}^{2}; W, W, W) \notag \\
&\quad
+2[-(u-m_{h}^{2})(D_{0}+D_{11})+uD_{13}](0, 0, m_{Z}^{2}, m_{h}^{2}, s, u; W, 0, W, G^{\pm}), \\
C^{4} &=
-\frac{1}{4}g^{3}g_{Z}m_{W}s_{W}^{2}, \\
F_{e}^{4} &=
-4(D_{12}-D_{13})(0, 0, m_{Z}^{2}, m_{h}^{2}, s, t; W, 0, W, G^{\pm}), \\
F_{\overline{e}}^{4} &= 0, \\
F^{4} &=
C_{0}(s, m_{Z}^{2}, m_{h}^{2}; W, W, W) \notag \\
&\quad
+2[-(t-m_{h}^{2})(D_{0}+D_{11})+tD_{13}](0, 0, m_{Z}^{2}, m_{h}^{2}, s, t; W, 0, W, G^{\pm}), \\
C^{5} &=
-(v_{\nu}+a_{\nu})\frac{2m_{W}^{2}}{v}g^{2}g_{Z}, \\
F^{5}_{e} &=
-2(D_{0}+D_{11}+D_{12}+D_{24})(0, m_{Z}^{2}, 0, m_{h}^{2}, t, u; W, 0, 0, W), \\
F^{5}_{\overline{e}} &=
-2D_{26}(0, m_{Z}^{2}, 0, m_{h}^{2}, t, u; W, 0, 0, W), \\
F^{5} &=
-C_{0}(t, 0, m_{h}^{2}; W, 0, W) \notag \\
&\quad
-[-2D_{27}+(t-m_{Z}^{2})(D_{0}+D_{11})+m_{Z}^{2}D_{12}](0, m_{Z}^{2}, 0, m_{h}^{2}, t, u; W, 0, 0, W), \\
C^{6} &=
-\frac{4m_{Z}^{2}}{v}g_{Z}^{3}, \\
F^{6}_{e} &=
-2(D_{0}+D_{11}+D_{12}+D_{24})(0, m_{Z}^{2}, 0, m_{h}^{2}, t, u; Z, 0, 0, Z), \\
F^{6}_{\overline{e}} &=
-2D_{26}(0, m_{Z}^{2}, 0, m_{h}^{2}, t, u; Z, 0, 0, Z), \\
F^{6} &=
-C_{0}(t, 0, m_{h}^{2}; Z, 0, Z) \notag \\
&\quad
-[-2D_{27}+(t-m_{Z}^{2})(D_{0}+D_{11})+m_{Z}^{2}D_{12}](0, m_{Z}^{2}, 0, m_{h}^{2}, t, u; Z, 0, 0, Z).
\end{align}}

\begin{figure}[t]
\begin{minipage}{0.3\hsize}
\centering
\includegraphics[scale=0.25]{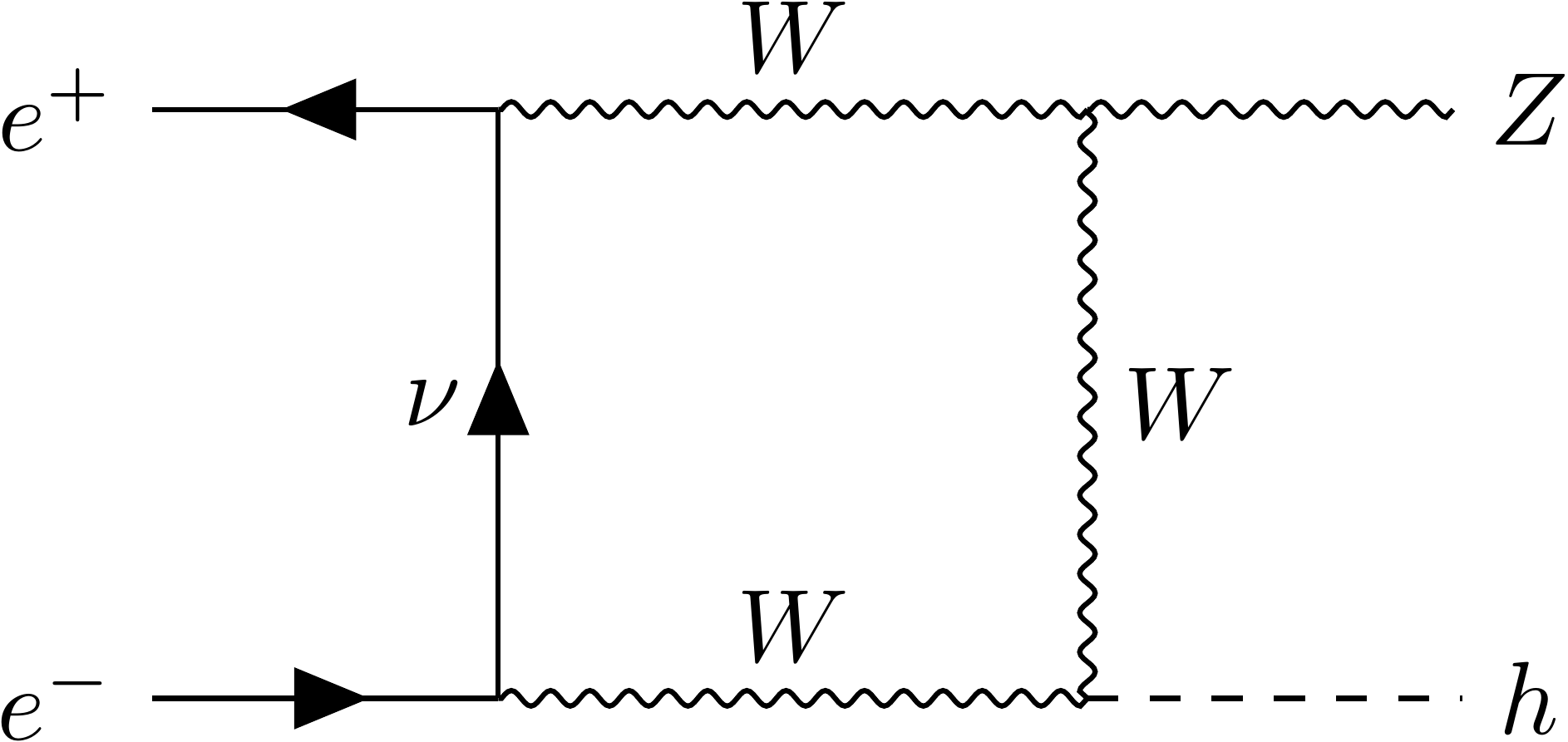}
\subcaption*{(1)}
\end{minipage}
\begin{minipage}{0.3\hsize}
\centering
\includegraphics[scale=0.25]{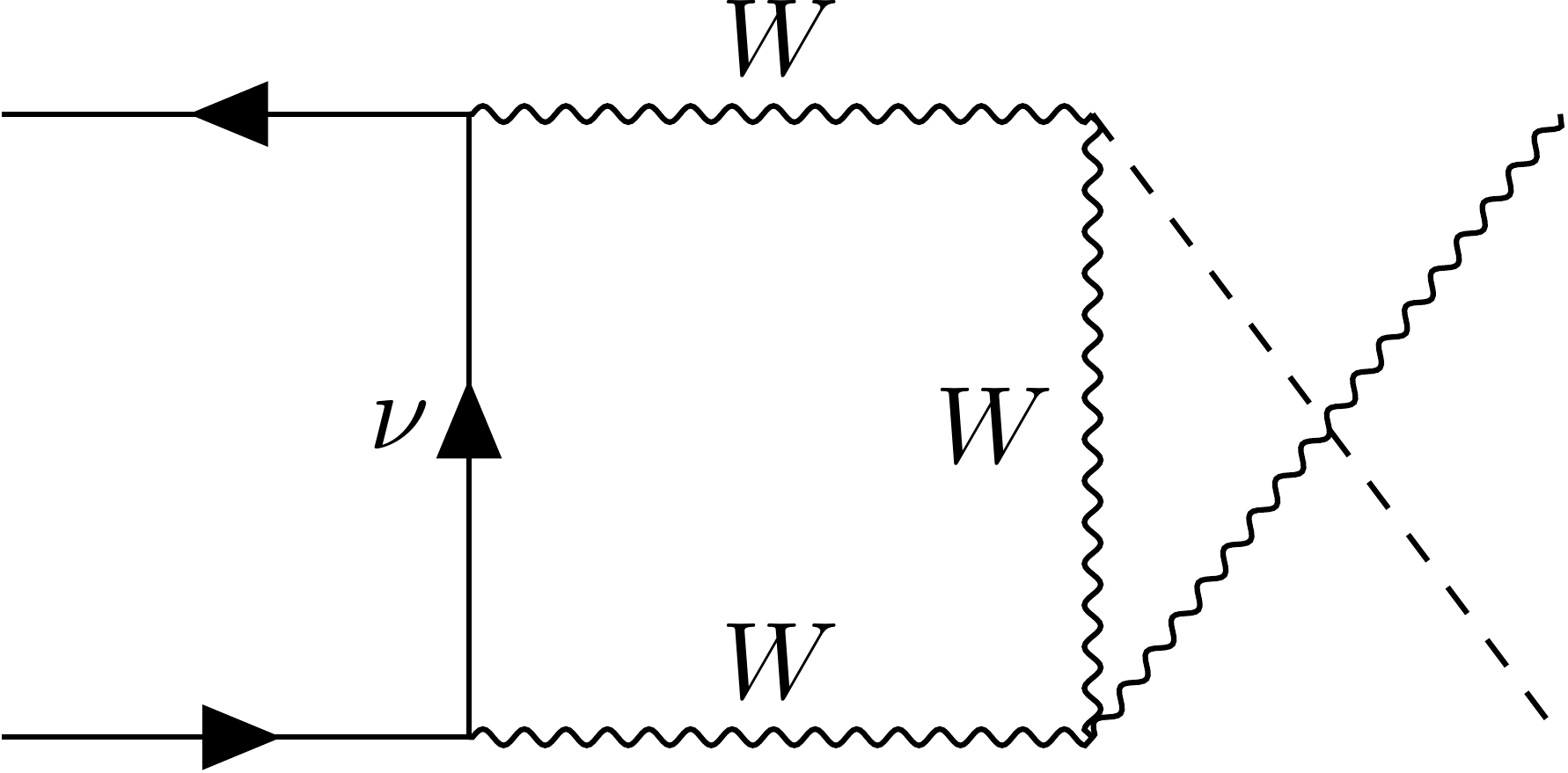}
\subcaption*{(2)}
\end{minipage}
\begin{minipage}{0.3\hsize}
\centering
\includegraphics[scale=0.25]{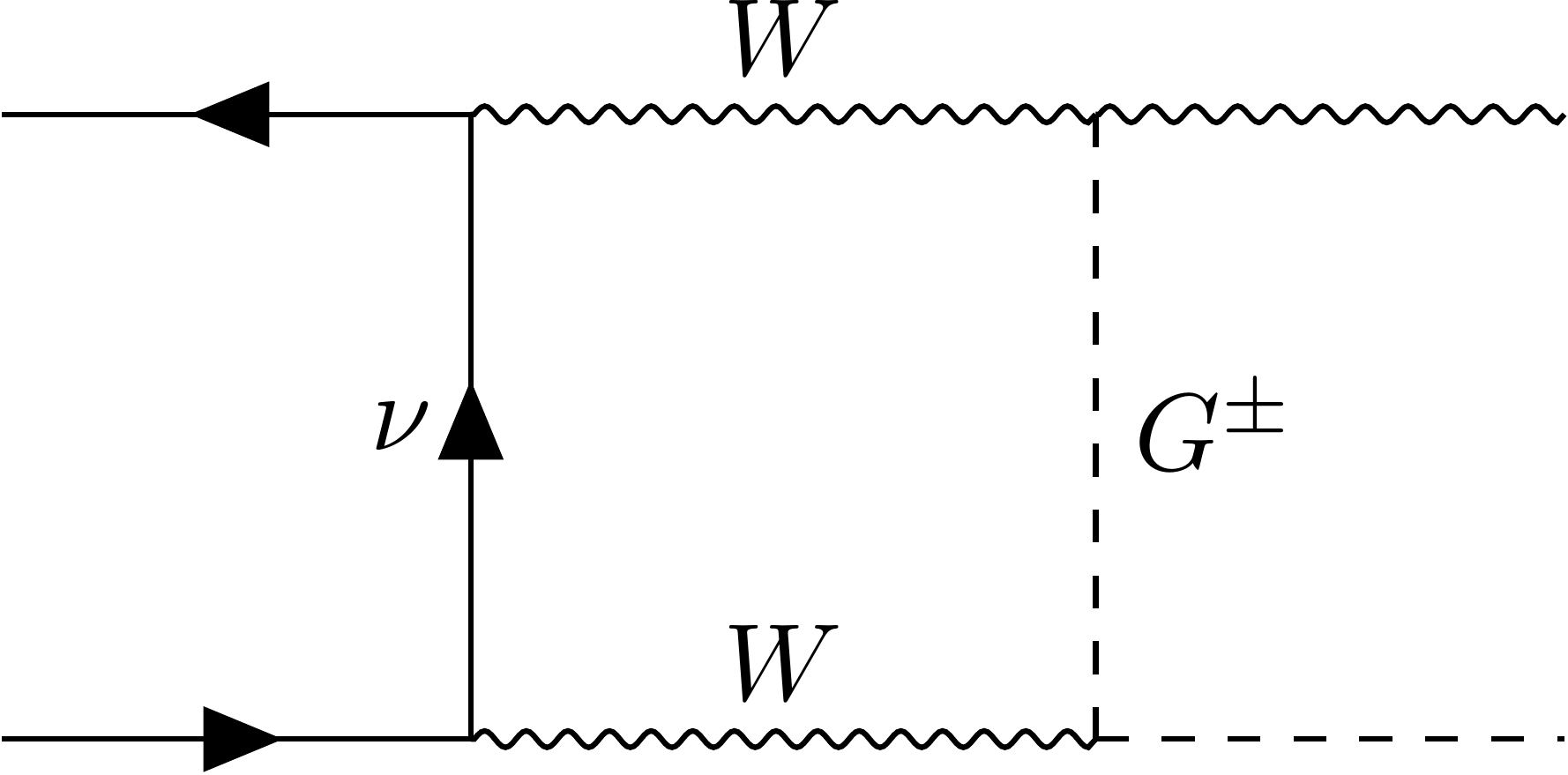}
\subcaption*{(3)}
\end{minipage} \\
\begin{minipage}{0.3\hsize}
\centering
\includegraphics[scale=0.25]{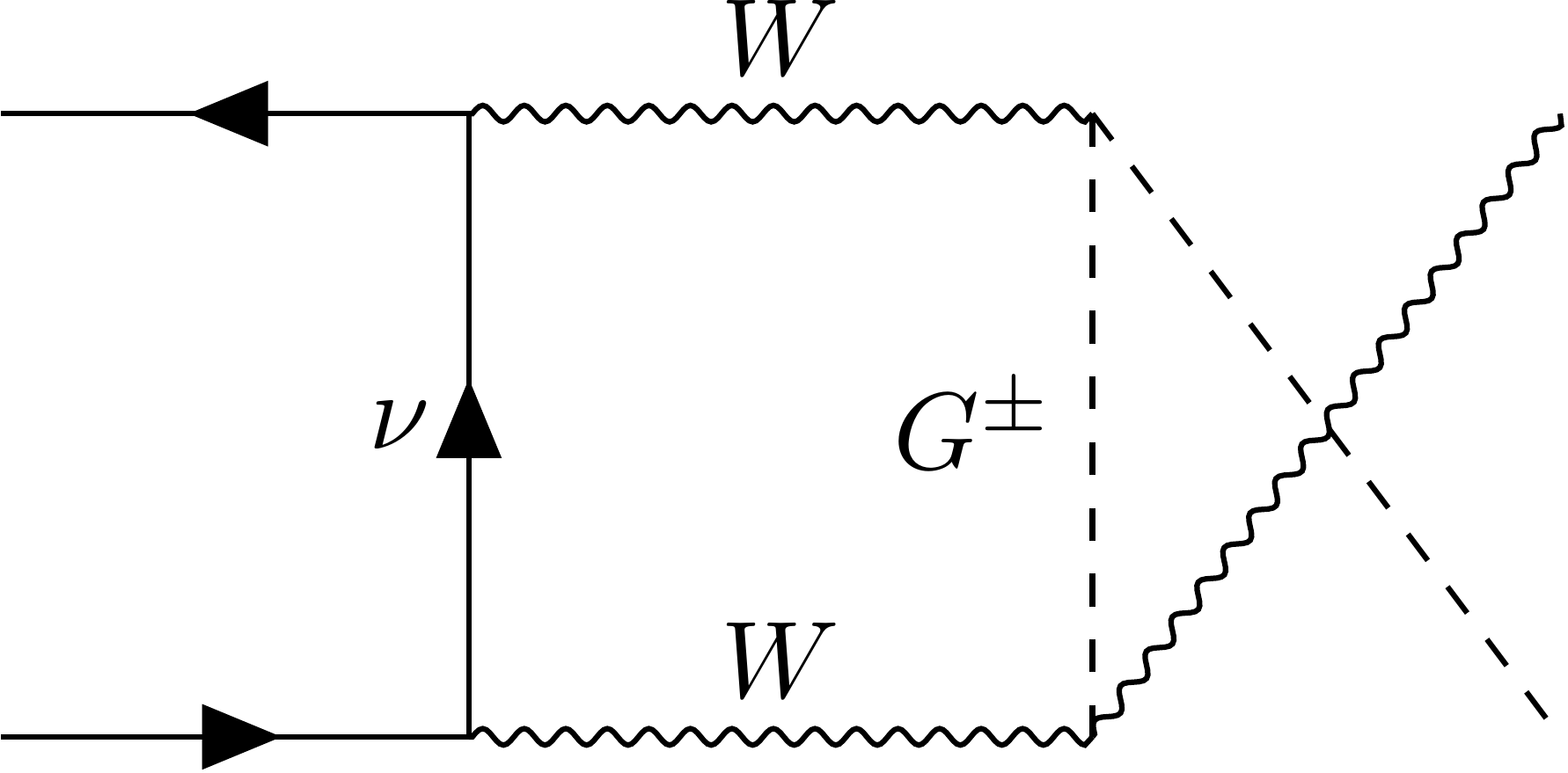}
\subcaption*{(4)}
\end{minipage}
\begin{minipage}{0.3\hsize}
\centering
\includegraphics[scale=0.25]{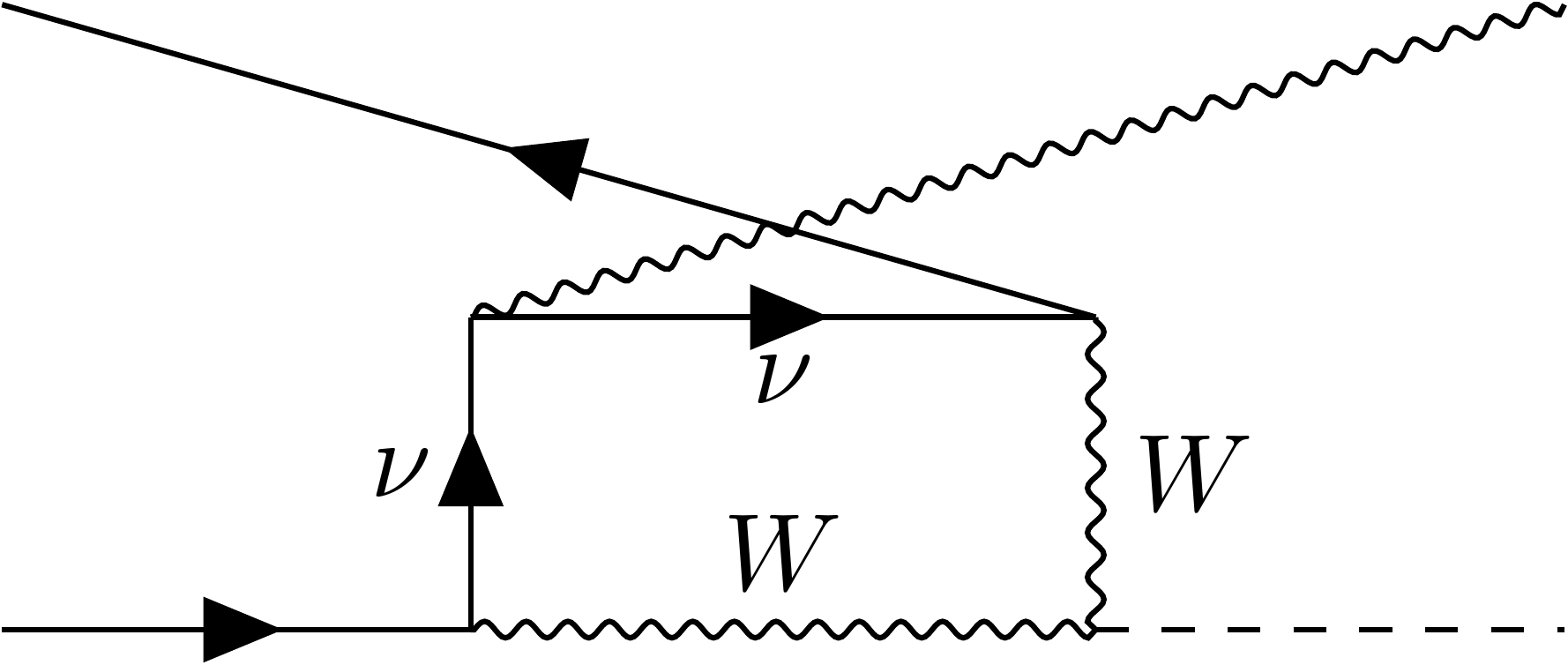}
\subcaption*{(5)}
\end{minipage}
\begin{minipage}{0.3\hsize}
\centering
\includegraphics[scale=0.25]{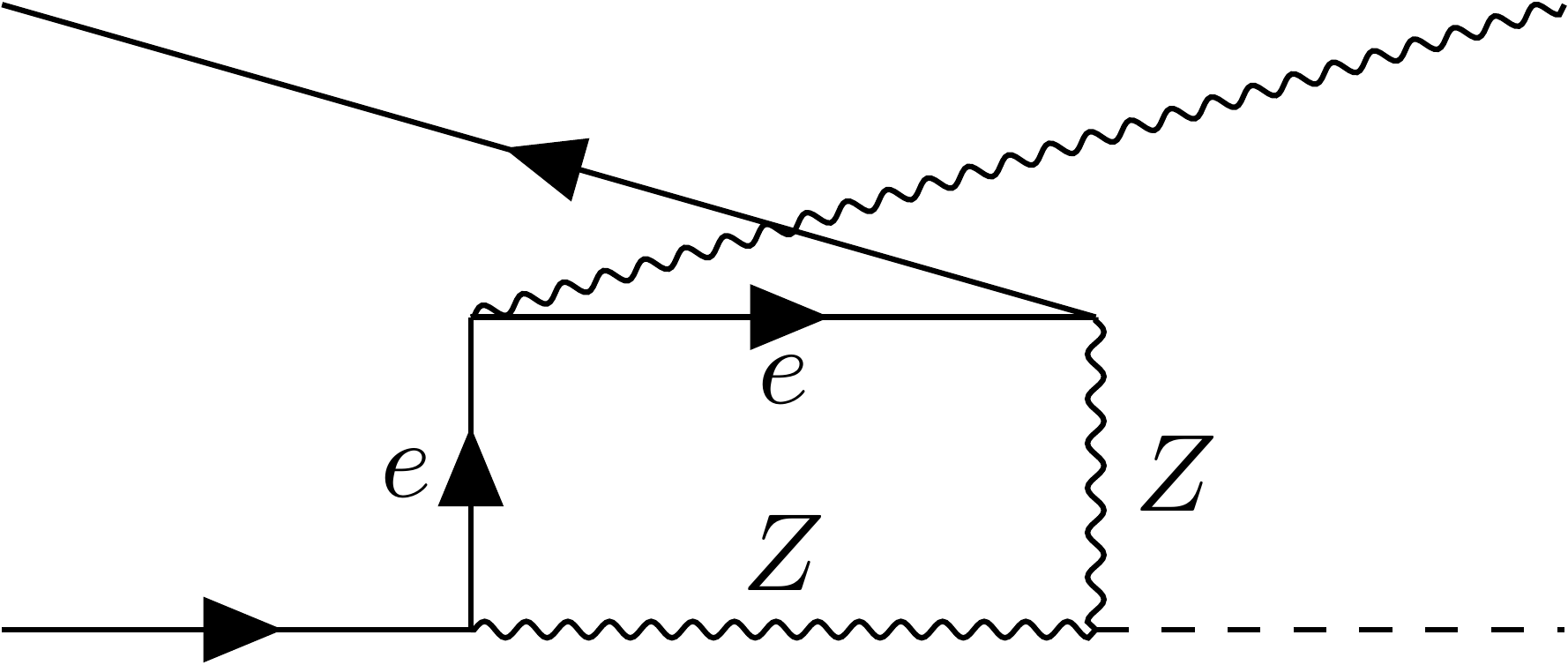}
\subcaption*{(6)}
\end{minipage}
\label{fig: Box_diagrams}
\caption{Box diagrams for $e^{+}e^{-}\to h Z$.}
\end{figure}

%